\documentclass{article}
\usepackage[margin=1in]{geometry}

\usepackage[title]{appendix}
\usepackage{enumitem}
\usepackage{titling}
\usepackage{amssymb}
\usepackage{amsmath}
\usepackage{graphicx}
\usepackage{gensymb}
\usepackage{url}
\usepackage{bm}
\usepackage[parfill]{parskip}
\usepackage{pdflscape}
\usepackage{soul}
\usepackage{hyperref}
\usepackage{fvextra}
\usepackage{tcolorbox}
\usepackage{authblk}
\usepackage{algorithm,algorithmic}% 
\usepackage[square,numbers]{natbib}
\newenvironment{code}%
 {\VerbatimEnvironment
  \begin{tcolorbox}[colback=gray, boxsep=0pt, arc=0pt, boxrule=0pt]
  \begin{Verbatim}[fontsize=\scriptsize, commandchars=\\\{\},
    breaklines, breakafter=*, breaksymbolsep=0.5em,
    breakaftersymbolpre={\,\tiny\ensuremath{\rfloor}}]}%
 {\end{Verbatim}%
  \end{tcolorbox}}

\title{Empirical Validation of a New Data Product from the Interstellar Boundary Explorer Satellite}

\author[1,*]{Kelly R. Moran}
\author[1]{Dave Osthus}
\author[1]{Brian P. Weaver}
\author[1]{Lauren J. Beesley}
\author[1]{Madeline A. Stricklin}
\author[2]{Paul H. Janzen}
\author[3]{Daniel B. Reisenfeld}
\affil[1]{Statistical Sciences Group, Los Alamos National Laboratory, Los Alamos, New Mexico, USA}
\affil[2]{Department of Physics and Astronomy, University of Montana, Missoula, Montana, USA}
\affil[3]{Space Science and Applications Group, Los Alamos National Laboratory, Los Alamos, New Mexico, USA}
\affil[*]{Corresponding author: Kelly R. Moran, krmoran@lanl.gov}
\date{}

\begin{document}

\maketitle

\abstract{Since 2008, the Interstellar Boundary Explorer (IBEX) satellite has been gathering data on heliospheric energetic neutral atoms (ENAs) while being exposed to various sources of background noise, such as cosmic rays and solar energetic particles. 
The IBEX mission initially released only a qualified triple-coincidence (qABC) data product, which was designed to provide observations of ENAs free of background contamination.
Further measurements revealed that the qABC data was in fact susceptible to contamination, having relatively low ENA counts and high background rates. 
Recently, the mission team considered releasing a certain qualified double-coincidence (qBC) data product, which has roughly twice the detection rate of the qABC data product.
This paper presents a simulation-based validation of the new qBC data product against the already-released qABC data product. 
The results show that the qBCs can plausibly be said to share the same signal rate as the qABCs up to an average absolute deviation of 3.6\%.
Visual diagnostics at an orbit, map, and full mission level provide additional confirmation of signal rate coherence across data products.
These approaches are generalizable to other scenarios in which one wishes to test whether multiple observations could plausibly be generated by some underlying shared signal.
}

%%--------------------------------------------------------------------------------------------------
%%--------------------------------------------------------------------------------------------------

\section{Introduction}\label{sec:intro}

%\hl{[Dan's comment: ``The intro should be written from the perspective of how the problem of confirming that two measurements of the same source is a problem of interest in space science in general, and that as a "case study" we're using the IBEX data to demonstrate how one goes about determining if two data sets area drawn from the same parent.  I think Paul and I can provide that.  We'll have to think about other situations where this might come up in space science.'']}

%\hl{A later comment addressing the additional-situations topic: ``Paul and I tried to think of other situations where one might want to combine data sets dominated by poisson statistics. Examples include combining events observed by spacecraft constellations (i.e., multiple spacecraft flying in close formation) hosting the same instruments (an example could be the upcoming Helioswarm mission), or instruments designed to observe the photon `helioglow' emanating from the helioshere. On IMAP: IMAP-Hi will have 4 types of doubles, 3 types of triples and 1 quad-coincidence data product.  We will want to combine all of these!  Also we will want to combine data from IMAP-Hi and IMAP-Lo, and IMAP-Hi and IMAP-Ultra.  Also, we will want to test if we can tell two dataset APART! Are two sets of observations statistically different??  Much of this should be put in the introduction.''}

In space science there are occasions when it is useful to test whether the data collected by two different instruments, or by different subsystems of the same instrument, represent statistically equivalent samples of the same source population.  This is particularly important in situations where the signal-to-noise or signal-to-background level is low, such that it is not straightforward by simple analysis to tell whether the separate observations are equivalent or not.  The situation we consider here is that of two separate but concurrent data sets collected by the IBEX-Hi energetic neutral atom (ENA) imager \citep{funsten2009interstellar} on board the NASA Interstellar Boundary Explorer (IBEX) \citep{mccomas2009interstellar}, which studies the outer heliosphere through the remote sensing of ENAs.  Recently, we prepared and are in the process of releasing a new ENA data product that complements the standard IBEX-Hi ENA data product.  The intent is to combine these two data products into a single data set with higher statistical accuracy than the currently available data set, but this is only reasonable if we can demonstrate that the two data sets are the same within their own intrinsic statistical and systematic uncertainties.  This is non-trivial to show because the ENA signal that IBEX measures is very weak, and it is collected in the presence of large backgrounds.  

Launched in 2008, IBEX is an ongoing mission to study the outer regions of the heliosphere, in particular, the heliosheath, which is the region where the solar wind interfaces with the plasma of the interstellar medium.  In this region, the solar wind is deflected from its outward radial trajectory from the Sun as the interstellar magnetic field exerts an inward pressure on the heliospheric magnetic field. Through charge exchange interactions with cold interstellar neutrals, the deflected solar wind is partially neutralized, and a portion of these neutrals, or ENAs, are able to travel back Sunward.  The Earth-orbiting IBEX spacecraft is designed to detect these ENAs as a function of arrival direction and energy, and using this data, all-sky maps of the ENA distribution can be made.  IBEX consists of two ENA imagers, IBEX-Lo, which measures ENAs in the energy range of $\sim10$ eV to $\sim2$ keV \citep{fuselier2009interstellar}, and IBEX-Hi, which measures hydrogen ENAs (originally solar wind protons) in the energy range of $\sim500$ eV to $\sim6$ keV \citep{funsten2009interstellar}.  Here, we consider only the data collected by the IBEX-Hi instrument. IBEX-Hi detects ENAs through use of a triple-coincidence detection method; that is, in the IBEX-Hi detector section, an incoming ENA can trigger three separate detectors within a narrow time window.  Specifically, if all three are triggered within 100 ns, an event is registered (see the next section for more details on the data collection method).  Although IBEX-Hi is the most sensitive ENA instrument flown to date, because the signal it is detecting is so weak, the typical triple-coincidence ENA count rate is of order 1 count every 20 seconds.  Prior to launch, it was expected that even though the signal rate was quite low, the triple-coincidence technique would effectively eliminate any background events.  Once in orbit, however, it was found that penetrating radiation was quite effective at producing triple-coincidence events; in fact the penetrating background rate is comparable to the heliospheric ENA rate, making it challenging to extract quality signal. 

Recently, we have demonstrated that a certain subset of the \textit{double-coincidence} events (where only two of the three detectors are triggered) has twice the signal rate as the standard triple-coincidence product, but an only slightly worse signal-to-background ratio.  This being such a signal-starved mission, we became motivated to develop the double-coincidence event set into a new IBEX data product, which when combined with the triple-coincidence product, can triple the total ENA signal rate, along with a significant improvement in the signal-to-background ratio.  Combining them, however, is not necessarily straightforward. Even though both event types are measured within the same detection system, they have different detection efficiencies, different background levels, and potentially different background sources.  Furthermore, the new data product was not well characterized prior to launch during ground calibration, nor has it undergone a decade's worth of scrutiny and validation as has the workhorse triple-coincidence product.  Thus, before combining the two data sets, it is necessary to carefully test whether the new product "gives the same answer" as the standard one; that is, to within error, are they measuring the same signal?  This paper lays out a statistical testing methodology for answering this question, not only showing that they are statistically equivalent to within the anticipated level of error, but also quantifying the degree to which they differ from one another.      

Before diving into the details of the IBEX data and the validation methodology, we describe other situations where this method can be applied. Of most similarity to IBEX, the upcoming Interstellar Mapping and Acceleration Probe (IMAP) mission \citep{mccomas2018interstellar} will host three advanced ENA imaging systems, IMAP-Lo, IMAP-Hi (the successors to IBEX-Lo and IBEX-Hi, respectively), and IMAP-Ultra (a near copy of the JENI instrument on the European JUICE mission to Jupiter \citep{mitchell2016energetic}).  As with IBEX-Hi, IMAP-Hi measures heliospheric ENAs using a coincidence method, but now via four detection channels, allowing up to quadruple-coincidence detection.  In fact there will be multiple double- and triple-coincidence combinations, that along with the quadruple-coincidence events, are to be combined into a composite signal rate.  As with IBEX-Hi, each of these event types will have different detection efficiencies and background rates, and it will be necessary to demonstrate that the individual coincidence combinations are statistically equivalent in order to validate the composite data product.  Furthermore, it is the intent to combine the signals from IMAP-Lo and IMAP-Ultra with IMAP-Hi where their energy ranges overlap. Again, it will be necessary to validate that the signal rates from the separate instruments are statistically equivalent before they are combined. 

Beyond the specific cases of IBEX and IMAP, there are other domains in which the method presented here can be of great utility.  In the arena of solar energetic particle (SEP) observations there are often situations where one would like to know if two data sets are statistically equivalent observations of the same source population.   For example, many spacecraft have energetic particle sensors with overlapping energy ranges.  In general, to measure $\sim$ keV to few hundred keV particles requires instruments that use vastly different detection techniques than those that measure in the 1 MeV to 100 MeV range, and it is often a challenge to reconcile the measurements in the middle of the spectrum where the energy ranges overlap. The challenges of this sort of comparison are discussed with regard to the EPI-Lo and EPI-Hi energetic particle sensors on Parker Solar Probe \citep{joyce2020energetic}. This situation arises with many low- and high-energy pairs of energetic particle instruments, such as ACE/SIS and ACE/CRIS, STEREO/LET and STEREO/HET, and even CRS and LECP on Voyager.

The method presented here can not only be used to test if two data sets provide measurements of the same underlying signal, but it can also be used to rigorously quantify the degree to which the measurements differ.  For example, solar energetic particle instruments often have multiple heads looking in different directions. There are situations in which one desires to evaluate the degree of anisotropy of a given SEP event, but such an analysis is complicated by low counting statistics and overlapping fields of view.  An extreme example of this is the EPI-Lo instrument which has 80 independent telescopes oriented in different directions \citep{mccomas2016integrated}.  Additionally, there are times when the Sun becomes very active and multiple events are coming off the Sun at the same time. In such instances, it can be difficult to tell whether different spacecraft are seeing the same event or if they are seeing different events, but at the same time.   This is yet another case in which the methodology presented here can be applied.

The goal for this paper is to develop and implement a validation protocol for the IBEX-Hi double-coincidence data that can be used to examine the assumption that the triple- and double-coincidence data products share an underlying common ENA signal rate. 
We will show both visual and quantitative metrics by which this assessment can be done at different time scales, from days to the full mission collection period.
We will also discuss how to learn margins of error about model components, e.g., about the double-coincidence background rate, and how to flag certain collection periods as likely deviating from the assumed shared-signal model.
The rest of this paper proceeds as follows: in Section \ref{sec:collection} we briefly describe how the IBEX-Hi instrument operates and collects data; in Section \ref{sec:methods} we show how to validate the double-coincidence data product against the triple-coincidence data product; in Section \ref{sec:res_hypothesis} we show that the double-coincidence data product has been validated against the triple-coincidence data product to within acceptable limits; finally, we discuss these results and their implications in Section \ref{sec:discussion}.
These methods can be applied more generally to problems of determining if there is a shared underlying common signal across multiple different datasets.

%%--------------------------------------------------------------------------------------------------
%%--------------------------------------------------------------------------------------------------

\section{IBEX Data Collection}\label{sec:collection}

%\hl{Dan's comment: After the intro, we should have a section decribing how IBEX collects data.  Again, I or Paul can draft that.}

%\hl{Kelly's comment: Paul mentioned perhaps trying to give the minimal amount of necessary data collection info here, since it's meant to be broadly applicable beyond IBEX. Maybe we could just include this as a paragraph in the intro?}

\begin{figure}[htpb]
    \centering
    \includegraphics[width=0.7\textwidth]{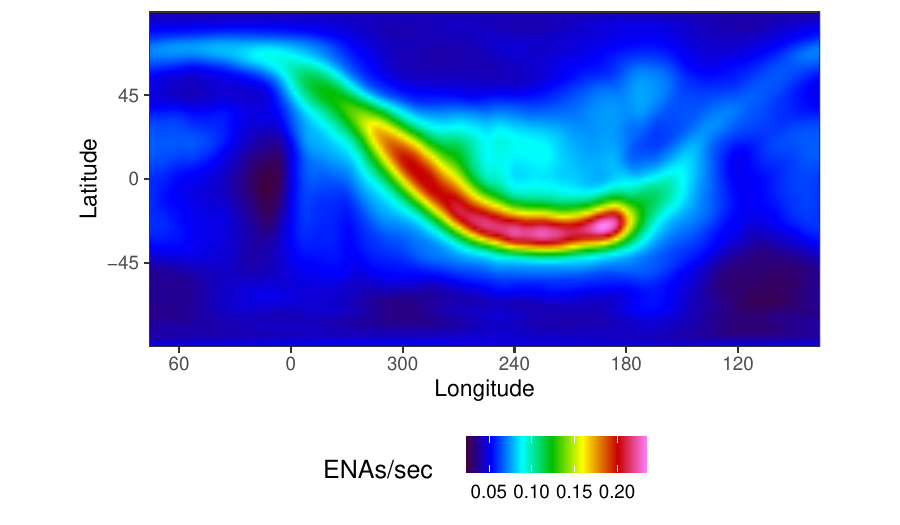}
    \caption{An example sky map (2010A ESA 3) created by applying the Theseus methodology (a newly developed sky map generation procedure) to the $1\degree$-binned triple-coincidence data product \citep{osthus2022towards}. One can see a curved structure moving across the map starting faintly at longitude and latitude (85, 70) and moving down towards (240, -30) before curving up and right again, which is the well-know IBEX ribbon, a region of enhanced ENA emission that encircles the heliosphere. }
    \label{fig:theseus_map}
\end{figure}

IBEX is a Sun-pointing spinning spacecraft ($\sim4$ rpm) that orbits the Earth with a period of 9.1 days.  The IBEX-Hi instrument is a single pixel ENA imager \citep{funsten2009interstellar} that views perpendicular to the spin axis with a field-of-view of $6.5\degree$ full-width at half-maximum. Twice per orbit, at perigee and apogee, IBEX repoints to maintain Sun pointing. Thus, over the course of $\sim 4.5$ days IBEX-Hi collects ENA events from a single $360\degree$ slice of the sky that is $\sim 6.5\degree$ wide.  The data set collected over the course of a single pointing direction is designated by orbit number and by the orbit segment (`a' for the ascending arc, and `b' for the descending arc); for example, the data collected in IBEX's 235th orbit during the ascending arc is labeled `Orbit 235a'.  Over the course of 6 months, IBEX-Hi views the entire sky as a series of $6.5\degree$-wide slices arranged by ecliptic longitude.  These slices are then assembled into a complete all-sky map of ENA emission. Each map is designated by year and whether it was collected during the first six months (`A' maps) or second six months (`B' map) of the year; for example, the map from the first half of 2010 is labeled `2010A'; an all-sky ENA map for 2010A is shown in Figure \ref{fig:theseus_map}. 

\begin{figure}[htpb]
    \centering
    \includegraphics[width=0.5\textwidth]{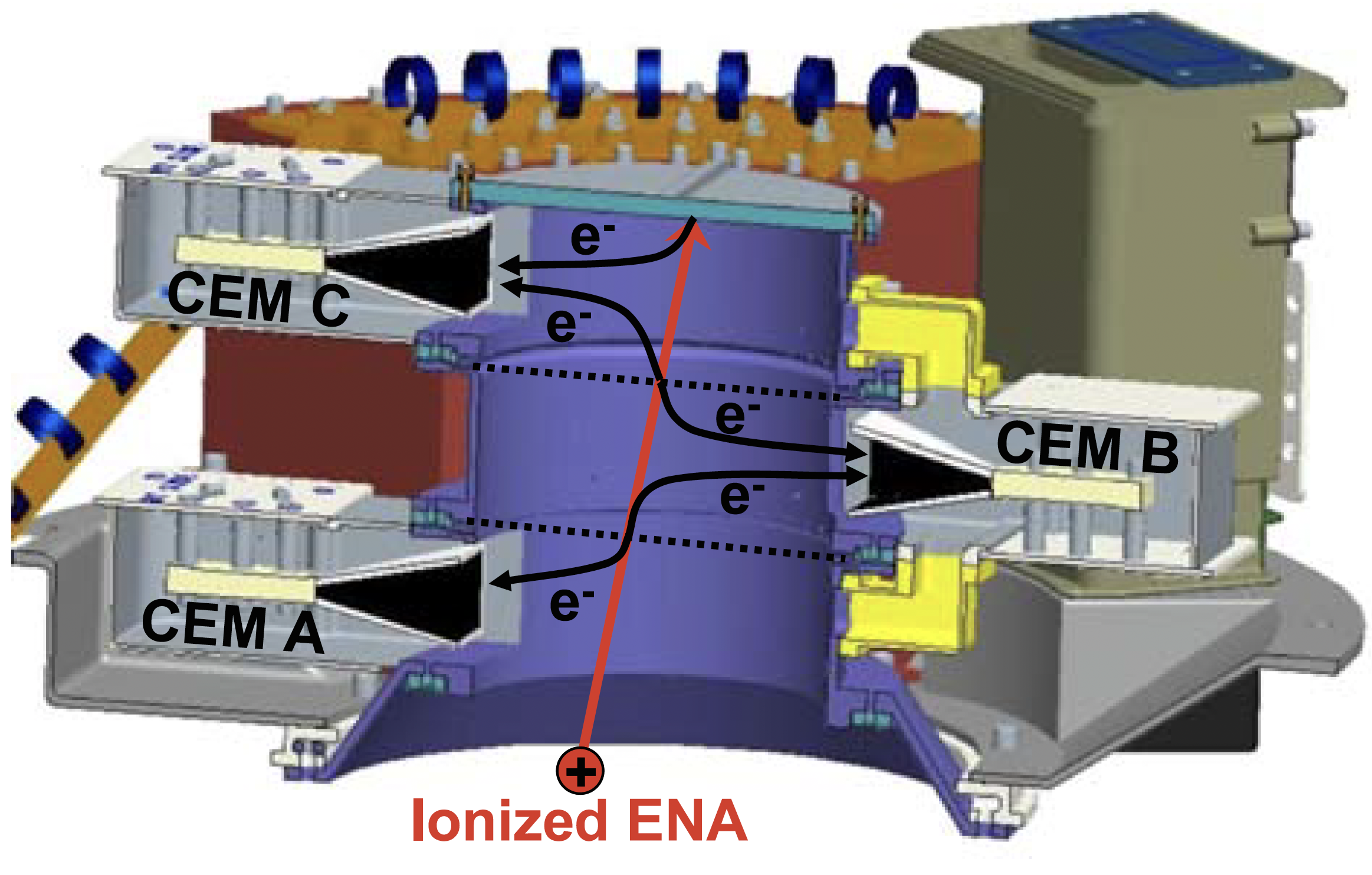}
    \caption{An illustration of the IBEX-Hi ENA detector subsystem (from \citep{funsten2009interstellar}). After energy selection by the ESA (not shown),  ionized ENAs enter the detector section and pass through two carbon foils (dotted lines) that generate secondary electrons (e$^-$) at the foils or the detector rear surface. These secondary electrons accelerated into channel electron multipiers (CEMs), enabling double (two chamber) or triple (three chamber) coincidence detection of a single ENA.}
    \label{fig:instrument_fig}
\end{figure}

A full description of the IBEX-Hi imager is given in \citep{funsten2009interstellar}; here, we briefly describe how IBEX-Hi detects hydrogen ENAs with sufficient detail to understand how the instrument detects the two types of ENA events under consideration in this study.  ENAs enter the instrument through a large 150 cm$^2$ collection aperture, and for the purpose of measuring the energy of the particles, they are ionized by passing through a thin carbon foil and then enter an electrostatic analyzer (ESA).  The ESA is sequentially tuned to allow passage of ionized ENAs in six broad energy passbands ($\Delta E/E \approx 0.65$) that cover the range $\sim0.5$ to 6 keV.  Generally only ESA steps 2 through 6, the passbands centered on 0.71, 1.11, 1.74, 2.73, and 4.29 keV, respectively, are used for scientific analysis.  After passage through the ESA, an ionized ENA enters a detector section composed of three stacked chambers (A, B, and C) separated by a pair of thin carbon foils (see Figure \ref{fig:instrument_fig}).  An incoming ion passes through the foils and strikes the rear of the detector section, triggering the release of secondary electrons along the way. These electrons are then guided into channel electron multipliers (CEMs) which have some probability of triggering a detection. Each chamber has its own CEM, and thus the CEMs are likewise labeled A, B and C.  If all three CEMs are triggered ($P \sim 0.07)$ this is referred to as an ABC triple-coincidence event.  Since it is nearly certain that an incoming ion will trigger CEM A or B before CEM C, only these events are counted as ENA events, and are referred to as qualified-ABCs (qABC) events. If CEM C is triggered first, it is most likely due to penetrating radiation, and is therefore rejected by the qualification scheme.  The qABCs have been the principle event type used for heliospheric science throughout the IBEX mission.  

More often, only two of the three CEMs are triggered, and these are referred to as double-coincidence events.  In general, double coincidences are significantly more susceptible to backgrounds than the qABCs, and so, previously, they have not been used for science. However, it was recently discovered that one double-coincidence event type, a qualified-BC (qBC), where CEM B is triggered at least 3 ns before CEM C, has twice the detection probability of a qABC ($P \sim 0.14$) and only a slightly higher background rate.  It is the qBC event type that we wish to use as a new data product and that we must validate against the qABCs. 

Whenever a qABC or qBC detection occurs, a ``direct event" datum is created and stored. A direct event includes the event type, the time the event occurred, the direction IBEX's spin axis was pointing, the look direction (spin phase angle) of IBEX-Hi's entrance aperture at the instant of detection, and the ESA setting. We will be working with a curated data product called the ``binned direct event data'', which are created via a standard binning process in which the direct events are grouped into 360 $1\degree$-look direction bins and five ESA steps\footnote{Note the currently released mission data product uses $6\degree$- rather than $1\degree$-look direction bins, but the binning process used is otherwise the same as that used here.}.

Not all direct events are heliospheric ENAs. 
Potential sources of background direct events include high-energy penetrating radiation particles, non-heliospheric ENAs (from the Earth or Moon), or other particles local to the spacecraft environment that can trigger coincidence events. For some types of background sources, the contribution to the total event rate can be independently determined from other coincidence combinations, in which case the background rate can be subtracted from the total rate further down the processing chain.  For other background sources, either no method has been found to independently determine their rate, or their rate is so high it completely overwhelms the signal. In such cases, those periods are simply excluded.  The exposure time for a binned direct event datum refers to the total duration of the time segments that are kept after the culling process.

To summarize, a binned direct event datum refers to a specific $1\degree$ bin, ESA setting, and chamber combination (here, qABC or qBC) for a given orbit. 
It includes information on the IBEX-Hi look direction, the exposure time, the number of measured direct events, an efficiency factor for how well the instrument can detect signal, and a background rate.
For a qABC binned direct event datum, the background rate is considered by the IBEX community to be validated.
A qBC binned direct event datum has its own background rate that is not yet considered validated and may potentially be biased, particularly at low ESA settings.
Although technically estimated with small uncertainties, the efficiency factors and background rates are treated as fixed in this paper.
Throughout the paper, the phrase ``qABCs'' and ``qBCs'' will be used loosely to mean a collection of qABC or qBC binned direct events (e.g., ``the 2012A qABCs'' refers to the qABC events from the first 6 months of 2012).

\begin{figure}[htpb]
    \centering
    \includegraphics[width=0.24\textwidth]{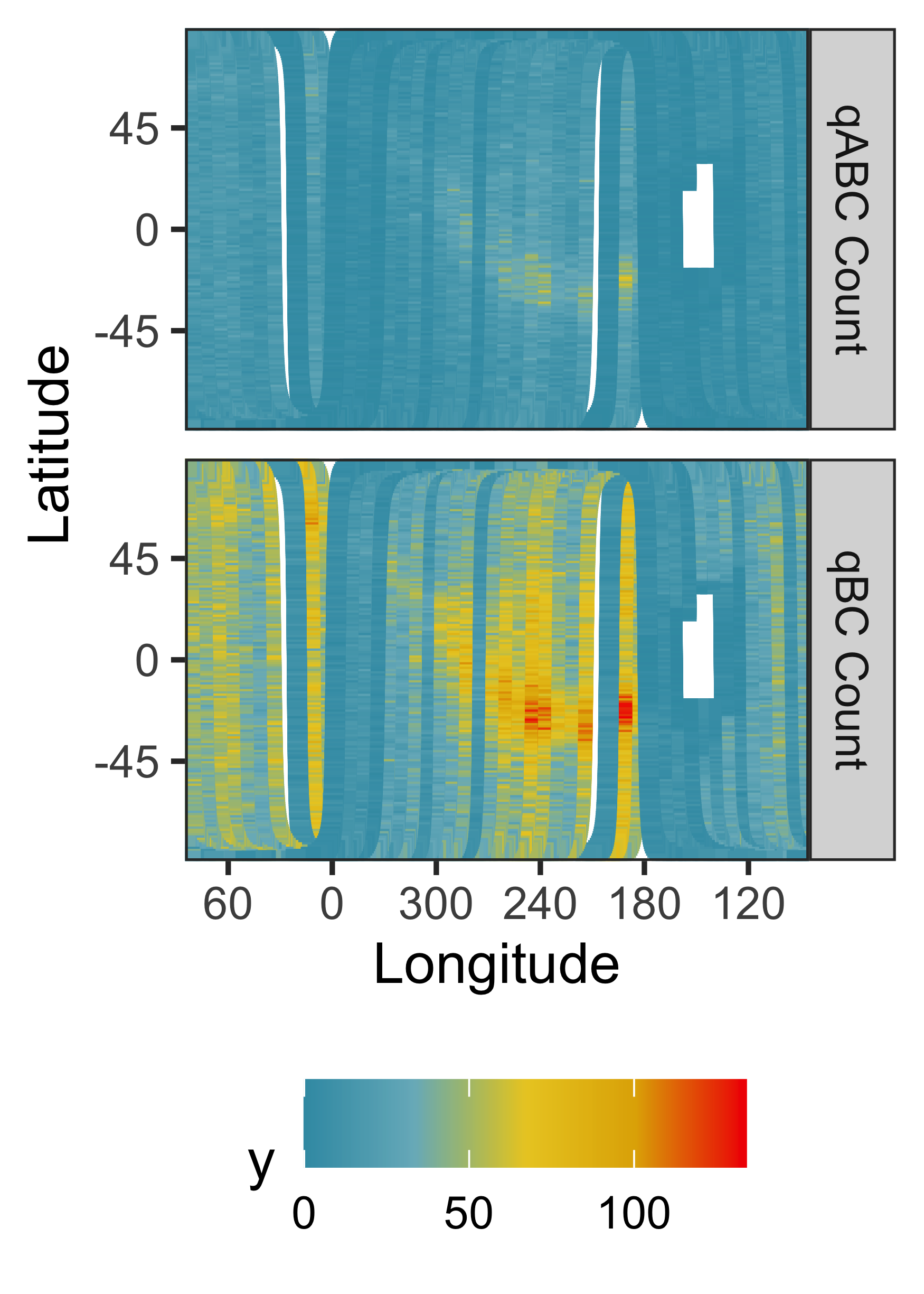}
    \includegraphics[width=0.24\textwidth]{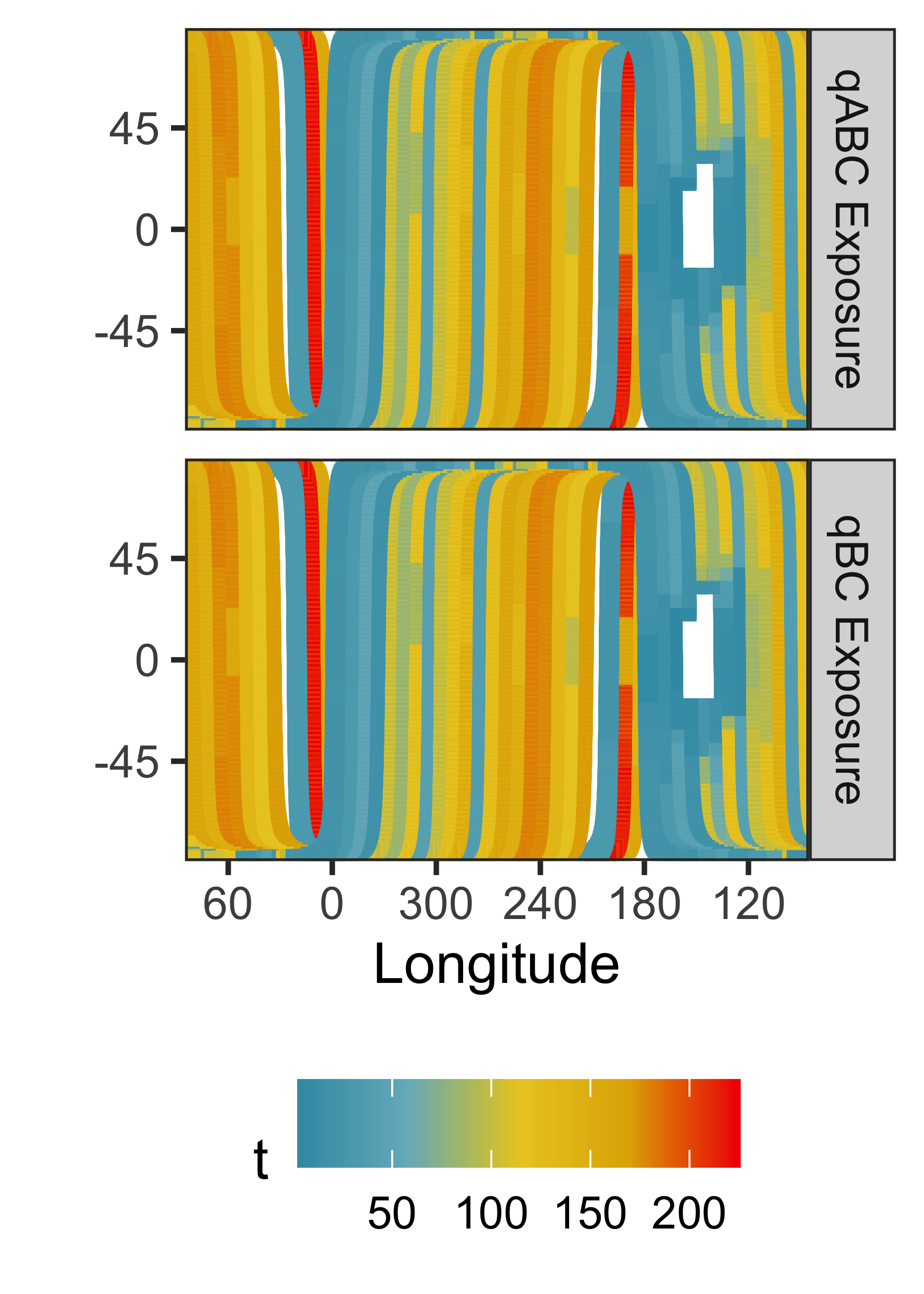}
    \includegraphics[width=0.24\textwidth]{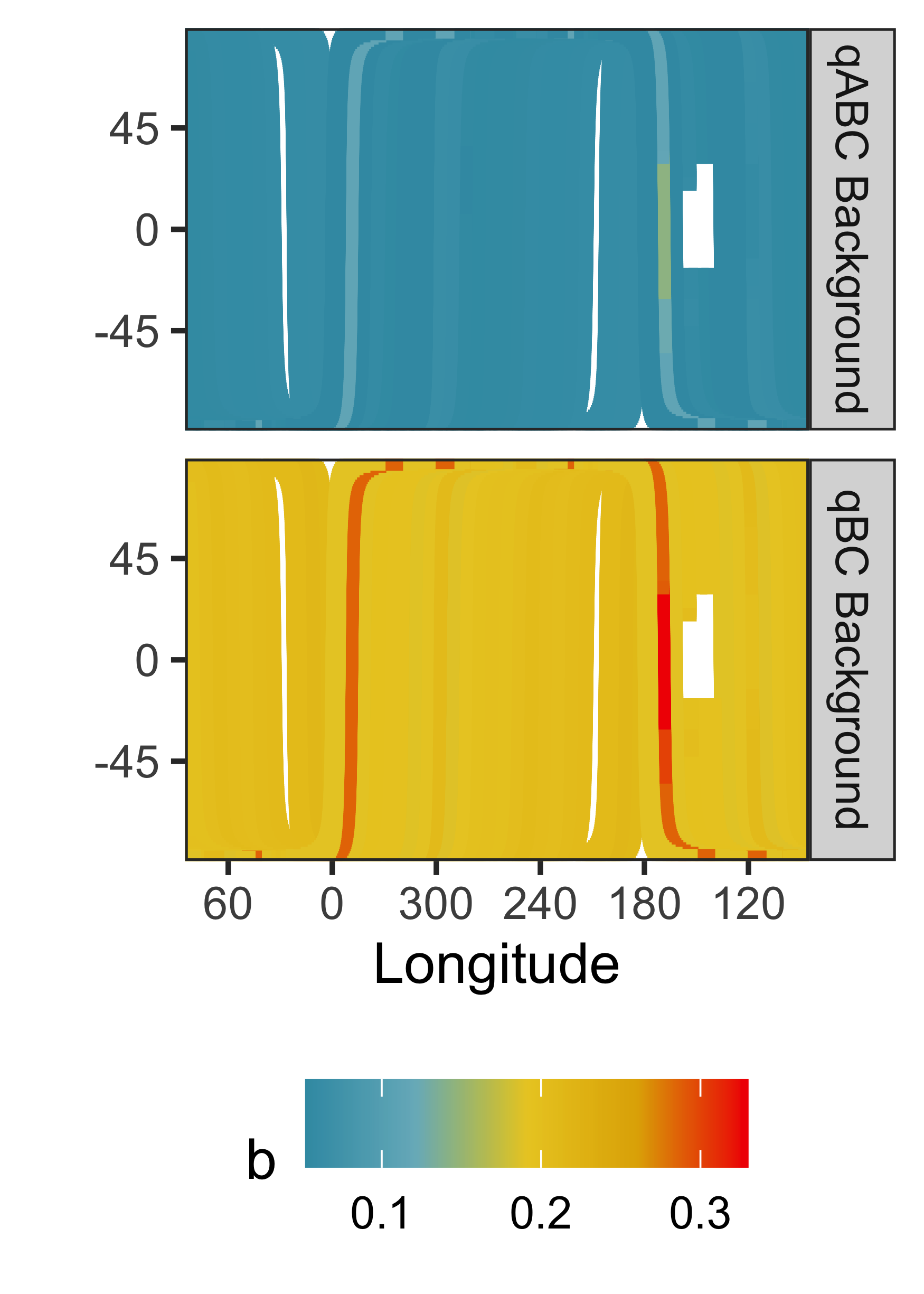}
    \includegraphics[width=0.24\textwidth]{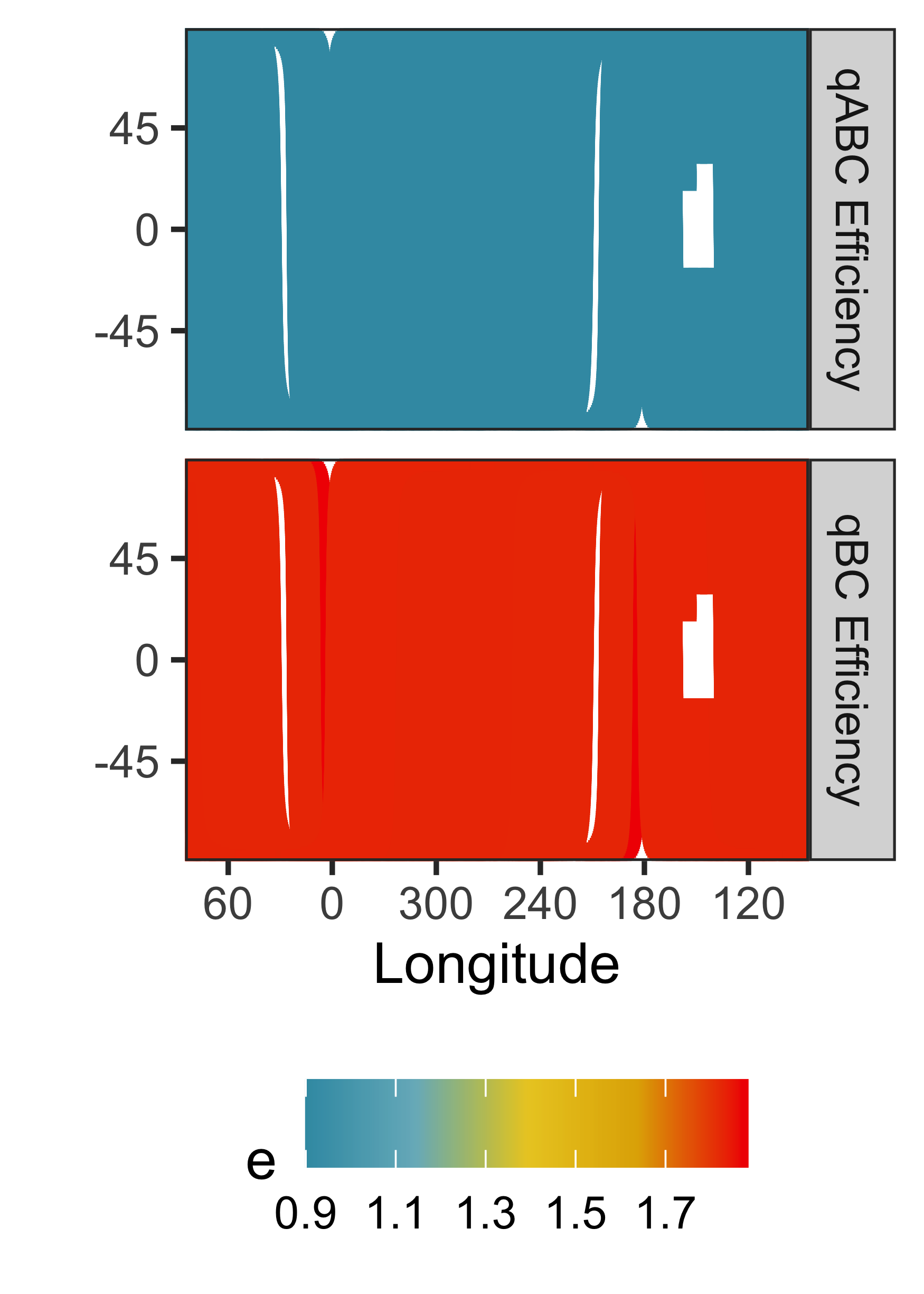}
    \caption{From left to right, the 2010A ENA counts (denoted $y$), exposure times ($t$), background rates ($b$), and efficiency factors ($e$) for the qABCs (top) and qBCs (bottom) for ESA 3. Recall a point here is a binned direct event datum, which potentially includes many direct events; the number of direct events in one binned direct event datum is the value $y$ in the left-most subplot.}
    \label{fig:eda}
\end{figure}

As mentioned above, the qBCs are characterized by higher total ENA counts and efficiency factors, but also higher background rates.
The qABCs and qBCs from 2010A at the ESA 3 setting are shown in Figure \ref{fig:eda}, which displays the number of direct events, the exposure time, background rate, and efficiency factor for each data type.
For 2010A, the ENA direct event counts range from 0 to 67 for the qABCs and from 0 to 134 for the qBCs per $1\degree$ bin, with the qBCs having 2.8 times more counts than the qABCs.
The counts panel depicts a coarse ENA sky map; one can make out a rough version of the ribbon, which was seen more clearly in Figure \ref{fig:theseus_map}. 
The exposure time varies depending on the look direction, here varying between 2.4 and 225.5 seconds, with values being the same for the qABCs and qBCs.
The number of direct events is positively correlated with exposure time; larger exposure times result in a higher number of recorded direct events.
Background rates are typically common across all bins for a given ESA and arc (i.e., half-orbit); here the background rate varies between 0.05 and 0.14 background particles per second for the qABCs, and between 0.19 and 0.33 for the qBCs.
On occasion, data gaps appear in the panels, either because data was lost due to spacecraft issues, or because large local background sources were present, making it impossible to extract a statistically useful ENA signal for an entire arc.

%%--------------------------------------------------------------------------------------------------
%%--------------------------------------------------------------------------------------------------

\section{Methods}\label{sec:methods}

\begin{figure}[htpb]
	\centering
	\includegraphics[width=0.95\textwidth]{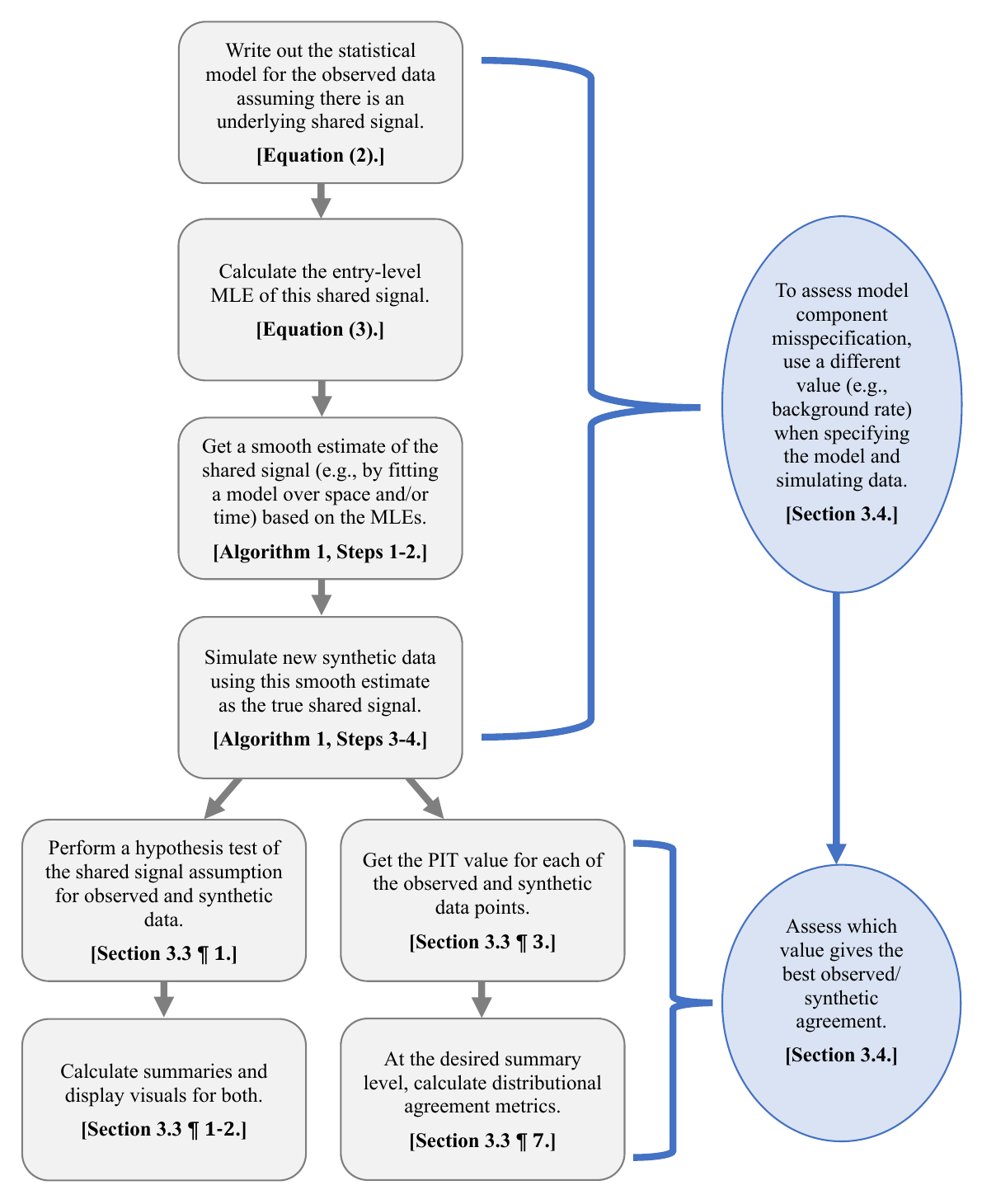}
	\caption{\small Summary of the full validation process, step by step. The statements describe the general step, while the bold parentheticals at the bottom of each step provide the  location in the main text of that step for the IBEX application.} \label{fig:flowchart}
\end{figure}

The broad question we are trying to answer is ``Could a shared set of underlying ENA signal rates have plausibly generated our sets of observed qABC and qBC binned direct events?''
%We make minimal assumptions about spatial or temporal relationships in the underlying ENA signal rate when answering this question because we want our findings to be agnostic to the specifics of what downstream models may be built using these data.
To answer this question, we compare the observed data to a synthetic reference data set for which there \textit{are} known shared signal rate maps, the generation of which is discussed below. 
Tools by which we diagnose the alignment between our assumed shared-signal model and reality, discussed in more detail in the following sections, include: visualizations of when, where, and how often the null hypothesis of a shared signal is rejected, the distribution of probability integral transform values associated with the null hypothesis, and an orbit-level metric for measuring and adjusting potential error in the qBC model components assuming the qABC background is specified correctly. A flow chart summarizing the entire validation process, which can be used in applications beyond this one, is shown in Figure \ref{fig:flowchart}.
We recommend that the reader not familiar with the methods mentioned above stop here and read the hypothesis testing and probability integral transform primers in Sections A and B of the supplemental materials before continuing through the rest of this section.

\subsection{Model Formulation}\label{sec:methods_model}

Consider $\mathbf{y}_i = (y_i^{abc}, y_i^{bc})$, an observation of a qABC and qBC binned direct event pair from a given IBEX-Hi look direction, ESA setting, and orbit; $i$ goes from $1$ to $N$, with $N$ being the total number of binned direct event data points in the IBEX-Hi data set. We assume independent Poisson distributions for each of $y_i^{abc}$ and $y_i^{bc}$. For $* \in \{abc,bc\}$ let $t_i^{*}$ denote exposure times, $s_i^{*}$ signal rates, $b_i^{*}$ background rates, and let $e_i^{*}$ denote the signal efficiency rate.
Mission scientists provide a standard deviation about each $b_i^{abc}, b_i^{bc}$, and $e_i^{bc}$; we treat the backgrounds and qBC efficiency factors as known during calculations and use these standard deviation as a reference against which to assess whether apparent model deviations are within the bounds of what we would expect based on these given error bounds (see supplemental materials Section C). Intuitively, larger values of $t_i^*$, $s_i^*$, or $b_i^*$ (longer exposure time, higher signal rate or background rate)  result in larger expected values and variance of $y_i^*$ (more expected ENA counts, and more variation thereof). Specifically, we model $y_i^{abc}$ and $y_i^{bc}$ as: 
\begin{equation}\label{eq:mod}
\begin{split}
&y_i^{abc} \sim \text{Poisson}(t_i^{abc}(e_i^{abc} s_i^{abc} + b_i^{abc})), \\
&y_i^{bc} \sim \text{Poisson}(t_i^{bc}(e_i^{bc} s_i^{bc} + b_i^{bc})).
\end{split}
\end{equation}
The maximum likelihood estimate (MLE) for the qABC signal rate $s_i^{abc}$, denoted by $\hat{s}_i^{abc}$, is $\max(0, (y_i^{abc}/t_i^{abc} - b_i^{abc})/e_i^{abc})$. Similarly, the MLE for the qBC signal rate $s_i^{bc}$, denoted by $\hat{s}_i^{bc}$, is $\max(0, (y_i^{bc}/t_i^{bc} - b_i^{bc})/e_i^{bc})$. 
Intuitively, the observed counts divided by exposure time minus the background rate gives us an uncorrected estimate of the ENA rate attributable to the signal per unit time. 
Dividing this number by the efficiency factor gives us the corrected signal rate estimate. 
Finally, because the signal rate cannot be negative, our MLE is the maximum of 0 and that corrected signal rate estimate. Figure \ref{fig:validation_shats} shows the $\hat{s}_i^{abc}$ for the 2010A map, the MLEs for the qABC data, in the left column. The $\hat{s}_i^{bc}$ for the 2010A map, the MLEs for the qBC data, are shown in the right column. 

Mathematically the MLE of the qABC or qBC signal can be found by maximizing its associated likelihood (equivalently, its log likelihood) with respect to $s$, with details following.\footnote{Note we drop sub- and superscripts for notational convenience.} The log likelihood $\ell(s)$ of the Poisson distribution having mean $t(e s + b)$ is 
$$\ell(s) = y \ln(t(e s + b)) - t(e s + b) - y!.$$
The derivative of $\ell(s)$ is 
$$\frac{d}{ds} \ell(s) = \frac{d}{ds}y \ln(t(e s + b)) - \frac{d}{ds}t(e s + b) - \frac{d}{ds}y! = \frac{y}{t(es+b)}te - te.$$ 
Setting the derivative to 0 allows us to find the points at which local extrema occur; 
$$0 = \frac{y}{t(es+b)}te - te \implies y=t(es+b) \implies s=(y/t-b)/e$$ 
is a critical point of the log likelihood. To assess whether a critical point is a local minimum or maximum, we check whether the second derivative is negative or positive. We have 
$$\frac{d^2}{ds^2} \ell(s) = \frac{d}{ds} \frac{y}{t(es+b)}te - \frac{d}{ds} te = -\frac{ye^2}{(es+b)^2} \leq 0$$ 
everywhere and $<0$ everywhere when $y>0$. Because the second derivative is negative when $y>0$, the log likelihood is concave in $s$ and $(y/t-b)/e$ is a local maximum. When the critical point $s=(y/t-b)/e$ is negative the constrained MLE will be 0, the closest possible value respecting the nonnegativity constraint, i.e., 
$$\hat{s}=\max(0,(y/t-b)/e).$$
When $y=0$, $\ell(s) = 0 \ln(t(e s + b)) - t(e s + b) - 0! = - t(e s + b)$ is maximized by setting $s$ to 0 as well; this is because $s = (0/t-b)/e = -b/e$ is negative and 0 is the closest nonnegative value $s$ can take.

\subsection{Hypothesis Testing and Evaluation}\label{sec:methods_hypothesis}

\begin{figure}[htpb]
	\centering
	\includegraphics[width=0.98\textwidth]{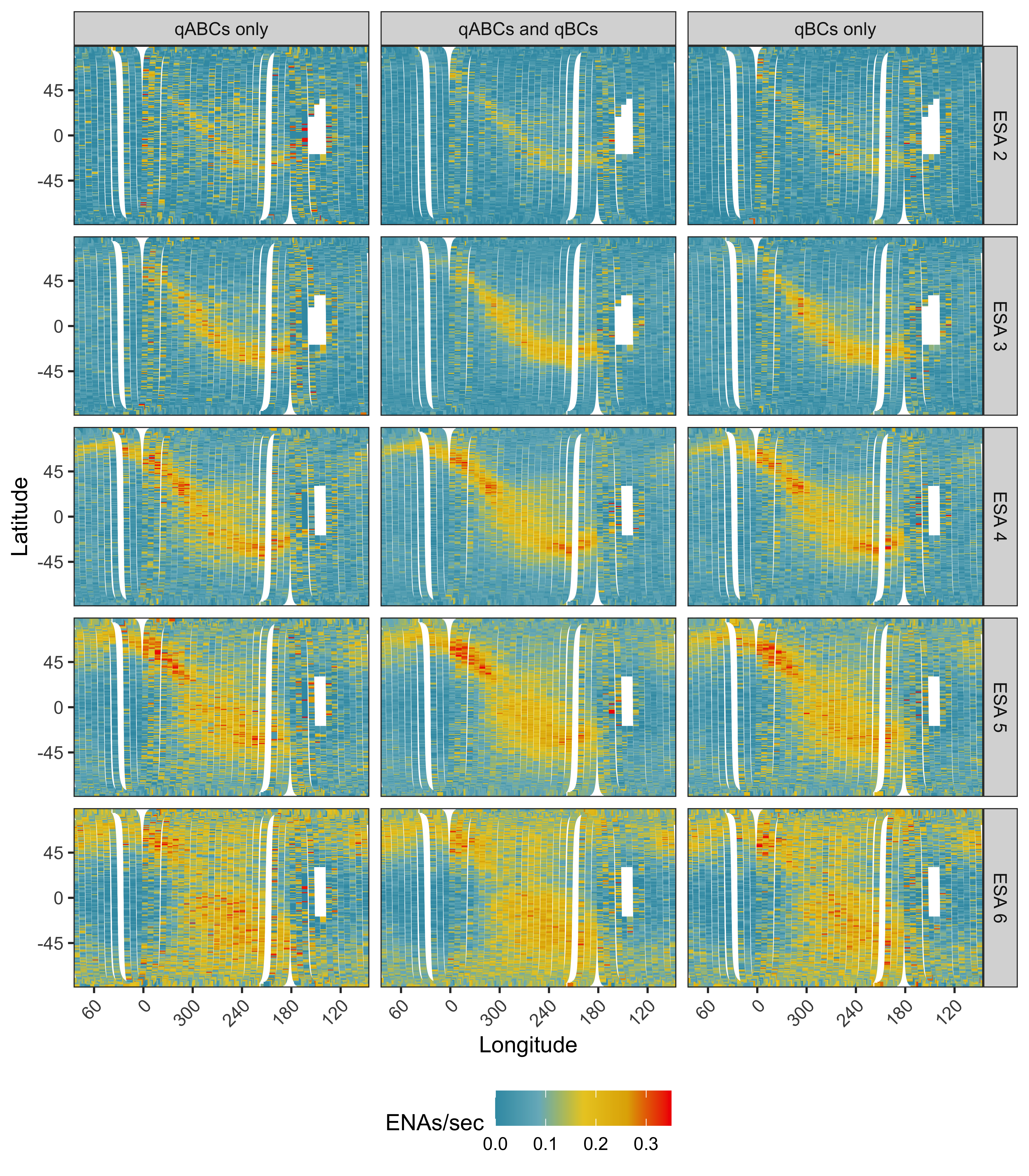}
	\caption{\small Signal rate MLEs at each look direction using only the qABC data ($\hat{s}_i^{abc}$, left), the combined qABC and qBC data ($\hat{s}_i$, middle), and the qBC data only ($\hat{s}_i^{bc}$, right) for ESAs 2 through 6 (in ascending order from top to bottom) for the 2010A orbits. The combined MLE is visually similar to the qABC MLE, but is less noisy. ENA/sec values above 0.35, comprising 0.3\% of look directions and occurring predominantly where there are very low exposure times, are omitted from the visualization.} \label{fig:validation_shats}
\end{figure}

Given the above model formation, we can more precisely rephrase our question for a \textit{single} count pair as: ``Under some shared signal rate $s_i$, could we have plausibly observed $\mathbf{y}_i$, given $e_i^{abc}$, $e_i^{bc}$, $t_i^{abc}$, $t_i^{bc}$, $b_i^{abc}$, and $b_i^{bc}$?'' This can be framed as a hypothesis test, with
\begin{equation}\label{eq:hypos}
\begin{split}
&\text{Null hypothesis } (\text{H}_0): \ s_i^{abc} = s_i^{bc} \\
&\text{Alternative hypothesis } (\text{H}_{\text{A}}): \ s_i^{abc} \neq s_i^{bc}.
\end{split}
\end{equation}
Under the null hypothesis $\text{H}_0$, Equation (\ref{eq:mod}) becomes the null model
\begin{equation}\label{eq:mod_h0}
\begin{split}
&y_i^{abc} \sim \text{Poisson}(t_i^{abc}(e_i^{abc} s_i + b_i^{abc})) \\
&y_i^{bc} \sim \text{Poisson}(t_i^{bc}(e_i^{bc} s_i + b_i^{bc})).
\end{split}
\end{equation}
The MLE for the shared signal rate $s_i$, denoted by $\hat{s}_i$, is 
\begin{equation}\label{eq:s_mle}
\hat{s}_i = \max(0, \tilde{s}_i), \quad \tilde{s}_i = (y_i^{abc}+y_i^{bc}-t_i^{abc}b_i^{abc}-t_i^{bc}b_i^{bc})/(t_i^{abc}e_i^{abc}+t_i^{bc}e_i^{bc}). 
\end{equation}
Figure \ref{fig:validation_shats} shows $\hat{s}_i$ for the 2010A map, the MLEs for the combined data, in the middle column.

The sum of two independent Poisson variables having means $\lambda_1$ and $\lambda_2$ is itself a Poisson with mean $\lambda_1 + \lambda_2$. Therefore the proof for the MLE of $s_i$ using the sum $y_i^{abc}+y_i^{bc}$ as the summary statistic is analogous to that shown in the previous subsection, where the mean of this sum-of-Poissons distribution is $t_i^{abc}(e_i^{abc} s_i + b_i^{abc}) + t_i^{bc}(e_i^{bc} s_i + b_i^{bc})$.\footnote{Specifically, we want to maximize the log likelihood $\ell(s)$ of the Poisson distribution having mean $t^{abc}(e^{abc} s + b^{abc}) + t^{bc}(e^{bc} s + b^{bc})$, which is $(y^{abc} + y^{bc}) \ln(t^{abc}(e^{abc} s + b^{abc}) + t^{bc}(e^{bc} s + b^{bc})) - t^{abc}(e^{abc} s + b^{abc}) - t^{bc}(e^{bc} s + b^{bc}) - (y^{abc} + y^{bc})!$. Its derivative $\frac{d}{ds} \ell(s)$ is $\frac{y^{abc} + y^{bc}}{t^{abc}(e^{abc} s + b^{abc}) + t^{bc}(e^{bc} s + b^{bc})}(t^{abc}e^{abc} + t^{bc}e^{bc}) -(t^{abc}e^{abc} + t^{bc}e^{bc})$. Setting the derivative to 0 allows us to find the points at which local extrema occur; $0 = \frac{d}{ds} \ell(s) \implies y^{abc} + y^{bc} = t^{abc}(e^{abc} s + b^{abc}) + t^{bc}(e^{bc} s + b^{bc}) \implies s = (y^{abc} + y^{bc} - t^{abc}b^{abc}-t^{bc}b^{bc}) / ( t^{abc}e^{abc}+t^{bc}e^{bc}) $ is a critical point of the log likelihood. Assessing the sign of the second derivative to ensure the critical value is a maximum then proceeds as shown in the previous subsection.}
Throughout the rest of the paper we let $\hat{\lambda}_i^* = t_i^{*}(e_i^{*} \hat{s}_i + b_i^{*})$ for $* \in \{abc,bc\}$ denote the expected value of the counts under $\text{H}_0$ using the MLE $\hat{s}_i$ in place of the true $s_i$. We provide an in depth hypothesis testing primer and additional details of our hypothesis testing framework in Section A of the supplemental materials.

\subsection{Visual and quantitative diagnostics}

We formally test each pair of points, i.e., designate whether the null hypothesis is rejected for each.
We use these binary test results in visual checks for orbit- map-, or ESA-specific bias.
To perform the tests, we create 80\%, 90\%, and 95\% highest density regions\footnote{The 100($1-\alpha$)\% highest density region is the smallest possible region having probability $1-\alpha$ for which every point inside the region has a probability at least as large as every point outside the region \citep{hyndman1996computing}.} about each pair of counts by using the joint $\hat{s}_i$ in place of $s_i$ in Equation (\ref{eq:mod_h0}).
Note this set of percentiles is meant to be illustrative rather than comprehensive and is not theoretically driven, our goal being to enable reporting and visualizing orbital, spatial, and temporal patterns in null rejection at different rejection thresholds.
Of concern would be, e.g., rejections that tend to occur predominantly along the ribbon or predominantly within a single map, as this would indicate possible regions in space and/or time for which the assumption of a shared signal is violated.

The use of the MLE $\hat{s}_i$ in place of $s_i$ in the hypothesis tests causes fewer rejections to occur than the nominal rejection rate would suggest (e.g., using a 95\% highest density region, we would expect to reject \textit{fewer} than 5\% of observations because we are using $\hat{s}_i$).
This effect is because the MLE is based on only two samples, so the $\hat{s}_i$ values are closer than the true $s_i$ to the observed counts, making us less likely to reject them. 
Because we no longer have a nominal rejection rate by which to compare our observed rejection rates, we turn to simulation to learn what MLE-adjusted rejection rate to expect.
Simply put, a set of binary reject or fail-to-reject results, while useful as a diagnostic, does not provide a holistic measure of the appropriateness of the assumed model.

We use the probability integral transform (PIT) values for the qABC counts and qBC counts as a diagnostic to assess the overall consistency of the observations with the null model \citep{diebold1998, gneiting2007probabilistic}.
In brief the PIT transform, also called the CDF transform, of some sample from a continuous distribution is generated by evaluating the CDF of that sample; Section B of the supplemental materials includes a PIT primer with more mathematical details.
We use the PIT histogram as a visual tool for evaluating the observation-null consistency; the more similar the distribution of observed PIT values is to that of the PIT values under the null hypothesis, the higher the consistency between the observations and the assumed null model.
The PIT values are calculated via a stochastic adjustment to the PIT calculation used for continuous random variables, necessary because our observations are counts and therefore the non-adjusted PIT can take only a discrete set of possible values \citep{dunn1996randomized}.\footnote{Specifically, the PIT value associated with the $i$th observation of the qABCs or qBCs is Uniform($l(\hat{\lambda}_i^*)$, $u(\hat{\lambda}_i^*)$), i.e., a random draw from a uniform distribution between $l(\hat{\lambda}_i^*)$ and $u(\hat{\lambda}_i^*)$, where $l(\hat{\lambda}_i^*)$ is the cumulative distribution function (CDF) of a Poisson evaluated at $y_i^* - 1$ with mean parameter $\hat{\lambda}_i^*$ if $y_i^*>0$ and 0 otherwise, and $u(\hat{\lambda}_i^*)$ is the CDF of a Poisson evaluated at $y_i^*$.}
We use $\mathbf{p}^{abc}$ and $\mathbf{p}^{bc}$ to denote the set of qABC- and qBC-specific PIT values, respectively.
If we had been using the true $s_i$ in our PIT calculation and not $\hat{s}_i$, we would expect the PIT histograms to be approximately $\text{Uniform}(0,1)$, and we could use deviations from uniformity as a diagnostic for model misspecification.

The use of the MLE $\hat{s}_i$ in place of $s_i$ causes the distribution of each $\mathbf{p}^*$ to be ``mounded'' (that is, it has fewer values close to 0 and 1 as a uniform distribution, and more values close to 0.5) and asymmetric. This mounding occurs because the MLE-based mean of the Poisson distribution tends to be closer to the observed count values than the true mean is. 
In effect, it is similar to what one would observe in the case of under-dispersion of the counts (i.e., counts with smaller variances than expected according to the Poisson distribution).
The asymmetry is a property of the skewness of the Poisson distribution; using the MLE in place of the true mean causes the Poisson skewness to propagate through to the PIT distribution. 
See the PIT primer in Section B of the supplemental materials for additional explanation and visuals of the above points.

The specific shape of the MLE-based PIT distribution differs with
changes in the other values informing the Poisson means: time, background rate, and efficiency.
Although we lack an analytic form for these MLE-based PIT distributions, we simulate data to create reference distributions; deviations from the reference PIT distributions, as opposed to the standard uniform distribution, can then be used as diagnostics of our model assumptions.
Seeing a visual shift in mean, skewness, or variance in the observed relative to simulated PIT histograms indicates an incorrect model assumption.

The basic idea behind simulating ``realistic'' synthetic data satisfying $\text{H}_0$ from which to get reference PIT distributions is to use the provided data to get a smooth shared signal rate estimate, then simulate \textit{new} qABC and qBC counts using that smooth shared signal rate. The details of the procedure for a given ESA-orbit combination are shown in Algorithm \ref{alg:synth_dat}.
\begin{algorithm}
  \caption{Generating synthetic count data having a shared underlying signal rate map.}
  \label{alg:synth_dat}
  \begin{algorithmic}[1]
    \STATE To get a smooth shared signal rate estimate, we fit a weighted generalized additive model (GAM) \citep{hastie1987generalized}, a flexible semiparametric nonlinear model that is commonly used for modeling continuous smooth data, to the shared signal rate estimates from the data. The predictors in this 1D GAM are the look direction indices, the outcomes are the $\tilde{s}_i$ defined in Equation (\ref{eq:s_mle}), and the weights are $w_i = \frac{t_i^{abc} + t_i^{bc}}{\sum_i( t_i^{abc} + t_i^{bc})}$. We let $ \dot{s}_i^{GAM}$ denote the GAM prediction at each look direction. \label{op0}
    \STATE We then get the smooth signal rate prediction at look direction $i$ as $\dot{s}_i^{smooth} = \max(0, \dot{s}_i^{GAM})$. \label{op1}
    \STATE We use these smoothed signal rate estimates to calculate the expected values of the counts for that ESA and orbit; specifically, $\dot{\lambda}_i^{*} = t_i^{*}(e_i^{*} \hat{s}_i^{smooth} + b_i^{*})$ for $* \in \{abc,bc\}$. \label{op2}
    \STATE Finally we simulate new Poisson counts having those expectations, $\dot{y}_i^{abc}$ from $\text{Poisson}(\dot{\lambda}_i^{abc})$ and $\dot{y}_i^{bc}$ from $\text{Poisson}(\dot{\lambda}_i^{bc})$. \label{op3}
  \end{algorithmic}
\end{algorithm}
In Algorithm \ref{alg:synth_dat}, $i$ are the indices such that the observations are from the considered orbit and ESA. We repeat this process for each unique ESA-orbit combination. Further details on the generation and properties of the synthetic reference data sets are provided in Section D of the supplemental materials.

We use the two-sample Cramér-von Mises (CvM) test \citep{anderson1962distribution} and the two-sample Kolmogorov-Smirnov (KS) test \citep{massey1951kolmogorov} to summarize the difference between our two empirical distributions – that of the simulated PIT values, and that of the observed PIT values.
At the ESA-orbit and ESA-map level (i.e. at levels below the mission level) we use CvM test.
This test can handle multivariate observations (i.e., it can test the qABC and qBC PIT values jointly), but is too computationally intensive to run at the mission level because of the volume of data involved.
At the mission level we use the KS test, another test meant to quantify empirical distribution similarity\footnote{Another, similar, test is the two-sample Anderson-Darling test \citep{pettitt1976two}.}, independently for the qABC and qBC PIT values.
We report the KS test statistic by ESA for the full data in Section \ref{sec:res_quant}; map- and orbit-level KS and orbit-level CvM values are available as a supplementary results file.
In all cases, lower statistic values indicate better distributional agreement, and higher statistic values indicate worse agreement. 

\subsection{Learning model component adjustments}\label{sec:meth_qbc}

Of interest to the IBEX mission is learning about systematic temporal trends in qBC background misspecification and the identification of potential ``orbits of concern'', i.e., orbits for which we see high levels of disagreement between simulation and reality. The Cramér-von Mises test statistic can also be used to accomplish these goals by doing the following:
\begin{enumerate}
    \item For a grid of possible qBC background adjustment levels $\rho\%$ ranging from $-75\%$ to $75\%$ in half percentage increments, we simulate a reference data set having the new ``adjusted'' qBC background level $\tilde{b}_i^{bc}$, where $\tilde{b}_i^{bc} = b_i^{bc}\cdot(1+\rho/100)$.
    \item For each $\rho\%$, we then calculate the CvM test statistic for each orbit and ESA combination assuming that the adjusted background is the truth, i.e., using $\tilde{\hat{\lambda}}_i^{bc} = t_i^{bc}(e_i^{bc} \hat{s}_i + \tilde{b}_i^{bc})$ in place of $\hat{\lambda}_i^{bc}$ in the PIT calculations.
    \item We call the $\rho\%$ value for which the CvM statistic is lowest the ``optimal'' adjustment for a given ESA-orbit.
    \item We fit an exposure time weighted GAM to this set of orbit-level optimal $\rho$ values and learn an ESA-and orbit-specific qBC background \textit{fitted optimal adjustment} that allows us to identify and quantify systematic trends over time in potential qBC background misspecification. We include a comparison of the GAM fits to fits estimated using a fully nonparametric weighted locally estimated scatterplot smoothing (LOESS) \citep{cleveland1979robust}, to ensure our findings are robust to choice of algorithm.
    \item We repeat the above for the synthetic data set, yielding synthetic optimal $\rho$ values and synthetic fitted optimal adjustments that can be compared to those found using the mission data.
    \item We then identify \textit{orbits of concern} as those for which the GAM-predicted CvM test statistic at the fitted optimal $\rho$ value falls above the 99th percentile value from the set of predicted CvM statistics at each fitted optimal $\rho$ value in the synthetic reference data set. These identified orbits can then be assessed by an IBEX mission scientist as desired. 
\end{enumerate}
Figure \ref{fig:bias_orbit} shows the result of performing steps 1 through 4 of the above for an example ESA-orbit.
Note that the choice of what percentile to use as a cutoff can be adjusted based on the desired level of sensitivity – a lower choice (e.g., using the 95\% quantile) will flag more orbits, while a higher choice (e.g., using the maximum value) will flag fewer.
% FIXME add a recommendation for what to do about those orbits / how to decide whether to drop the qBC/qABC/both? 
% If computational time were of no concern, an alternative strategy would be to get a bootstrap sample of 1,000 fully simulated comparison data sets and learn an individual cutoff for each orbit.

Intuitively, if the correction process shows that a given ESA-orbit would be expected to have more or less qBC counts, we can either adjust the background component or the signal component.
If we adjust the signal component, we can do that by changing the efficiency factor or the signal rate (since the overall signal contribution from the qBCs is the product of the efficiency factor and the signal rate, a \% change in one term is analogous to a \% change in the other term).
We perform the same adjustment and outlier identification process as above assuming that the correction should be made to the qBC signal component rather than the qBC background component.
Doing so allows us to get an estimate of how much the signal rate of the qBC may differ from that of the qABC if we assume \textit{zero error} in the specification of all other model components (i.e., the assumption that will lead to the highest discovered deviation between the qABC and qBC signal rates).

Finally, we show the results of both of these optimal-adjustment fitting and outlier identification processes run using a synthetic reference data set as the original input for comparison. 
This simulation process gives us a heuristic for how much error and variation of results may be expected due to the process itself. 
We expect some slight deviations between simulation and reality because we don't know the true signals $\{s_i\}$ generating the data and are using an estimate. 
If, e.g., we were finding fitted optimal adjustments of around a percent or so using a synthetic reference set as our original input, this would tell us that a fitted absolute $\rho$ of up to around 1\% found using the mission data may just be due to the modeling and fitting process and any remainder above 1\% is likely a ``true'' correction stemming from the data and not the fitting process.

%%--------------------------------------------------------------------------------------------------
%%--------------------------------------------------------------------------------------------------

\section{Results}\label{sec:res_hypothesis}

If we assume any lack of agreement between the qABCs and qBCs is due to errors in the provided qBC backgrounds, we find that the same ENA rate sky map could have plausibly generated the observed qABC and qBC data products to within a tolerance of a few percentage points on average. %; on average across all ESAs, the fitted qBC background rate is 1.4\% larger than the provided qBC background under this assumption.
Per the background standard deviations provided by mission scientists, the median \textit{anticipated} percent adjustment on the backgrounds is approximately $+/-$5\%, with slight temporal variation in expected adjustment bounds.
Our \textit{learned} fitted qBC background errors are predominantly within these bounds, with the exception being the observations from 2009 through early 2011 at ESA 2; the fitted qBC background rate for ESA 2 is 10.8\% smaller than the given background at the start of the mission and approaches the given background as time goes on. % with a worst-case suggested adjustment of -10.8\% at the beginning of the mission that improves and reaches approximately zero by 2018
See Section C in the supplemental materials for more information about and visualizations of these anticipated fitted adjustments.

If on the other hand we attribute data product disagreement solely to errors in the qBC efficiency factor (or, analogously, to deviations in qBC signal rate from the qABC signal rate), the fitted adjustment levels for ESA 2 and 6 reach worst-case learned adjustment values of -22.6\% and -9.1\% respectively.
However, the average absolute adjustment across all ESAs under this assumption is still relatively low, at 3.6\%.
If, more realistically, we allow the qBC background to account for the data product disagreement up to the error bounds anticipated by mission scientists, the only identified deviations in the qBC efficiency factor occur in ESA 2 before 2011.
In this scenario, the worst-case learned qBC efficiency factor adjustment value for this ESA and time window is -11.6\% at the start of 2009, and it drops within the anticipated qBC efficiency factor bounds of $+/-$3\% by mid-2010.

These findings can be thought of as an implicit validation of the relative efficiency values and backgrounds for nearly all ESAs and orbits, with the exception being the data from 2009 through mid-2010 at ESA 2.
There are similar numbers of observations rejected between the non-synthetic and synthetic data, and no systematic patterns of rejection by look direction, ESA, or time.
More detailed results follow.

\subsection{Hypothesis test results}\label{sec:res_quant}

\begin{figure}[htpb]
	\centering
	\includegraphics[width=0.85\textwidth]{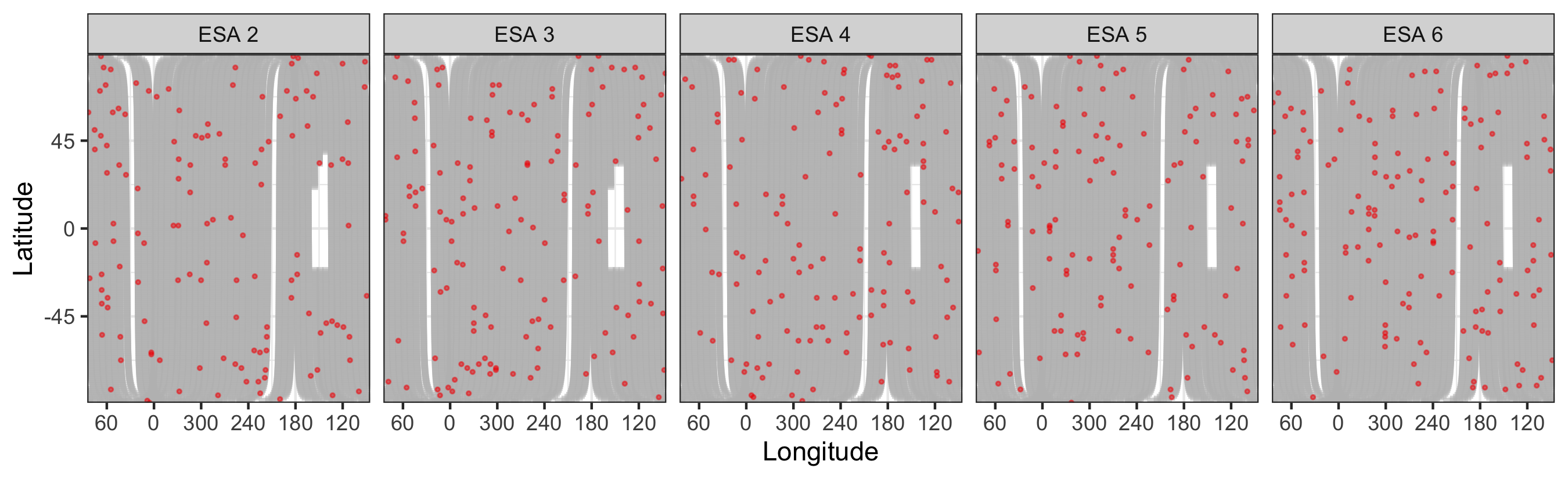}
	\caption{\small Look directions for the 2010A orbits for which $\text{H}_0$ was rejected using a significance level of 0.05, plotted by latitude and longitude as red points against a background of grey non-rejected points. Of concern would be if there were clearly visible rejection patterns, e.g., if the observations for which the null was rejected always occurred along the ribbon, or in a single orbit. No such rejection patterns are visually apparent. From left to right, the percent of observations for which we reject the null is 1.6, 1.7, 1.7, 1.6, and 1.9\%.} \label{fig:validation_mappoints}
\end{figure}

\begin{figure}[htpb]
	\centering
	\includegraphics[width=0.95\textwidth]{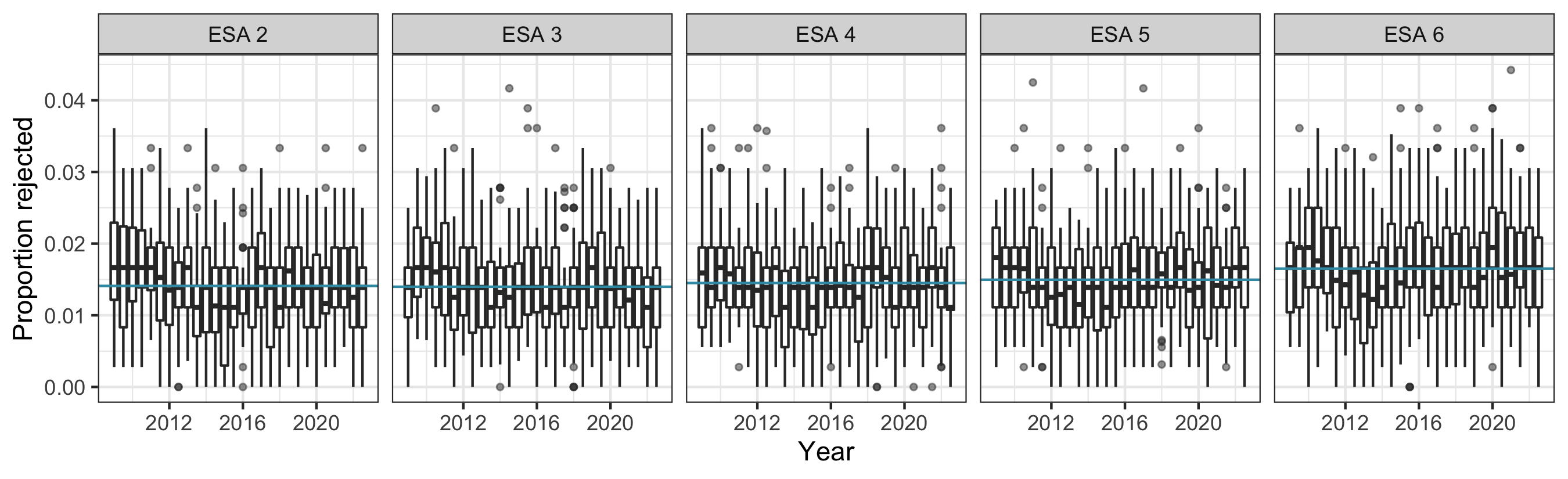}
	\caption{\small Temporal rejection patterns shown by map and ESA. Here a single point (i.e., contributor to the box plots) is the proportion of look directions for which we reject $\text{H}_0$ within a given ESA and orbit using a significance level of 0.05, the horizontal line shows the median of the points, the lower and upper box values are the 25th and 75th percentiles of the points, the end of the lines show the farthest point no more than 1.5 times the interquartile range from the box value, and the plotted circles are those points not falling within the span of the lines. The overall ESA-specific average proportion of observations for which the null hypothesis is rejected is shown as a horizontal line. The highest value in 2011 for ESA 5 was orbit 0126, in which we reject $\text{H}_0$ for 4.3\% of observations; this orbit was one we identified as an outlier, the signal rate estimates for which are shown in Figure \ref{fig:htest_oneOrbit}.} \label{fig:validation_esa_yr}
\end{figure}

With a significance level alpha of 0.05\footnote{The significance level specifies the probability of rejecting the null hypothesis when it is true. It is how one defines the strength of the evidence that will be required from one's sample in order to reject the null hypothesis; a smaller alpha means stronger evidence will be needed.}, we reject $\text{H}_0$ for 1.48\% of all $(y_i^{abc}, y_i^{bc})$ pairs across all orbits and ESAs in the observed data, and for 1.43\% of synthetic observations.
With a difference in rejection rates of only 0.05\%, we are rejecting the null for roughly the same amount of observations as we would expect to under simulation.
Broken down by ESA, we reject the null for 1.41, 1.40, 1.45, 1.50, and 1.65\% of observations for ESAs 2 through 6, respectively. In simulation ESAs 2 and 6 have the largest (albeit still small) difference in rejection proportions relative to observations, with null hypotheses being rejected for 1.29 and 1.50\% of observations, respectively.
Note that having rejection rates that are consistent with what we would expect via simulation does not necessarily imply that the rejection patterns at an orbit-, spatial-, or temporal-level are consistent with what we would expect via simulation.
We use the binary hypothesis test results to assess whether there are any patterns suggestive of spatial-, temporal-, or ESA-specific structure in the misalignment between the qABC and qBC signal.
Figure \ref{fig:validation_mappoints} shows the spatial rejection pattern by ESA for an example map, and Figure \ref{fig:validation_esa_yr} shows the temporal- and ESA-level rejection patterns at a significance level alpha of 0.05. 
There are no visible spatial, temporal, or ESA-level patterns in how many observations are rejected. 
Analogous results for two other significance levels alpha, as well as for the synthetic reference data, are shown in Section E of the supplemental materials.
Although the above findings are suggestive of a shared source generation, they are not conclusive without further evidence in support of the model assumptions.

\begin{figure}[htpb]
	\centering
	\includegraphics[width=0.95\textwidth]{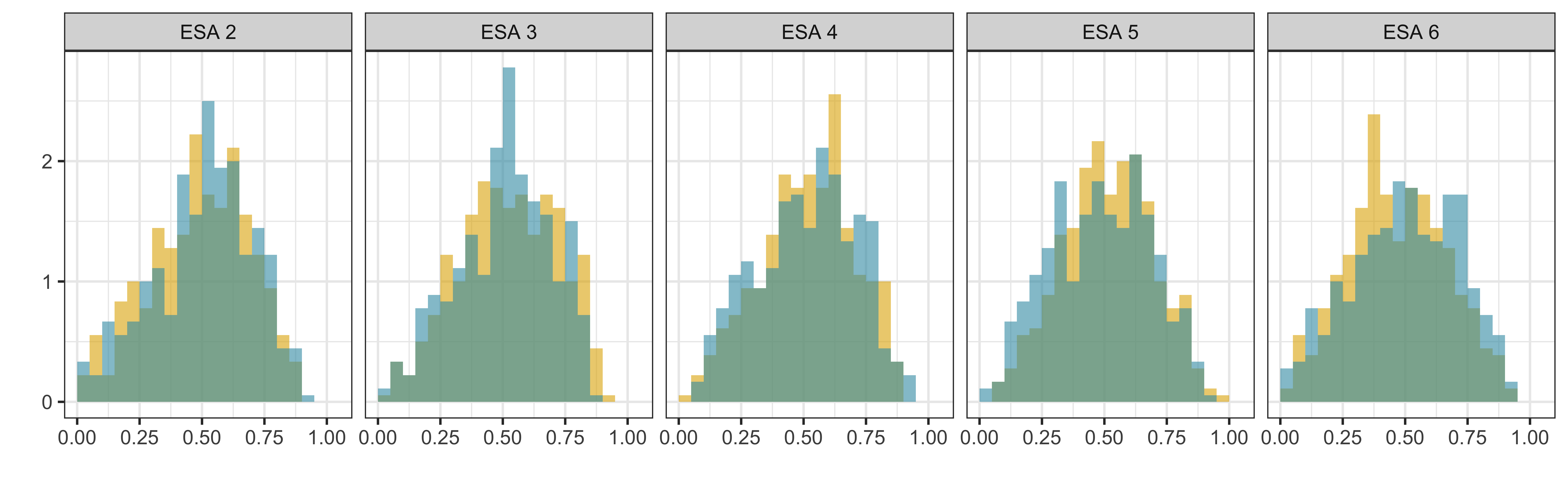}
	\includegraphics[width=0.95\textwidth]{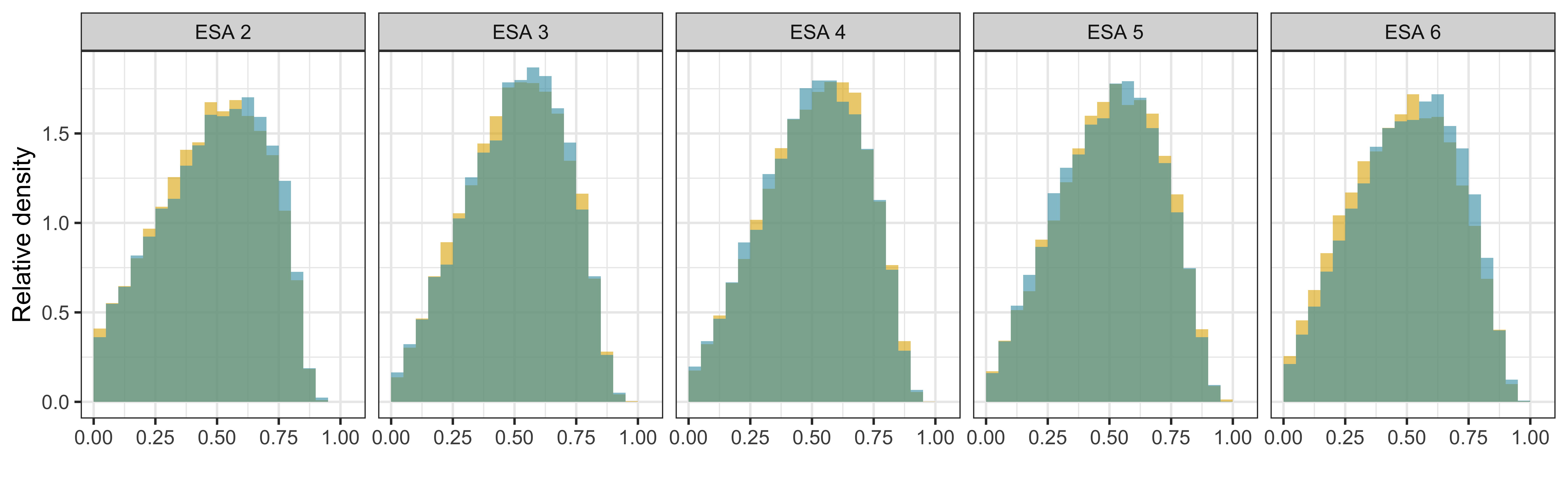}
	\includegraphics[width=0.95\textwidth]{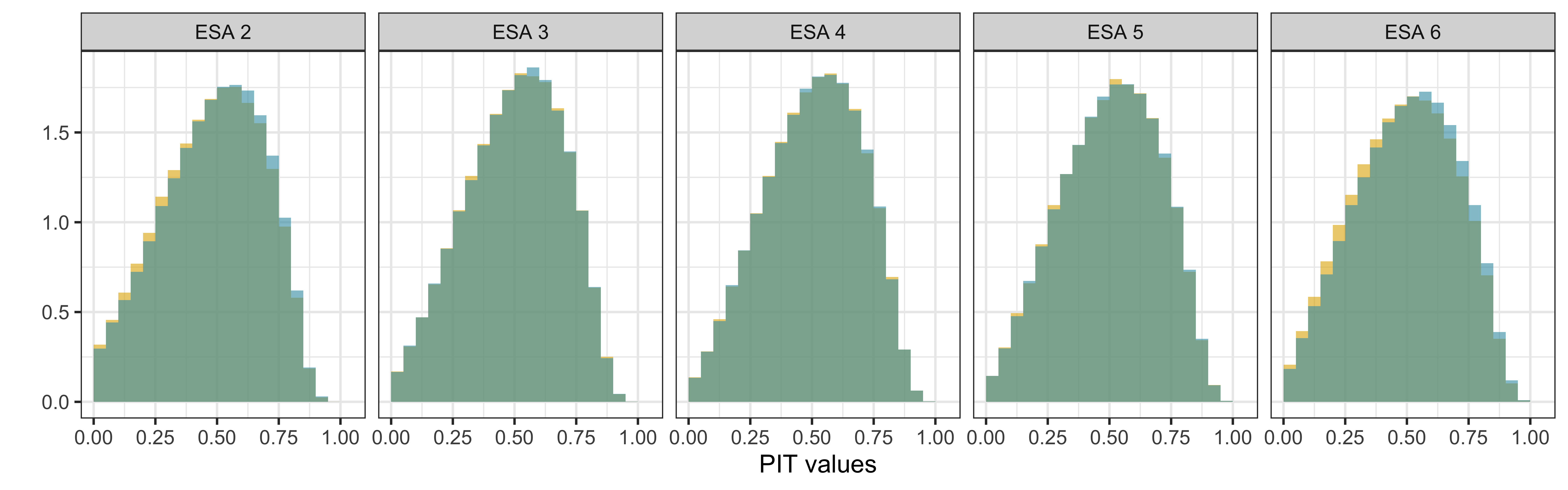}
	\includegraphics[width=0.6\textwidth]{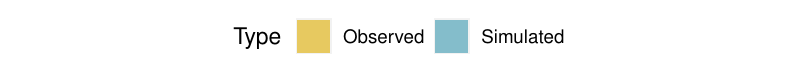}
	\caption{\small PIT histograms for the qBCs from an example orbit (orbit 0276b, top), an example map (2010A, middle), and the full mission (bottom). At the mission level, the KS statistic values for ESA 2 through 6 are 0.016, 0.003, 0.002, 0.003, and 0.024, respectively, with lower values indicating better agreement between observed and simulated data.} \label{fig:res_pvals_bc}
\end{figure}

We use the PIT histogram as a single orbit-, single map-, and full mission-level assessment of model fit that provides more detailed insight than rejection patterns alone. 
The distribution of observed-data PIT values is visually similar to that of synthetic-data PIT values at the orbit, map, and full mission level (i.e., for the full set of data), with deviations at the orbit level not suggestive of anything beyond random variability in most cases. At the map or full mission level we are able to see clearer patterns of deviations emerge. Figure \ref{fig:res_pvals_bc} shows the observed and synthetic qBC PIT histograms for each ESA from an example orbit, an example map, and the full mission. As the amount of observations increases, we see better resolved (i.e., less noisy) histograms. The largest deviations between observation and simulation occurs in ESA 6, with ESA 2 also exhibiting relatively larger deviations than the rest of the energy passbands.
An analogous plot for the qABC PIT histograms is provided in Section E of the supplemental materials.

At an ESA-orbit level, for which the CvM statistic is computationally tractable, 89.8\% and 92.2\% of ESA-orbits have observed PIT distributions that are statistically indistinguishable at a 0.05 level from their synthetic counterparts before and after background adjustment, respectively.
At a mission level, for which we rely on the KS statistic, the qABC PIT values are significantly different from their synthetic counterparts at a 0.05 level for ESA 2, 4, 5, and 6; the qBCs for ESA 2 and 6.
Upon adjusting the qBC background per the method discussed in the next section, the PIT values at the full mission level improve on average but are still statistically distinguishable from their synthetic counterparts; the PIT values for the qABCs and qBCs after qBC background adjustment are shown in Section F of the supplemental materials.
However, even very minor imperfections in the model assumptions would yield statistically significant differences with such a large data set; in total, 301,218, 306,036, 313,524, 317,820, and 317,040 pairs of PIT values are used in the tests for ESA 2 through 6, respectively.

\subsection{Model component adjustments}\label{sec:res_background}

\begin{figure}[htpb]
\centering
\includegraphics[width=0.75\textwidth]{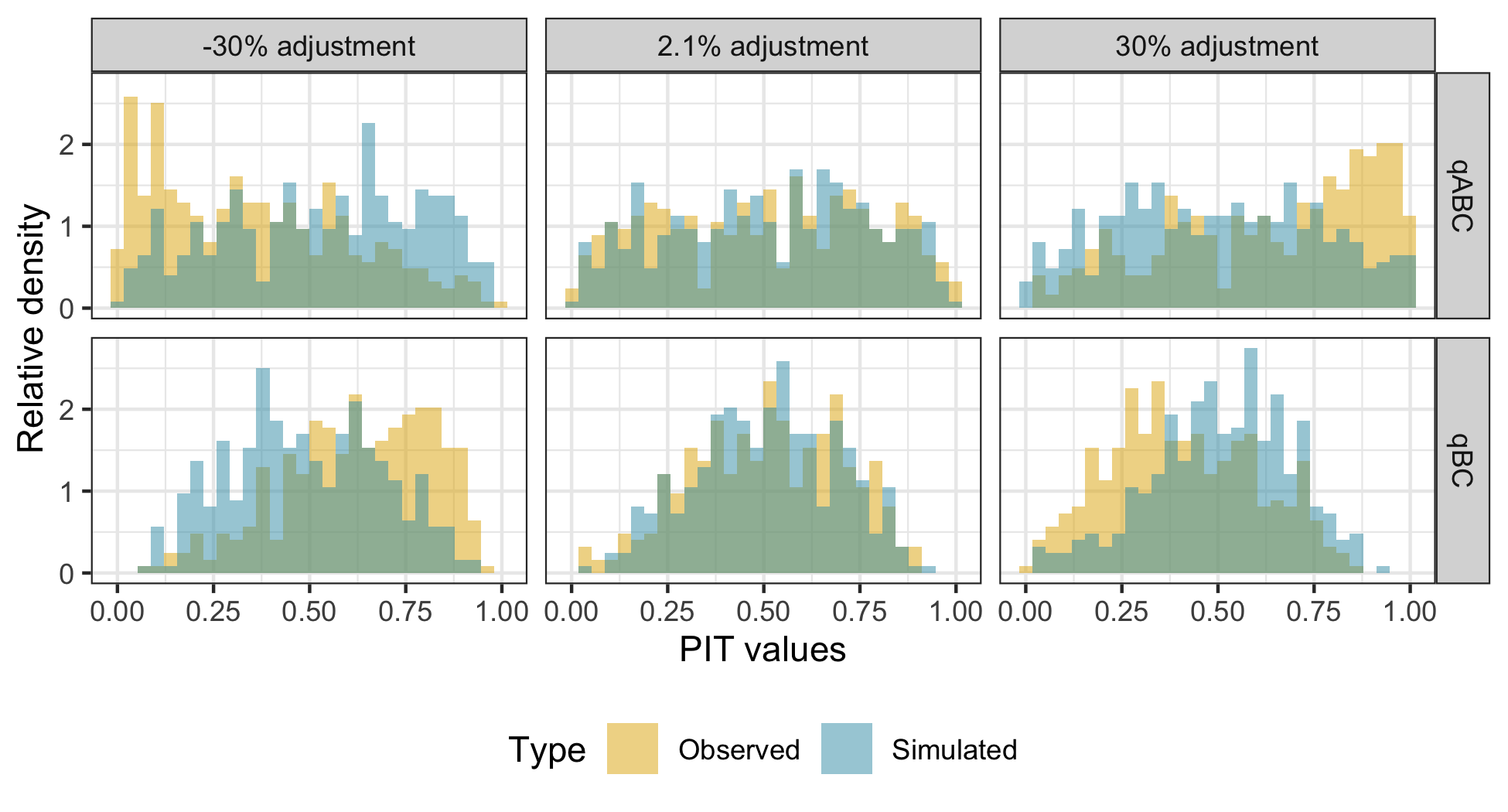}
\includegraphics[width=0.75\textwidth]{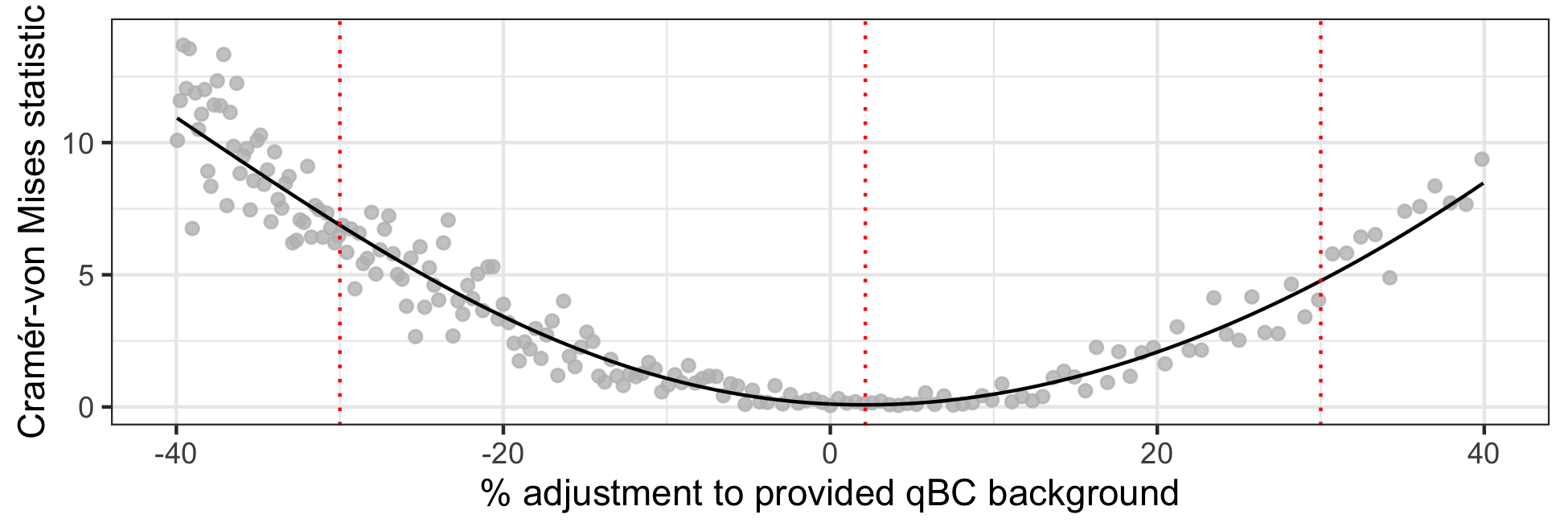}
\caption{\small Top: histograms of the PIT values for the observed and synthetic data sets for ESA 3 of orbit 0276b assuming different \% adjustments (i.e., $\rho$ values) to the provided qBC background. Of the adjustment levels shown, the 2.1\% level (middle column) was that found to have the lowest fitted optimal Cramér-von Mises (CvM) test statistic value (i.e., the background adjustment yielding the highest agreement between simulation and observation according to the smooth fit). Bottom: observed CvM test statistic values under different $\rho$ values (grey points) and the GAM fit to those points (black line) for ESA 3 of orbit 0276b. A lower CvM statistic value implies closer distributional agreement. The vertical dashed lines indicate where the $\rho$ values shown in the top plot occur.}\label{fig:bias_orbit}
\end{figure}

The inconsistencies between the synthetic- and observed-data PIT histograms may be caused by systematic biases in the provided qBC backgrounds. Figure \ref{fig:bias_orbit} shows the bivariate PIT distributions for orbit 0276b at three qBC background adjustment levels $\rho$, each associated CvM test statistic, and the test statistic values across the entire grid of considered $\rho$ values. For this orbit, to achieve the best agreement between the observed and synthetic distributions one would need to increase the provided background by 2.1\%. Note, however, that there is a window of adjustment levels between roughly -2\% and 7\% that seem to produce similar agreement between simulation and reality, as measured by the Cramér-von Mises test statistic. 

\begin{figure}[htpb]
\centering
\includegraphics[width=0.98\textwidth]{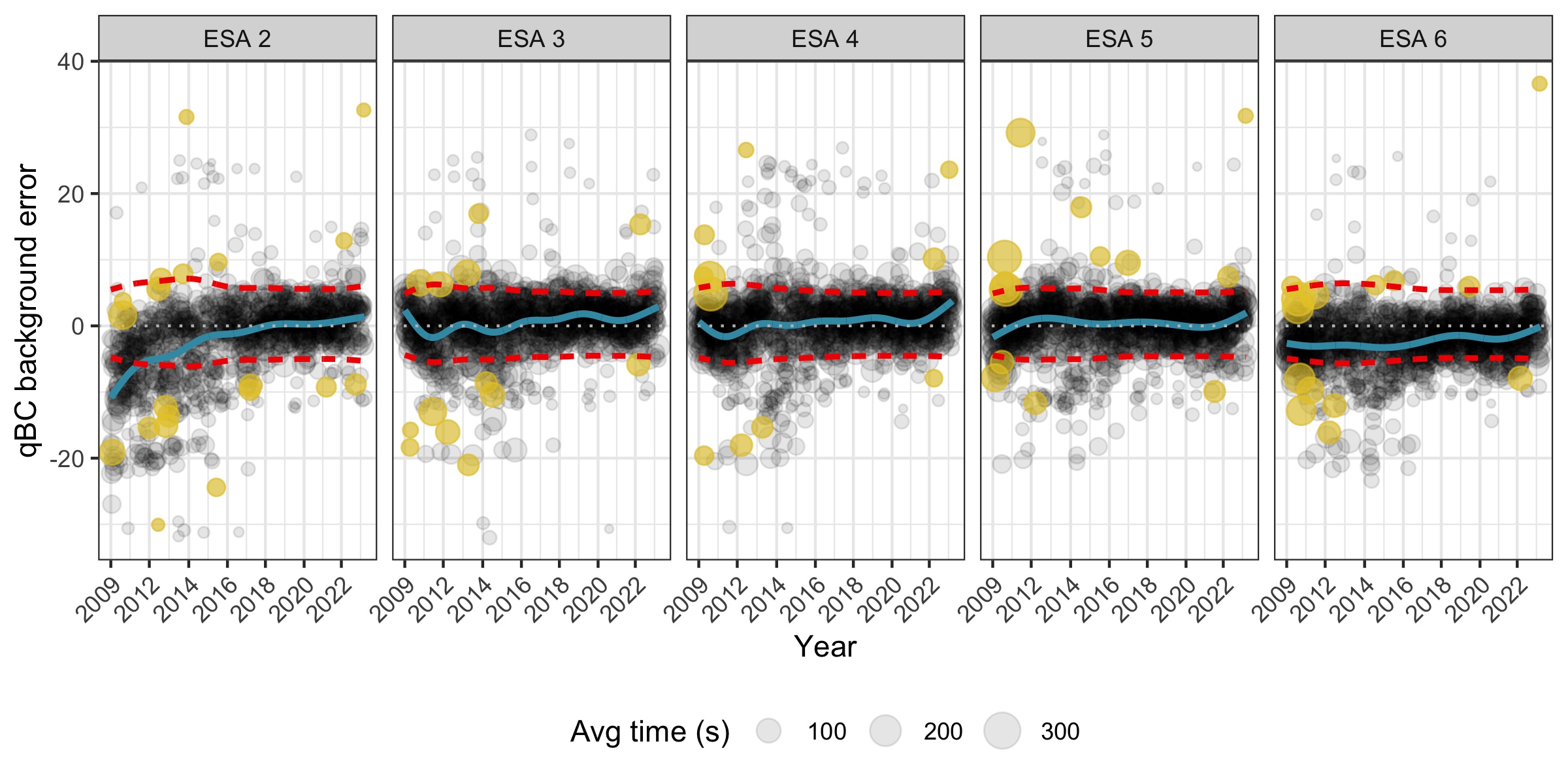}
\caption{\small Each point corresponds to a single ESA-orbit and shows the identified ``best'' qBC background adjustment, i.e., that minimizing a GAM fit to the CvM statistic across all $\rho$ values. The temporally fitted qBC background adjustment is shown as a blue line, and the flagged ``ESA-orbits of concern'' are shown as colored yellow points. The size of each point corresponds to the average exposure time in that ESA-orbit. The red dashed lines spanning zero show the temporally smoothed anticipated adjustments provided by mission scientists; of concern are regions where the fitted background adjustment falls outside of the bounds of the anticipated adjustments, i.e. where the blue line crosses over the red lines. The only window in which this occurs is from 2009 through early 2011 in ESA 2.}\label{fig:berr_fit}
\end{figure}

Figure \ref{fig:berr_fit} shows the ESA-specific temporal qBC background adjustment $\rho$ learned via a GAM fit to the CvM statistic across adjustments as well as the flagged ``ESA-orbits of concern''; at worst, the fit to the learned qBC background adjustment reaches -10.8, 2.9, 3.8, 2.0, and -3.3\% for ESAs 2 through 6, respectively. 
A LOESS fit, by comparison, yields worst-case fitted adjustments of -10.8, 2.9, 3.9, 2.4, and -3.5\% for ESAs 2 through 6, respectively; that is, the fit is consistent to under half a percent across these two choices of model.
Relative to the background standard deviations provided by mission scientists, these worst-case fitted adjustments are all within two-sigma of a level of adjustment we would expect for all but ESA 2, for which the fitted adjustment falls outside this range before early 2011.
In the simulation-based run of this assessment, meant to diagnose how much of an adjustment finding may be due to the fitting process itself, the learned absolute adjustment stays under 2\% across all ESAs and is on average 0.7\%; details are provided in Section F of the supplemental materials.
This simulation-based run was used to determine a threshold for categorizing an ESA-orbit as being of concern.
Specifically, (1) a GAM was fit in which the input was the full set of qBC background adjustment values and the output was the CvM statistic, (2) we recorded the predicted CvM value at the fitted ``best'' qBC background adjustment (i.e., the blue line from supplement Figure F.3), and (3) we recorded the 99th quantile of this set of predicted CvM values to use as the threshold in the mission data background adjustment fitting process, which is analogous to steps 1 and 2 here.
We find that 1.4\% of ESA-orbits should be flagged as potential ``ESA-orbits of concern'' for the science team to investigate based on the above threshold. Note that the orbits designated as of concern are not always the largest outliers; sometimes they are fairly close to the fitted line, while many orbits that are farther from the fitted line are not marked as of concern. This apparent inconsistency is due to exposure time; many of the outliers that are not marked as of concern have low exposure times and therefore large uncertainties about their optimal adjustment.

The highest absolute fitted adjustments and the majority of ESA-orbits of concern can be found in the first half of the data; at later times, the fitted adjustments are closer to 0 and there are fewer identified abnormal ESA-orbits.
There are several reasons why disagreements between qABC and qBC fluxes are more likely earlier in the mission.
Detector-section efficiencies changed most in the first year or two of the mission.
A suspected out-gassing event occurred in Orbits 55 through 57 (late 2009) and likely caused a sudden change in detector-section behavior.  If so, it becomes difficult to determine the detector section's performance to high accuracy beforehand.
Solar maximum conditions limited the amount of data (exposure time) useful for heliospheric studies around 2012-2014, and contributed increased penetrating backgrounds.
Finally, IBEX-Hi operating voltages were modified in mid-2013 to significantly reduce the ion gun background in ESAs 2, 5, and 6, and thus any issues associated with mischaracterization of this background.
%[Additionally, ESA 2 data shows signs consistent with an additional, unidentified, unremoved background.]
%(Compton-Getting correction of ESA 2 data suggests the presence of an additional, unidentified background.)

\begin{figure}[htpb]
\centering
\includegraphics[width=0.98\textwidth]{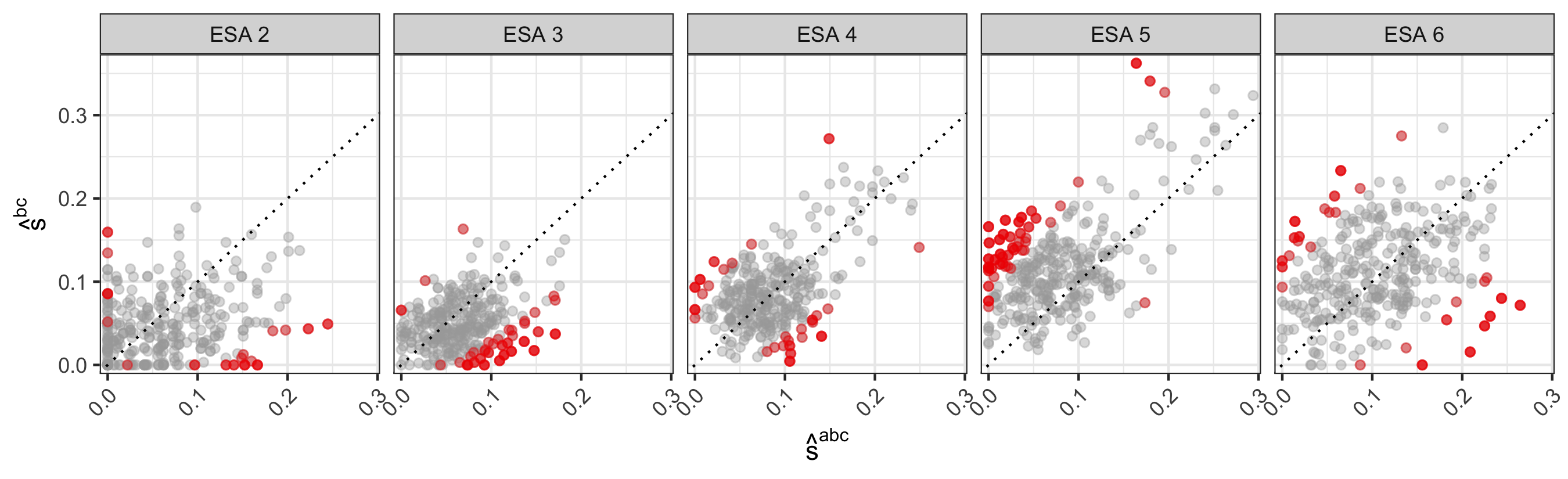}
\caption{\small The hypothesis test results by ESA for orbit 0126 shown with each individual MLE value; rejections are shown in red and the dashed line shows the $y=x$ relation. Overall, we reject $\text{H}_0$ for 2.4\% of observations from orbit 0126. From left to right by ESA we reject the null for 1.0, 2.3, 1.6, 4.3, and 2.6\% of observations. It is visually clear that for ESA 3, the pattern of rejections is biased in such a direction as to suggest that the qBCs are under-estimating the signal relative to the qABCs, while the reverse is true for ESA 5 (and ESA 6, although the effect is more subtle for this ESA).}\label{fig:htest_oneOrbit}
\end{figure}

Figure \ref{fig:htest_oneOrbit} shows the independently estimated signal rate MLEs and hypothesis test results for all points in an example orbit, orbit 0126, for which ESA-orbit 3-0126, 5-0126, and 6-0126 were flagged as ESA-orbits of concern (orbit 0126 is the top-most identified outlier point near 2011 in the ESA 5 subplot of Figure \ref{fig:validation_esa_yr}). 
The spatial rejection pattern, and the observed and simulated PIT histograms for this orbit before and after background adjustment are shown in Section F of the supplemental materials.
The optimal learned qBC background adjustment for the most extreme misfit of these ESA-orbits, 5-0126, is 27.4\% (that is, the optimal qBC background value is almost 28\% larger than the given background); this orbit can be seen in yellow at around 2011 near the top of the ESA 5 sub-plot of Figure \ref{fig:berr_fit}.
This adjustment would lead the qBC background for this ESA-orbit to be, on average, 0.020 as opposed to 0.016.
There is an identiable quirk, in hindsight, in the processing of orbit 126 data involving the merging of the count rate during exposure times with slightly enhanced isotropic backgrounds to the count rate during the quietest exposure times \citep{mccomas2012first}.  This outlier identification process can serve as a validation step to identify and rectify the impact of such quirks on background estimation.

Statistically, we can’t tell whether an adjustment should be made to the background vs. the efficiency factor because the terms are not separately identifiable. 
Section G of the supplemental materials show an example of this identifiability issue via simulation in which we show how similar the PIT histograms look under different types of multiplicative misspecification of the qBC background, efficiency factor, and signal rate.
Supplemental materials Section H shows the results of the correction and outlier identification process under the assumption that the signal component is the culprit of some or all of the apparent misalignment between expected and observed counts.
At worst, the learned qBC signal deviation relative to the assumed shared signal reaches -22.6, 4.8, 4.8, -3.4, and -9.1\% for ESAs 2 through 6, respectively.
The average absolute adjustment is 7.3, 2.2, 1.5, 0.9, and 6.3\% for ESAs 2 through 6, respectively.
A worst-case adjustment of -22.6\% for ESA 2 sounds quite troublesome for the shared-signal hypothesis. 
However, the magnitude of this finding is in part due to the signal component being closer to zero for ESA 2; it takes a larger percent change in the signal component than the background to lead to the same change in expected counts for ESA 2 because the signal component is only about 40\% that of the average background.\footnote{The average background rates are similar across ESAs and larger in magnitude than the average $e^{bc}*s$ for ESA 2. If, e.g., you want to increase the expected counts by 2 and have model components $t=200, e*s=0.1,$ and $b=0.2$, you would need a 5\% increase in $b$ because $t(es + 1.05b) = t(es + b) + 0.05tb = t(es + b) + 2$. On the other hand, you would need a 10\% increase in $s$ because $t(1.1es + b) = t(es + b) + 0.1sb = t(es + b) + 2$.}
Additionally, recall that for these calculations we were making the unlikely assumption that all other model components need \textit{zero adjustment}.
It is much more scientifically likely that the bulk of the apparent model misspecification is due to misspecification of the backgrounds, particularly because the largest apparent deviations occur at the beginning of the mission.
This is because qABC/qBC signal calibration is performed via the bright magnetospheric signal every 6 months, in which there is rarely much temporal drift. % FIXME will cite Paul's paper here!!
If we allow the qBC background to account for the data product disagreement up to the error bounds anticipated by mission scientists then no further adjustments are needed for ESAs 3 through 6, and the worst-case learned qBC efficiency factor adjustment value for ESA 2 is -11.6\% at the start of 2009, dropping within the anticipated qBC efficiency factor bounds of $+/-$3\% by mid-2010.
Finally, we note that although we have treated the background rates and efficiency factors for the qABCs as known and unbiased because they are an existing and released data product, were this not the case it would be difficult to distinguish which of the qABCs or qBCs may be leading to apparent misspecification without making additional model assumptions.

%%--------------------------------------------------------------------------------------------------
%%--------------------------------------------------------------------------------------------------

\section{Discussion}\label{sec:discussion}

In this document, we developed a statistical validation procedure for paired count observations and performed this validation process for the new IBEX-Hi qBC data product.
We showed that the qBCs can plausibly be said to share the same signal rate as the qABCs up to the anticipated adjustment levels for all ESAs and time windows except for ESA 2 during 2009 through mid-2010.
Allowing the qBC background to account for the data product disagreement up to the error bounds anticipated by mission scientists, the average learned qBC efficiency factor adjustment value for this ESA and time window is -7.2\%, with a worst-case adjustment of -11.6 at the start of 2009.
Optimally adjusting the qBC background rates up to the anticipated error bounds leads to observed data that are visually indistinguishable from synthetic data under an assumed shared signal rate, and statistically indistinguishable for 92.2\% of ESA-orbits.
That is, we can plausibly say there is an underlying true shared signal rate underlying the nearly all of the qABCs and qBCs.
Finally, we are able to provide a myriad of visual diagnostics for mission scientists to explore at an orbit, map, and full mission level that allow additional visual confirmation of model coherence.

There are a number of advantages to the proposed validation framework. 
First, the processing steps are minimized since the analysis is done so early on in the IBEX pipeline. 
Not only are we using binned direct events instead of fitted or transformed data, but our question involves ENA rates, rather than fluxes, our findings are at the observation-level, and so do not depend on map making choices, and no pooling across spatial locations, time, or ENAs is needed.
Second, we can specify the level of error to which we validate the data, i.e., you can fail to validate the qBC data under this paradigm if the error is not within a pre-specified tolerance.
Finally, there are several convenience factors: it’s easy to report results at different levels of aggregation, the method is fast, and the hypothesis testing framework is a defensible choice commonly used in statistics.

Confirming that the two counts can be used as measures of the same underlying ENA signal rate allows us to pool the data, i.e., use the qBCs along with the qABCs, in estimation pipelines and downstream modeling.
We already saw in Figure \ref{fig:validation_shats} that combining the data stabilizes the maximum likelihood estimates of the signal rate.
Doing so also leads to improved estimation of the underlying signal rate map, with estimates that are more accurate, more precise, and visually less contaminated with artifacts.
Supplemental materials Section I provides synthetic and mission data comparisons of using a single data stream vs. data pooling in a map estimation model.

The concept of making assumptions about how data are generated, simulating data under that assumption, and comparing those simulations to reality is quite general.
It is also not restricted to two observations because the CvM statistic supports higher dimensional inputs, nor is it restricted solely to equidispersed counts, with alternative count distributions (e.g., the Conway-Maxwell Poisson distribution \citep{conway1962queuing, huang2017mean}) available.
This validation procedure could be used in other space science situations in which one might want to combine data sets dominated by Poisson statistics. 
Examples include combining events observed by spacecraft constellations (i.e., multiple spacecraft flying in close formation) hosting the same instruments such as the upcoming Helioswarm mission, or single instruments recording multiple measurement types such as the IMAP-Hi instrument.

This procedure could also be extended to non-space applications
For example, in the measuring of nuclear material, such as Californium 252, singles, doubles, and often triples coincidences are observed.  
A commonly used physical model in this case are the point-mass equations \citep{langner1998application} which shows how these three count rates are related physically if the assumptions of this model hold.  
In other words, the data should be generated from a Poisson distribution whose expected value follows the point-mass equations. 
This procedure could validate the assumptions of this model, and if they are not valid, describe how the nuclear material is distributed. 
For example, if the material is in the form of a liquid, then the point-mass equations will not hold and we should see a departure from the model in the measured singles, doubles, and triples count rates.

%%--------------------------------------------------------------------------------------------------
%%--------------------------------------------------------------------------------------------------

\section{Acknowledgements}
Research presented in this manuscript was supported by the Laboratory Directed Research and Development (LDRD) program of Los Alamos National Laboratory (LANL) under project number 20220107DR and by the NASA IBEX Mission as part of the NASA Explorer Program (80NSSC20K0719).
Approved for public release: LA-UR-23-29873.

%%--------------------------------------------------------------------------------------------------
%%--------------------------------------------------------------------------------------------------

\bibliography{references}

%%--------------------------------------------------------------------------------------------------
%%--------------------------------------------------------------------------------------------------

\clearpage

\appendix 

\counterwithin{figure}{section}

\title{Supplementary Materials}
\maketitle

\section{Hypothesis testing primer}\label{supp:hypo}

Hypothesis testing is performed in statistics when one wishes to test an assumption regarding some parameter. There are many ``named'' hypothesis tests, including T and Z tests, ANOVA tests, Chi square goodness of fit tests, and the Komolgorov-Smirnov test.\footnote{The Komolgorov-Smirnov test may be tempting to consider, but this test is inappropriate because we are not testing for equality of probability mass functions (PMFs) but rather for equality of a \textit{parameter} appearing within said PMFs.} The appropriateness of a named test depends on the nature of the data used and the reason for the analysis. 
See \url{https://www.datasciencecentral.com/hypothesis-tests-in-one-picture/}, e.g., for a decision chart on when to use some common named hypothesis tests.
However, some data sets and questions do not fall cleanly into an existing named hypothesis testing framework. Our problem is one example of such a case, but can still be addressed within the general hypothesis testing framework.

As mentioned above, hypothesis testing is used to assess the (im)plausibility of a hypothesis given observed data. The steps involved in performing a hypothesis test, whether or not the test is named, are:
\begin{enumerate}
\item State two hypotheses, a null hypothesis and an alternative hypothesis, so that only one can be true.
\item Set a significance level alpha (commonly 0.05). The significance level specifies the probability of rejecting the null hypothesis when it is true. It is how one defines the strength of the evidence that will be required from one's sample in order to reject the null hypothesis; a smaller alpha means stronger evidence will be needed.
\item Determine and calculate the appropriate test statistic -- the test statistic is some function of the observed data that will be used to decide whether to reject the null hypothesis. In our examples and real test procedure, the observed data itself is the test statistic.\footnote{That is, the function mapping the data to a test statistic is just an identity function.}
\item Either calculate a p-value (chance of seeing something at least as extreme as the observed test statistic if the null is true) or determine a critical region (the region in which you will reject the null) for the above significance level and test statistic.
\item Reject the null hypothesis if the p-value is less than alpha or, analogously, if the test statistic falls in the critical region.
\end{enumerate}
A few notes on the basic process. First, the significance level alpha can be made higher or lower depending on the relative risks and rewards of more or less stringent significance requirements. As alpha increases  we have more power to reject the null in the case that it is not true, but we also have a higher chance of rejecting the null hypothesis when we shouldn't have; conversely, as alpha decreases we have a higher chance of not rejecting the null hypothesis when we should have. Second, the decision of whether to reject the null hypothesis will be the same whether one uses a p-value or a critical region approach, but these approaches differ in \textit{how} one makes that decision. The p-value approach asks ``Under the null hypothesis, how likely am I to see a test statistic at least as extreme (i.e., unlikely) as the one I observed?'' E.g., a p-value of 0.03 means that you only have a 3\% chance of seeing a test statistic at least as extreme as that observed if the null hypothesis is true. The critical region approach asks ``Under the null hypothesis, where will the least likely alpha\% of the test statistics fall?'' If your test statistic is in this region, you know it is unlikely to have been seen under the null hypothesis. In some cases, one of the two may be easier to calculate or communicate than the other. We chose to use the critical region approach for our full analysis, but will show how to perform tests using both approaches here.

Below, we first show an example of performing a simple hypothesis test. We then move on to the details of the hypothesis test we use in our analysis.

\subsection{Testing the mean of a Poisson variable using the above framework}

Suppose that we have a bag full of tiles labelled with counting numbers (i.e., $0,1,2,\ldots$). We are interested in whether the average across all numbers in the bag is equal to 2.2. There are infinite number of tiles in the bag. Intuitively, if the mean \textit{were} 2.2 we wouldn't be too surprised if we pulled out a tile and saw a 2 or a 3 label (which are both close to 2.2), but we may be surprised to see a 7 or an 8 (which are both far from 2.2). We assume a Poisson distribution for these counts and we call the mean (average value) $\lambda$; the Poisson distribution is a way for us to specify how likely we are to see each count, given the average count.\footnote{Formally, we assume $y \sim \text{Poisson}(\lambda)$, with $\lambda \geq 0$.} Say we pull one tile from the bag and call its label $y$. We consider two cases: (C1) suppose we observe $y=3$ and (C2) suppose we observe $y=7$.

\subsubsection{State two hypotheses} 

\begin{figure}[htpb]
	\centering
	\includegraphics[width=0.475\textwidth]{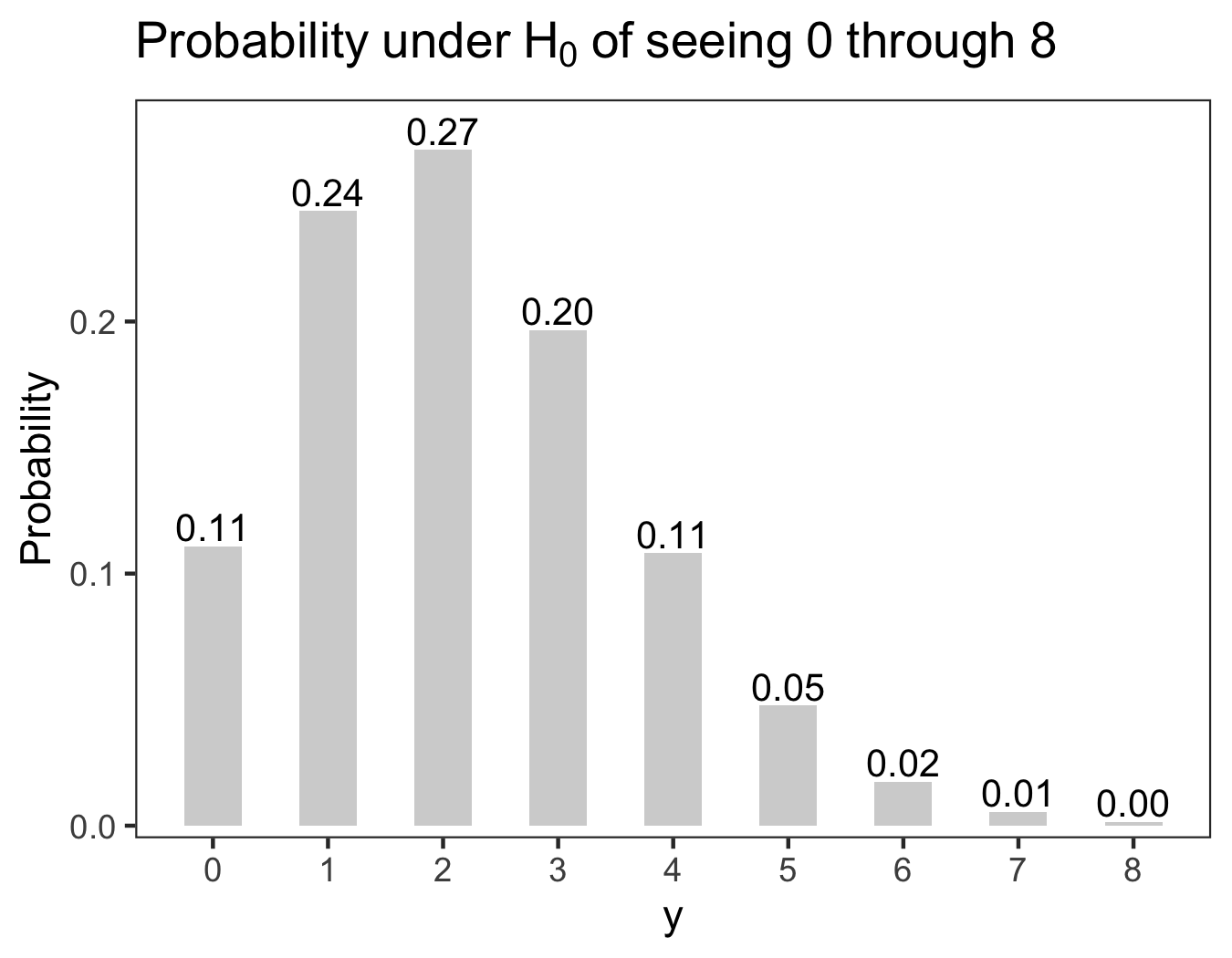}
	\caption{\small The probability of seeing possible counts $y$ assuming a Poisson distribution with mean 2.2. Taller bars indicate a greater chance of seeing that count. (Note the total here is 1.01 instead of 1.0 due to rounding.)\addtocounter{footnote}{-1}\protect\footnotemark\addtocounter{footnote}{+1}}\label{fig:h_simple_1}
\end{figure}

Here the null hypothesis is $\text{H}_0: \ \lambda = 2.2$ and the alternative hypothesis is $\text{H}_{\text{A}}: \ \lambda \neq 2.2$. The probability of seeing a given count $y$ under $\text{H}_0$, that is assuming the mean is 2.2, is shown in Figure \ref{fig:h_simple_1}.

\subsubsection{Set a significance level alpha and define the test statistic} We set alpha to be 0.05. Our test statistic is just the observed count, $y=3$ (in C1) or $y=7$ (in C2).

\subsubsection{Either calculate a p-value or determine a critical region, and } 
\addtocounter{footnote}{-1}\footnotetext{Note any count $y$ is possible but we truncate our visualization past $y=8$.}

\begin{figure}[htpb]
	\centering
	\includegraphics[width=0.475\textwidth]{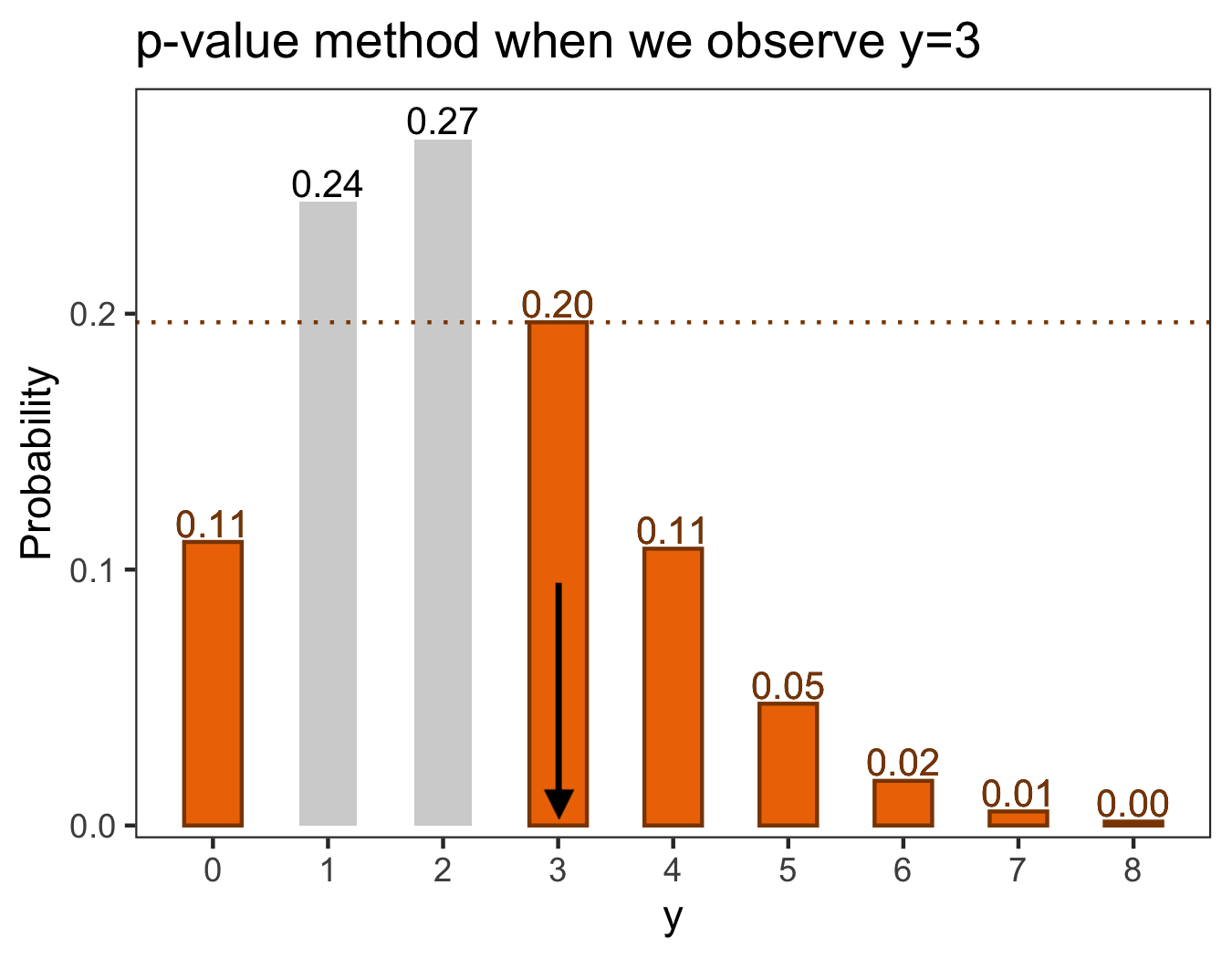} \hspace{5mm}
	\includegraphics[width=0.475\textwidth]{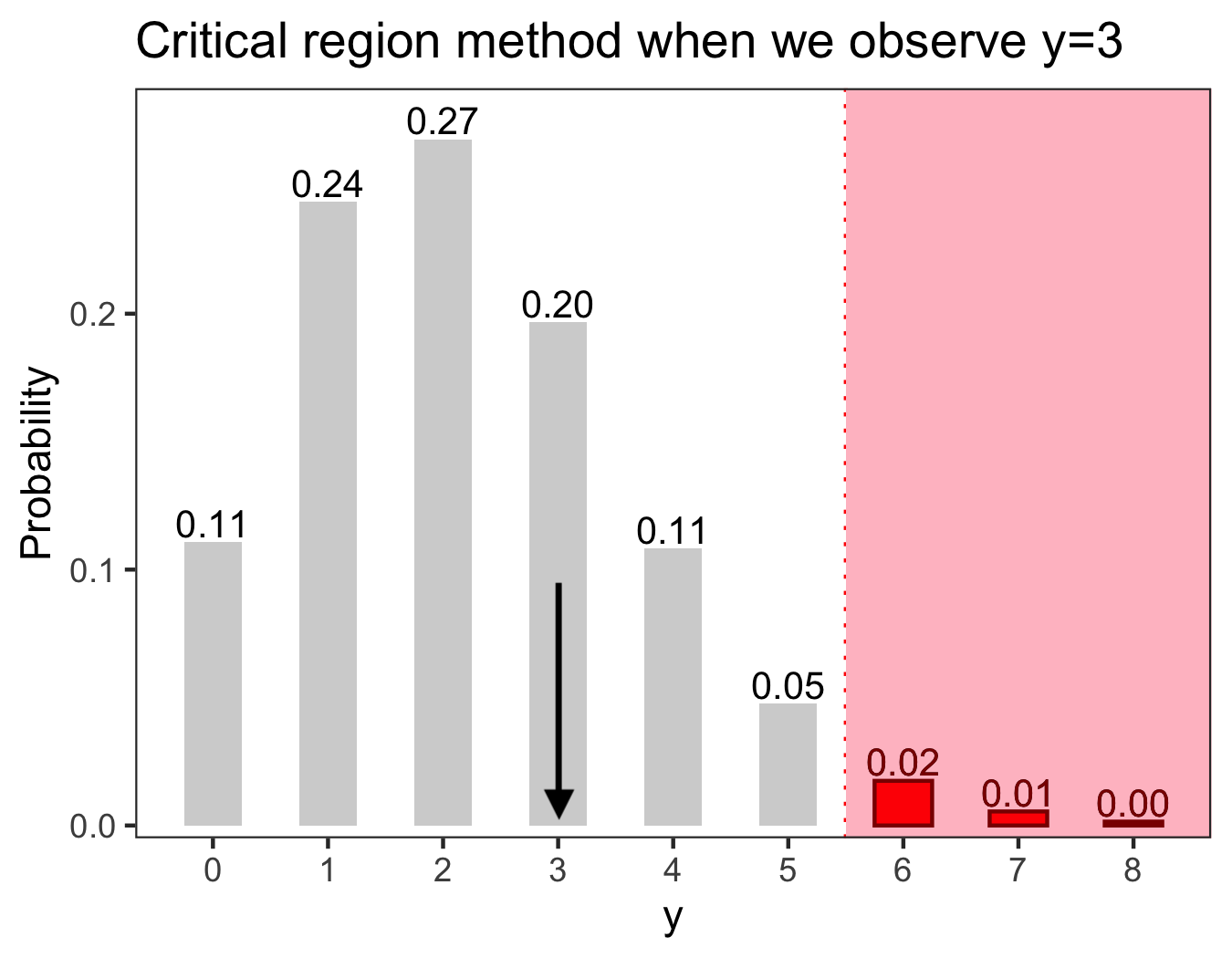}
	\caption{\small Results for C1 when we observe a 3. Left: The probability of seeing $y=3$ is marked by a horizontal dashed line, and the orange highlighted bars show counts we are at least as unlikely to see as 3. Right: The pink shaded background and red highlighted bars indicate the critical region. Note the observed $y$ falls outside of this region.} \label{fig:h_simple_2}
\end{figure}

First, consider the p-value method. Values ``at least as extreme as that observed'' under the null include all counts that are as unlikely to see as a 3. All counts other than 1 and 2 are as unlikely as 3 (see the left sub-plot of Figure \ref{fig:h_simple_2}). If we sum all of these probabilities, we get our p-value: 0.49. That is, we have a 49\% chance of observing a value at least as unlikely as a 3 under the null hypothesis.

Now consider the critical region method. The critical region, where the least probable 5\% of the data fall, includes all counts $y \in \{6, 7, 8, \ldots\}$ (see the right sub-plot of Figure \ref{fig:h_simple_2}).\footnote{Operationally, we get the critical region by finding where the most probable 95\% of the data fall assuming the null is true; the critical region is the complement of that most probable region.}\footnote{We overshoot 95\% here -- the probability of one of 0 through 5 being observed is 0.975. But the probability of one of 0 through \textit{4} being observed is 0.928; due to the discrete nature of the data we need to overshoot in order to get to \textit{at least} 95\%.}

\begin{figure}[htpb]
\centering
\includegraphics[width=0.475\textwidth]{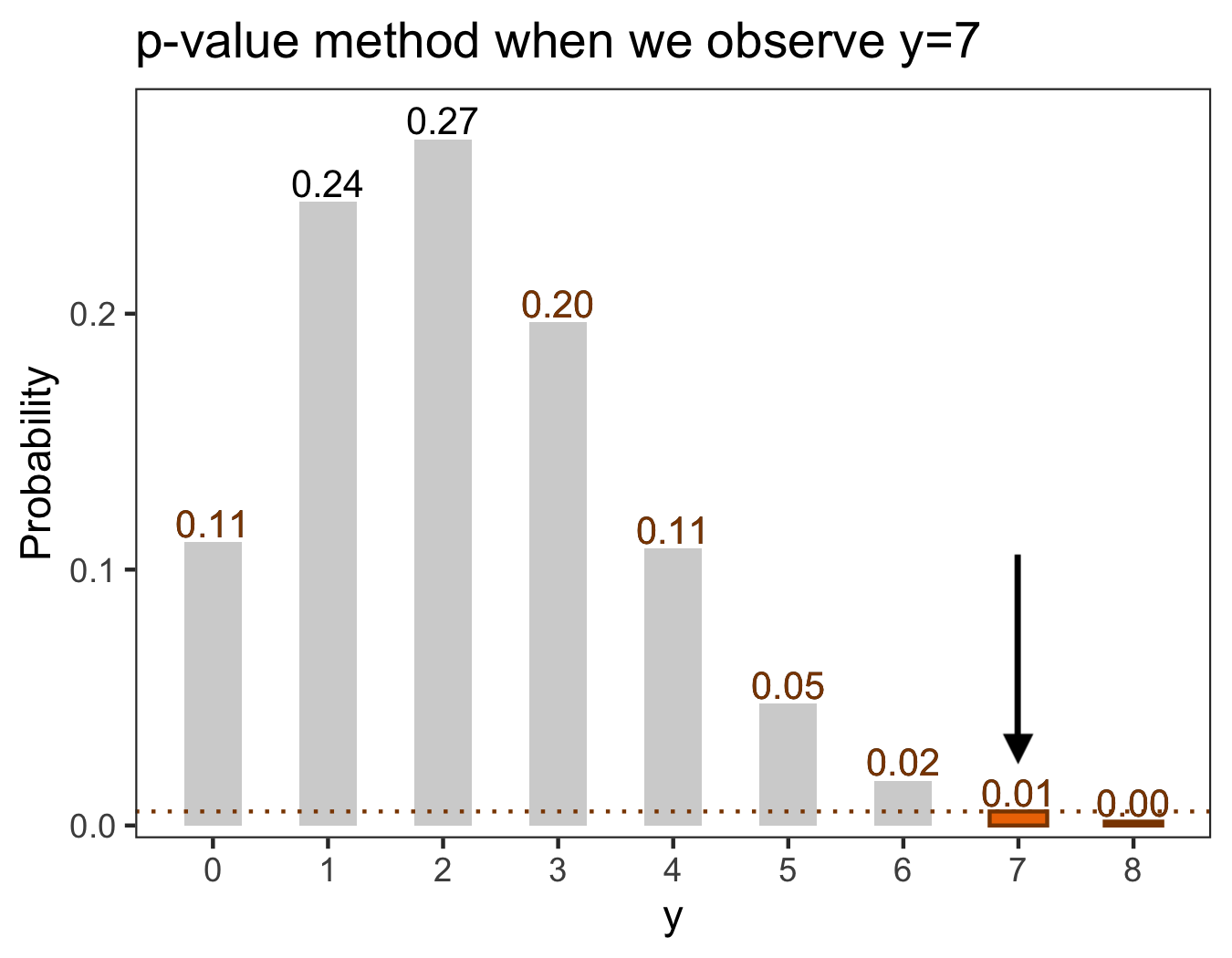} \hspace{5mm}
\includegraphics[width=0.475\textwidth]{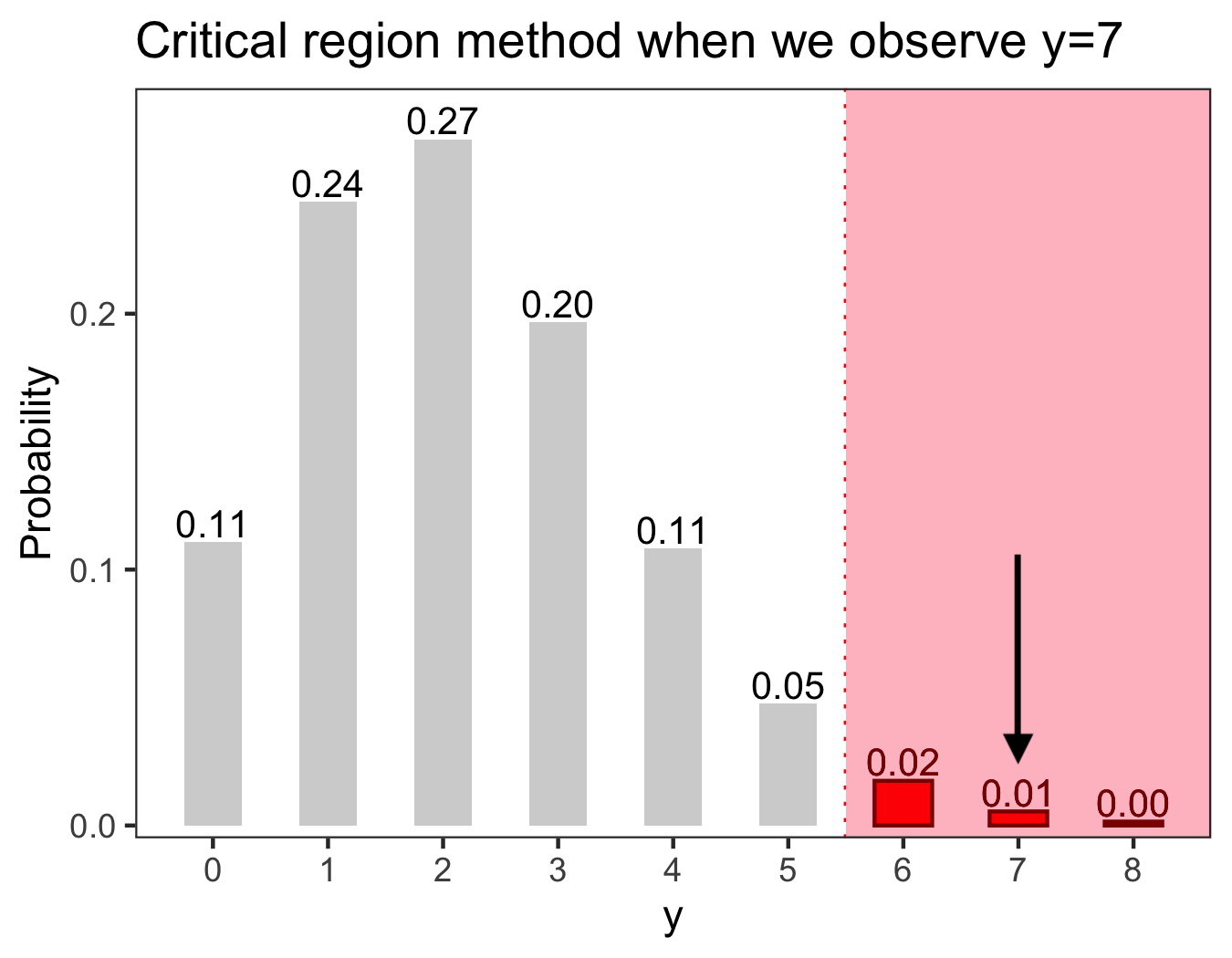}
\caption{\small Analogous results to those shown in Figure \ref{fig:h_simple_2} for C2 when we observe a 7. Note in this case the observed $y$ falls inside of the critical region.} \label{fig:h_simple_2b}
\end{figure}

An example of these same approaches for if we had seen $y=7$ is shown in Figure \ref{fig:h_simple_2b}. Here the p-value is 0.01, and the observed count \textit{does} fall in the critical region. \footnote{Note that seeing a 7 is not \textit{impossible} under the null hypothesis, it's just unlikely.}

\subsubsection{Decide whether you reject the null hypothesis} 

In C1, when $y=3$, we \textit{fail to reject} the null using either means of assessment (which is to be expected -- if the means yield contradictory results, something is wrong!). Our p-value is \textit{greater} than 0.05. Our observed count $y=3$ \textit{does not} fall in the critical region.

In C2, when $y=7$, we \textit{reject} the null using either means of assessment. Our p-value is \textit{less} than 0.05. Our observed count $y=7$ \textit{does} fall in the critical region.

\subsection{Our hypothesis testing process}\label{supp:hypo_vis}

Now suppose that we have \textit{two} bags full of counting numbers (i.e., $0,1,2,\ldots$). We are interested in whether the average across all numbers in the first bag is equal to some value, say $\lambda_i^{abc}$, \textit{and} the average across all numbers in the second bag is equal to another value, say $\lambda_i^{bc}$. There are infinite numbers in both bags. For the sake of exposition, suppose $\lambda_i^{abc} = 3.4$ and $\lambda_i^{bc} = 14.1$. Intuitively, if these were the true means we wouldn't be too surprised if we pulled out a tile from each and saw the pair of counts (3,14) or (4,15), but we may be surprised to see, e.g., (12,14) with a first count far from its mean, (3,45) with a  second count far from its mean, or, most surprisingly, (12,45) with both counts far from their means.

Formally, we observe a \textit{pair} of counts $(y_i^{abc}, y_i^{bc})$, the qABC and qBC counts for some specific orbit, look direction, and ESA. We assume each individual count here comes from a Poisson distribution; these distributions specify how likely we are to see any given count and together can tell us how likely we are to see a \textit{pair} of counts. However, the Poisson distributions, specified in Equation (1) in the main text, have somewhat more complicated mean terms specified by exposure times $t$, signal rates $s$, backround rates $b$, and multiplicative calibration factor $e$. %\ref{eq:mod}
Namely, we have $\lambda_i^{abc}=t_i^{abc}(e_i^{abc} s_i^{abc} + b_i^{abc})$ and $\lambda_i^{bc}=t_i^{bc}(e_i^{bc} s_i^{bc} + b_i^{bc})$. Consider the qABC count, $y_i^{abc}$. Intuitively, we will expect to see a larger number of counts $y_i^{abc}$ when the satellite points in a given direction for longer (i.e., for larger $t_i^{abc}$ is), when the signal rate $s_i^{abc}$ is higher, or when the backround rate $b_i^{abc}$ is higher. The same logic is true for the qBC count $y_i^{bc}$. Suppose we see a specific count pair $(y_i^{abc}, y_i^{bc}) = (2, 13)$ from orbit file 0100.\footnote{Specifically here $i=34981$, which is the 13th observed look direction from orbit file 0100.} We want to know whether the counts could have been generated from some shared signal $s_i$, that is whether the means are $\lambda_i^{abc}=t_i^{abc}(e_i^{abc} s_i + b_i^{abc})$ and $\lambda_i^{bc}=t_i^{bc}(e_i^{bc} s_i + b_i^{bc})$ with $s_i$ common and not qABC- or qBC-specific.

\subsubsection{State two hypotheses} 

Here the null hypothesis is that there \textit{is} some shared signal, i.e., $\text{H}_0: \ s_i^{abc} = s_i^{bc}$. The alternative hypothesis is that there \textit{is not} some shared signal, i.e., $\text{H}_{\text{A}}: \ s_i^{abc} \neq s_i^{bc}$. Under $\text{H}_0$ the pair of counts follows the distribution outlined in Equation (3), i.e., $\lambda_i^{abc}=t_i^{abc}(e_i^{abc} s_i + b_i^{abc})$ and $\lambda_i^{bc}=t_i^{bc}(e_i^{bc} s_i + b_i^{bc})$. %\ref{eq:mod_h0}
We know the times $t_i^*$, backgrounds $b_i^*$, and the multiplicative factors $e_i^*$. We don't actually know the shared signal rate $s_i$ so we use our ``best guess'' at $s_i$ based on the data, i.e., the maximum likelihood estimate $\hat{s}_i$, in order to calculate the means $\lambda_i^{abc}$ and $\lambda_i^{bc}$ assuming the null is true.\footnote{Under the null, we find the maximum likelihood estimate of $s_i$ as
$\underset{s_i}{\mathrm{argmax}} \ \ell(s_i; y_i^{abc}, y_i^{bc}, t_i^{abc}, t_i^{bc}, b_i^{abc}, b_i^{bc}, e_i^{abc}, e_i^{bc}),$
where $\ell(s_i; \cdot)$ is the log likelihood of the joint distribution for $y_i^{abc}$ and $y_i^{bc}$. Specifically, 
$\ell(s_i; \cdot) = \ln(P(y_i^{abc}|s_i,t_i^{abc},b_i^{abc},e_i^{abc}) \times P(y_i^{bc}|s_i,t_i^{bc},b_i^{bc},e_i^{bc})),$
where $P(y_i^{*}|\cdot)$ with $* \in \{abc,bc\}$ is a Poisson probability mass function with mean specified as in Equation (1). Operationally, if we didn't know the analytic form of this MLE we could find the $s_i$ that minimizes $-\ell(s_i; \cdot)$ using an optimization routine available in a software package (e.g., the \texttt{optim()} function in \textbf{R} with settings \texttt{method=`L-BFGS-B'} and \texttt{lower=0}).}
This gives us $\lambda_i^{abc}=4.2$ and $\lambda_i^{bc}=10.4$.\footnote{The values for each of the terms included above are as follows: \{Lat$_i$=32.5; Lon$_i$=130.6; ESA$_i=2$;  $y_i^{abc}=2$;  $y_i^{bc}=13$;  $t_i^{*}=32.1$; $e_i^{abc}=1.0$; $e_i^{bc}=2.1$;  $b_i^{abc}=0.07$;  $b_i^{bc}=0.2$;  $ \hat{s}_i=0.06$;  $\lambda_i^{abc}=4.2$;  $\lambda_i^{bc}=10.4$\}. The shared signal rate estimate $\hat{s}_i$ is calculated using Equation (4) in the main text.}
The null hypothesis that we had framed in terms of our signal rates can now be rewritten in terms of our means as $\text{H}_0: \ \lambda_i^{abc}=4.2$ and $\lambda_i^{bc}=10.4$. The alternative hypothesis is $\text{H}_{\text{A}}: \ \lambda_i^{abc} \neq 4.2$ or $\lambda_i^{bc} \neq 10.4$.

\subsubsection{Set a significance level alpha and define the test statistic} We set alpha to be 0.05. Here our test statistic is just the observed count pair $(y_i^{abc}, y_i^{bc}) = (2, 13)$.

\subsubsection{Either calculate a p-value or determine a critical region} 

\begin{figure}[htpb]
	\centering
	\includegraphics[width=0.5\textwidth]{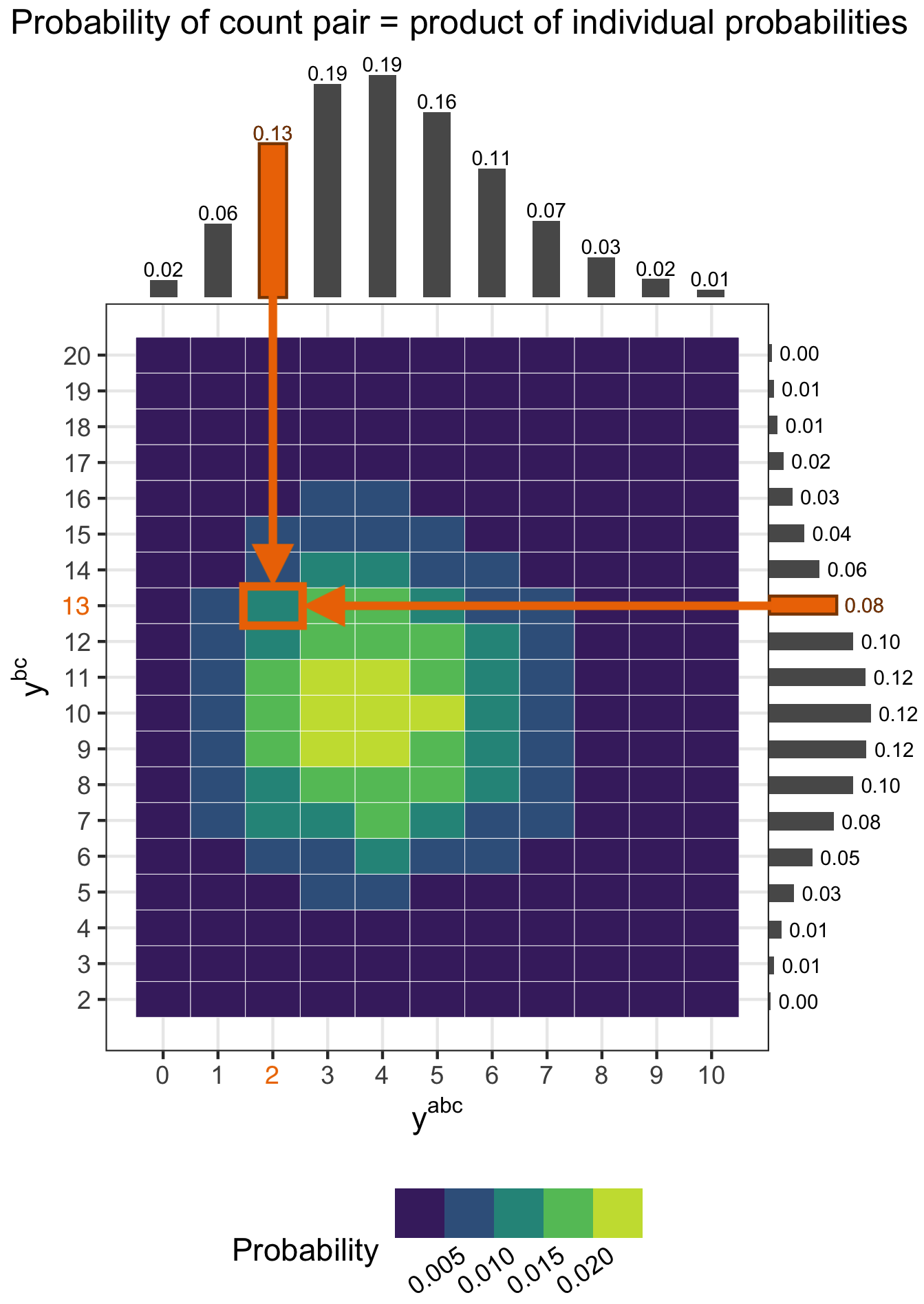}
	\caption{\small Probabilities for $y^{abc}$ and $y^{bc}$ individually (bars on top and right, respectively, analogous to Figure \ref{fig:h_simple_1}) and together (center) under $\text{H}_0$. Higher bars (individual) and lighter color (together) indicate higher probability. The probability of seeing our observed pair, $(y_i^{abc}, y_i^{bc}) = (2, 13)$ marked by the orange box, is the probability that $y_i^{abc}=2$ (top orange bar) times the probability that $y_i^{bc}=13$ (right orange bar), i.e., $0.13 \times 0.08 = 0.01$.\addtocounter{footnote}{-1}\protect\footnotemark\addtocounter{footnote}{+1}}\label{fig:h_real_1}
\end{figure}

A visual of the probabilities of $y^{abc}$ and $y^{bc}$ individually, as well as the probability of seeing each $(y^{abc},y^{bc})$ \textit{pair} (the so-called ``joint probability''), under the null hypothesis is shown in Figure \ref{fig:h_real_1}.\footnote{The joint probability under independence is just the product of each individual (i.e., marginal) probability. Formally, let $P(y^{abc}=j|s_i,t_i^{abc},b_i^{abc},e_i^{abc})=p_j^{abc}$ denote the probability that $y^{abc}=j$ for $j=0,1,2,\ldots$. Similarly, let $P(y^{bc} = k|s_i,t_i^{bc},b_i^{bc},e_i^{bc}) = p_k^{bc}$ denote the probability that $y^{bc}=k$ for $k=0,1,2,\ldots$. Because $y^{abc}$ and $y^{bc}$ are independent, the probability of observing a given pair of values can be calculated as the product of each individual probability. That is, the probability that $y^{abc}=j$ and $y^{bc}=k$ is equal to $p_j^{abc} \times p_k^{bc}$.} These joint probabilities will form the basis for both the p-value method and the critical region method, below.

\addtocounter{footnote}{-1}\footnotetext{Note any pairs of counts are possible but we truncate our visualization to focus around the higher probability pairs.}

\begin{figure}[htpb]
	\centering
	\includegraphics[width=0.9\textwidth]{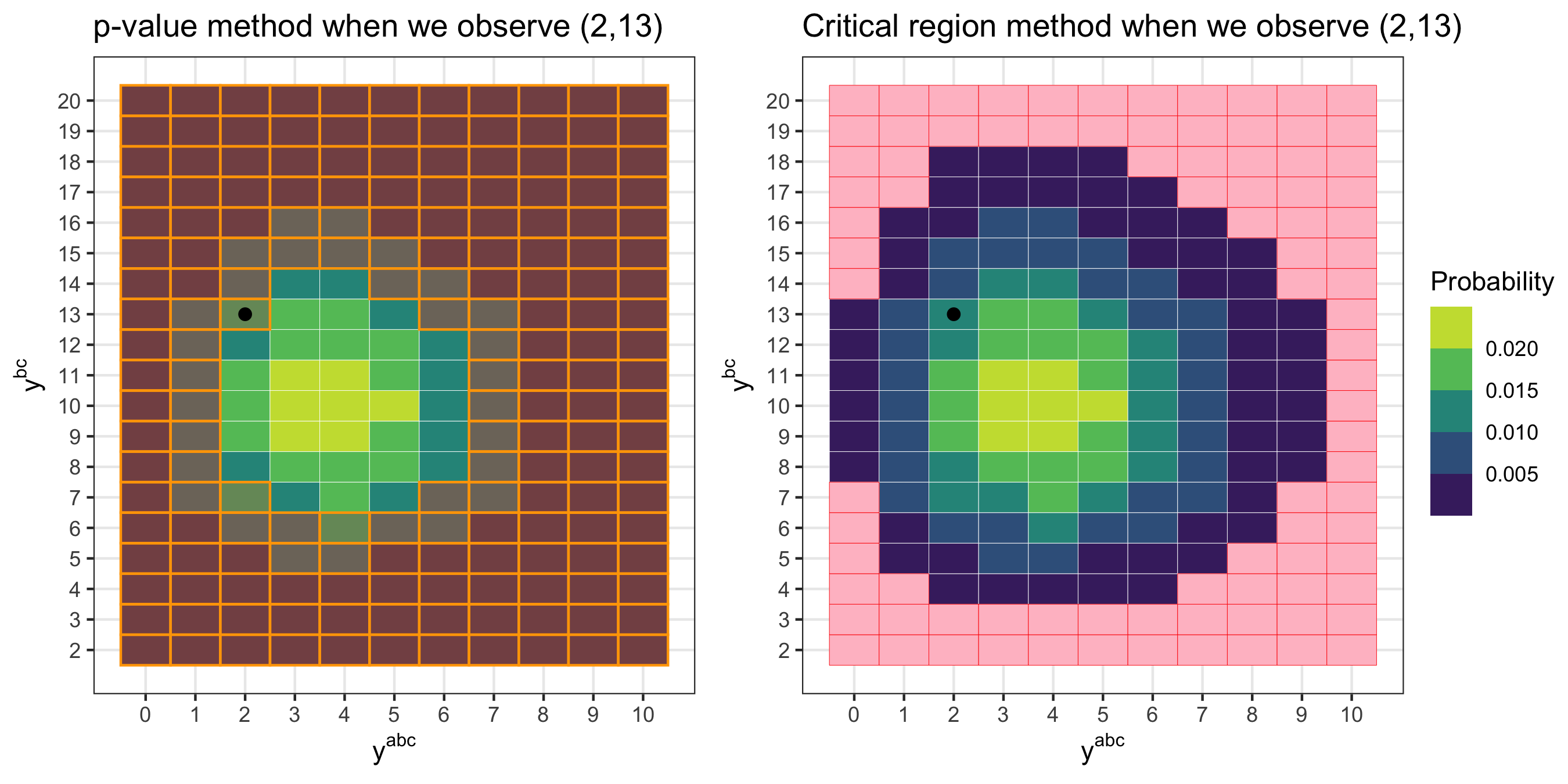}
	\caption{\small Left: The probability of seeing $(y_i^{abc}, y_i^{bc}) = (2, 13)$, the location of which is marked by the black circle, is 0.01. The orange shaded boxes indicate $(y^{abc}, y^{bc})$ pairs having probability less than or equal to 0.01; we sum these probabilities to get our p-value: 0.44. Right: The pink shaded squares with red borders denote the critical region. Note the observed $y$ falls outside of the critical region.}\label{fig:h_real_2}
\end{figure}

First, consider the p-value method. Values ``at least as unlikely as that observed'' under the null include all count pairs having probability equal to or less than the probability of observing (2, 13) in Figure \ref{fig:h_real_1}; these pairs are marked with orange shaded boxes on the left sub-plot of Figure \ref{fig:h_real_2}. If we sum all of these probabilities, we get our p-value: 0.44. That is, we have a 44\% chance of seeing a count pair at least as unlikely as (2, 13) under the null hypothesis.

Now consider the critical region method. The critical region, where the least probable 5\% of the data fall, is shown as the pink region on the right sub-plot of Figure \ref{fig:h_real_2}.\footnote{Operationally, we get the critical region by first finding where the most probable 95\% of the data fall under $\text{H}_0$. To do so we begin at the most probable value(s) and work our way through all possible $(y^{abc}, y^{bc})$ pairs in descending probability order until we reach a cumulative total of least 0.95. The critical region is the complement of this most probable region.} The observed count pair (2, 13) falls outside of the critical region.

\subsubsection{Decide whether you reject the null hypothesis} 

We fail to reject the null using either means of assessment. Our p-value is greater than 0.05. Our observed count pair, $(y_i^{abc}, y_i^{bc}) = (2, 13)$, does not fall in the critical region.

\subsubsection{Making decisions for all observations}

If we kept drawing count pairs from our two bags full of counting numbers \textit{and the null hypothesis were true} we would expect to reject the null about 5\% of the time (because we set alpha to be 0.05). That is, even when the null is true we will sometimes reject it because of chance. We perform this hypothesis testing process for all observations across all orbits, expecting to reject the null for around 5\% of observations if the counts truly could have each been generated from some shared signal. Note that each count pair has its own specific signal rate and pair of means to test because it comes from a different point direction, orbit, and ESA -- that is, for each count pair we imagine a different pair of infinite bags that we are drawing from.

\begin{figure}[thpb!]
	\centering
	\includegraphics[width=0.98\textwidth]{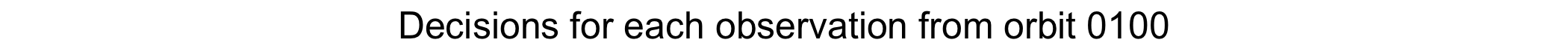}
	\includegraphics[width=0.98\textwidth]{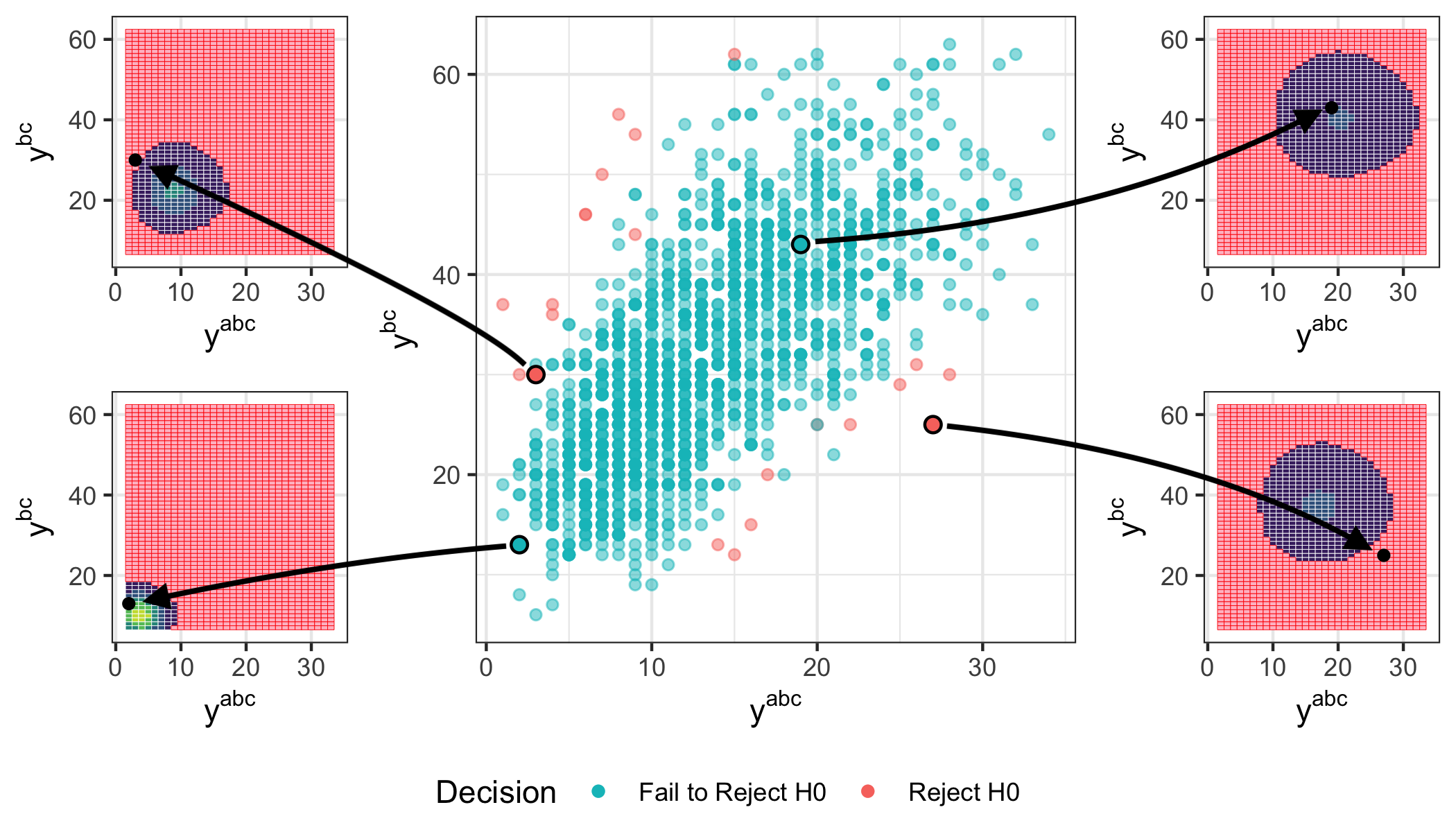}
	\caption{\small For each observation from orbit 0100 we perform a hypothesis test by identifying the critical region \textit{for that observation}; see the four highlighted observations for examples. The result is a fail-to-reject or reject designation for each observation; here, we fail to reject the null for 98.6\% of observations. Note that these designations broken down by ESA were shown in Figure 11.}\label{fig:h_all_1}
\end{figure}

Figure \ref{fig:h_all_1} shows the results for all observations from orbit 0100; the particular observation we used as a vignette above is shown in the bottom left sub-plot. We fail to reject 98.6\% of the observations from this particular orbit.

\clearpage

%%--------------------------------------------------------------------------------------------------
%%--------------------------------------------------------------------------------------------------

\section{Probability integral transform primer}\label{supp:pit}

\begin{figure}[thpb]
	\centering
	\includegraphics[width=0.98\textwidth]{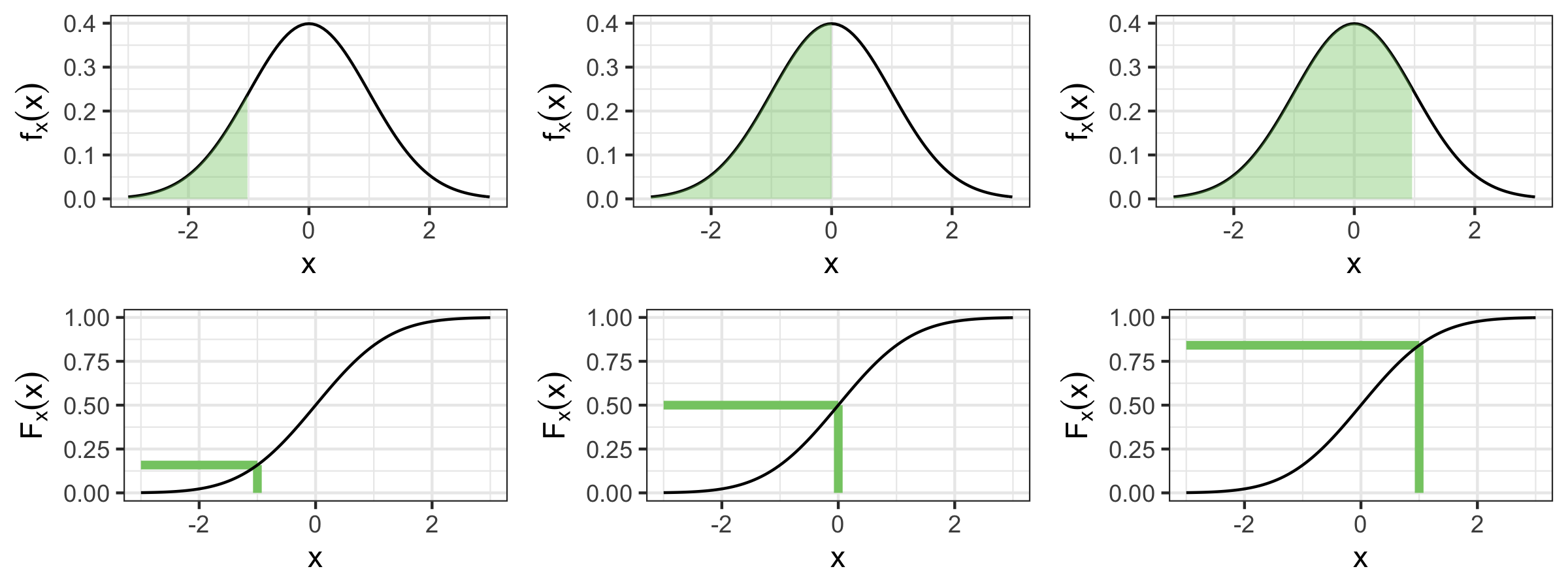}
	\caption{\small PDF (top) and CDF (bottom) of the standard normal distribution. The area of the shaded region under the PDF equals the value of $F_X(x)$ in the CDF plot; from left to right the $x$ values shown are -1, 0, and 1.}\label{fig:pit_cdfs}
\end{figure}

First, some definitions for continuous variables (i.e., variables that can take on an infinite set of possible values within some range). 
An example of a continuous variable is centimeters of rainfall per year, which can take on as its value any real nonnegative number.
A \textbf{continuous distribution} describes the probabilities of the possible values of a continuous random variable. A \textbf{probability density function (PDF)} is a function that integrates to 1 providing the relative likelihood that a continuous random variable takes on a given value. The \textbf{cumulative distribution function (CDF)} is the probability that a continuous random variable $X$ will have a value less than or equal to some value $x$; analogously, it is the integral of the PDF from negative infinity up to $x$. The CDF can only take values between 0 and 1, inclusive. A common continuous distribution is the standard normal distribution (i.e., a normal distribution with mean 0 and variance 1), which has PDF $f_X(x) = \frac{e^{-x^2/2}}{\sqrt{2\pi}}$ and CDF $F_X(x) = {\frac{1}{\sqrt{2\pi}}}\int_{-\infty }^{x}e^{-t^{2}/2}\,dt$, and support over all real numbers. 
Figure \ref{fig:pit_cdfs} shows the PDF and CDF of the standard normal with three different example $x$ values evaluated.

Informally, the \textbf{probability integral transform (PIT)}, also called the CDF transform, of some sample $x$ from a continuous distribution is generated by plugging that sample back into its own CDF. 
If you sample many times from a continuous distribution and plug those samples into the CDF of that distribution, the output is a set of uniform draws between 0 and 1. 
Formally, the PIT of $x$ is found by calculating $F_X(x) = p(X \leq x)$, and a property of the PIT is that $F_X(X) ~ \text{Unif}(0,1)$ for $X$ having some continuous distribution.

\begin{figure}[thpb]
	\centering
	\includegraphics[width=0.9\textwidth]{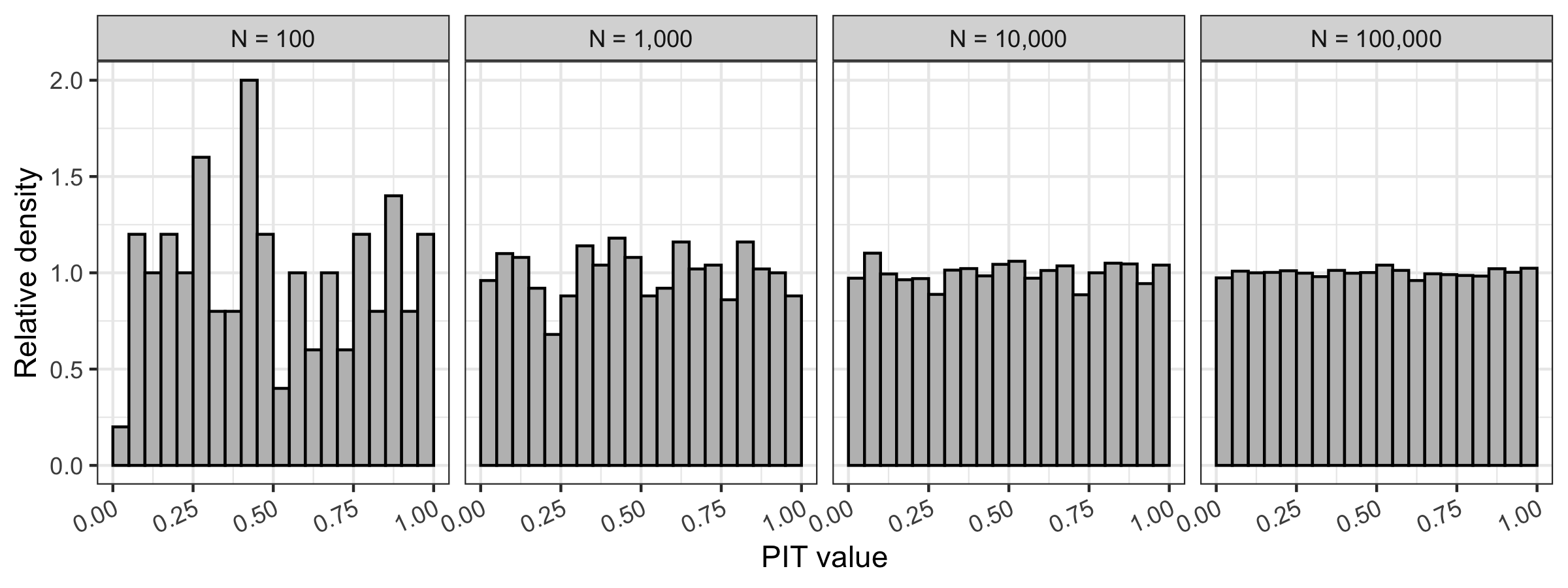}
	\caption{\small PIT histograms of the standard normal distribution for $N$ samples for four different choices of $N$, beginning with $N=100$ and increasing by a factor of 10 within each sub-plot from left to right.}\label{fig:pit_nsamps}
\end{figure}

To illustrate this concept, we sample $x_1,\ldots,x_N$ from a standard normal distribution and calculate each $F_X(x_1), \ldots, F_X(x_N)$. We expect the resulting histogram of PIT samples (i.e., the $F_X(x_1), \ldots, F_X(x_N)$ values) to look approximately uniform. Figure \ref{fig:pit_nsamps} shows these PIT histograms for different choices of $N$. Notice that sampling randomness leads to bumpier looking histograms for smaller values of $N$.

\begin{figure}[thpb]
	\centering
	\includegraphics[width=0.9\textwidth]{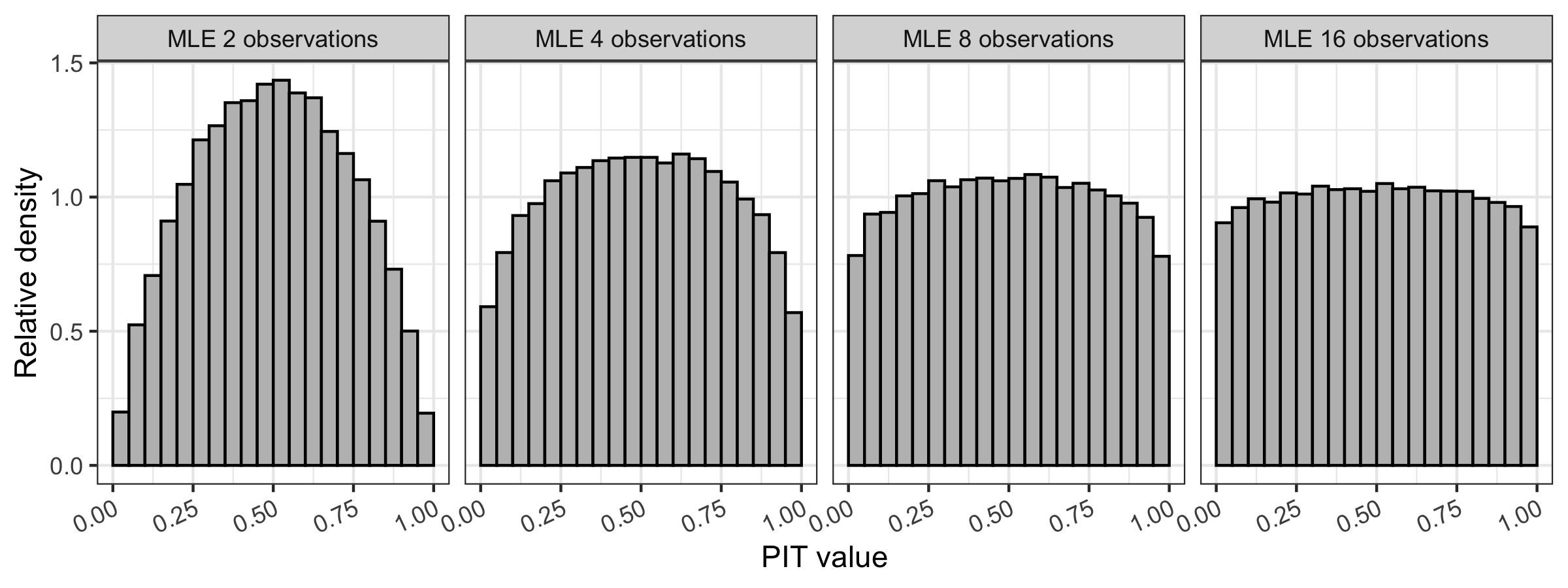}
	\caption{\small PIT histograms of the standard normal distribution where the sample mean calculated from $M$ samples is used in place of the true mean. From left to right, $M=2,4,8,$ and $16.$ The fewer samples are used, the more ``mounding'' is present in the distribution of PIT values. As $M$ increases, the distribution approaches that of a standard uniform. In each case, $N=100,000$ samples from the PIT distribution were drawn.}\label{fig:pit_mles}
\end{figure}

The above examples have assumed we knew the parameters of the distribution in question (that is, we assumed we knew our samples came from a normal distribution with mean 0 and variance 1). 
Suppose instead we didn't know the mean of the distribution and used an estimate based on the data -- specifically, the mean of $M$ samples.
To illustrate the effect this mean estimation step has on the distribution of PIT values, we simulate $N=100,000$ PIT realizations under different choices of $M$.
To get a single PIT draw, we sample $x_1,\ldots,x_M$ from a standard normal distribution, get the sample mean $\Bar{x} = \frac{1}{M}\sum_{j=1}^M x_j$ and calculate $\Bar{F}_X(x_1)$ where $\Bar{F}_X$ is the CDF of a normal distribution having mean $\Bar{x}$ and variance 1.
That is, we imagine that we have observed these samples $x_1,\ldots,x_M$ for which we \textit{know the variance} but we \textit{don't know the true mean}, so we use the sample mean in place of this unknown true mean in the calculation of $\Bar{F}_X$.
The sample mean becomes a better and better estimate of the true mean as the number of samples increases.
The result of doing the above $N$ times for $M=2,4,8,$ and $16$ is shown in Figure \ref{fig:pit_mles}.
The fewer samples are used, the more ``mounding'' is present in the distribution of PIT values.
That is, \textbf{when you use the MLE in place of the true mean parameter there are fewer PIT values close to 0 or 1 than one would expect under uniformity, and more close to 0.5}.
This effect occurs because using the MLE in place of the true mean will lead any given sample within the set of draws contributing to the mean to be closer to it (on average) than they are to the true mean.
As $M$ increases, the MLE approaches the truth and the distribution approaches that of a standard uniform.

\begin{figure}[htpb]
	\centering
	\includegraphics[width=0.98\textwidth]{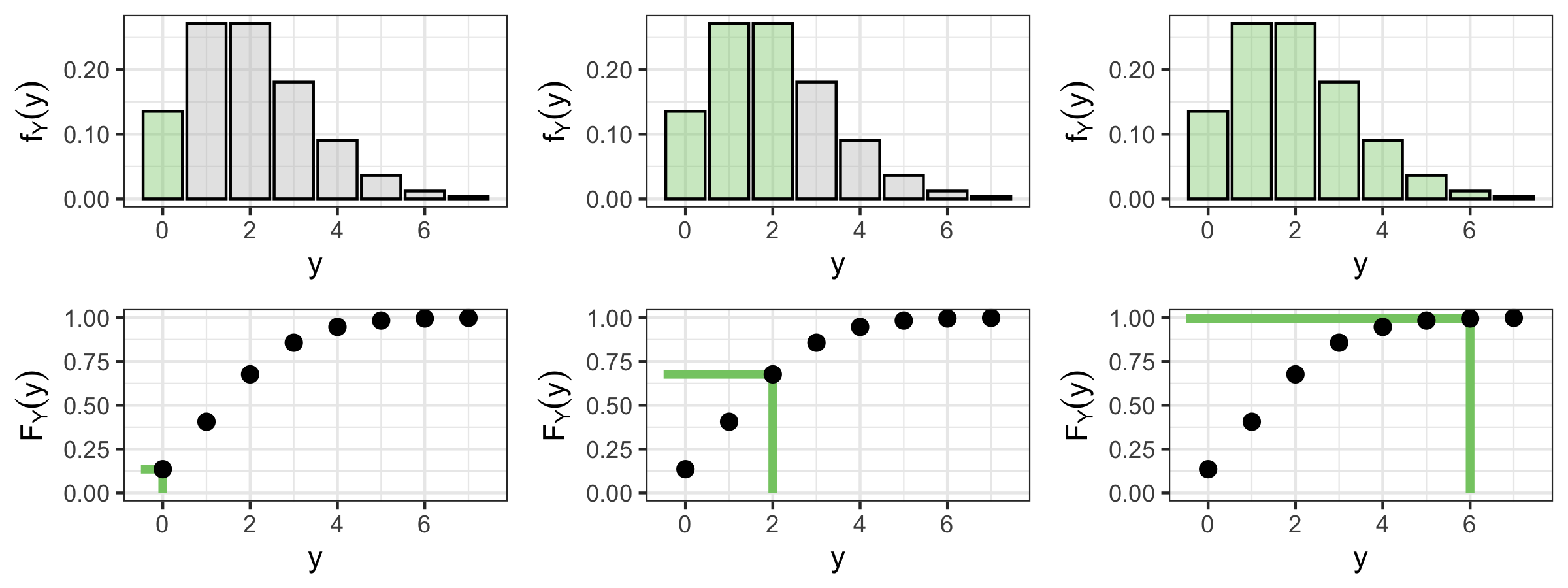}
	\caption{\small PMF (top) and CDF (bottom) of the Poisson distribution having mean 2. The sum of the shaded PMF values equals the value of $F_Y(y)$ in the CDF plot; from left to right the $y$ values shown are 0, 2 and 6.}\label{fig:pit_cmfs}
\end{figure}

We now turn to the discrete variable setting. As in the continuous variable setting, we begin with some definitions for discrete variables (i.e., variables that can take on a finite or countably infinite set of possible values within some range). 
An example of a discrete variable is number of particles observed by a sensor, which can take on as its value any nonnegative integer.
A \textbf{discrete distribution} describes the probabilities of the possible values of a discrete random variable. A \textbf{probability mass function (PMF)} provides the probability that a discrete random variable is exactly equal to some value; the sum across all possible values taken is 1. The CDF at a value $y$ in the discrete case is the sum of the PMF from negative infinity up to and including $y$.
We focus on the Poisson distribution, which is the discrete distribution considered in the main text, with some mean $\lambda$.
Call this random Poisson variable $Y$, having PMF $f_Y(y) = \frac{e^{ - \lambda } \lambda ^y }{y!}$ and CDF $F_Y(y) = p(Y \leq y) = e^{ - \lambda } \sum_{j=0}^y \frac{\lambda ^j }{j!}$.
Figure \ref{fig:pit_cmfs} shows the PMF and CDF of the Poisson distribution with mean 2 having three different example $y$ values evaluated.

\begin{figure}[thpb]
	\centering
	\includegraphics[width=0.6\textwidth]{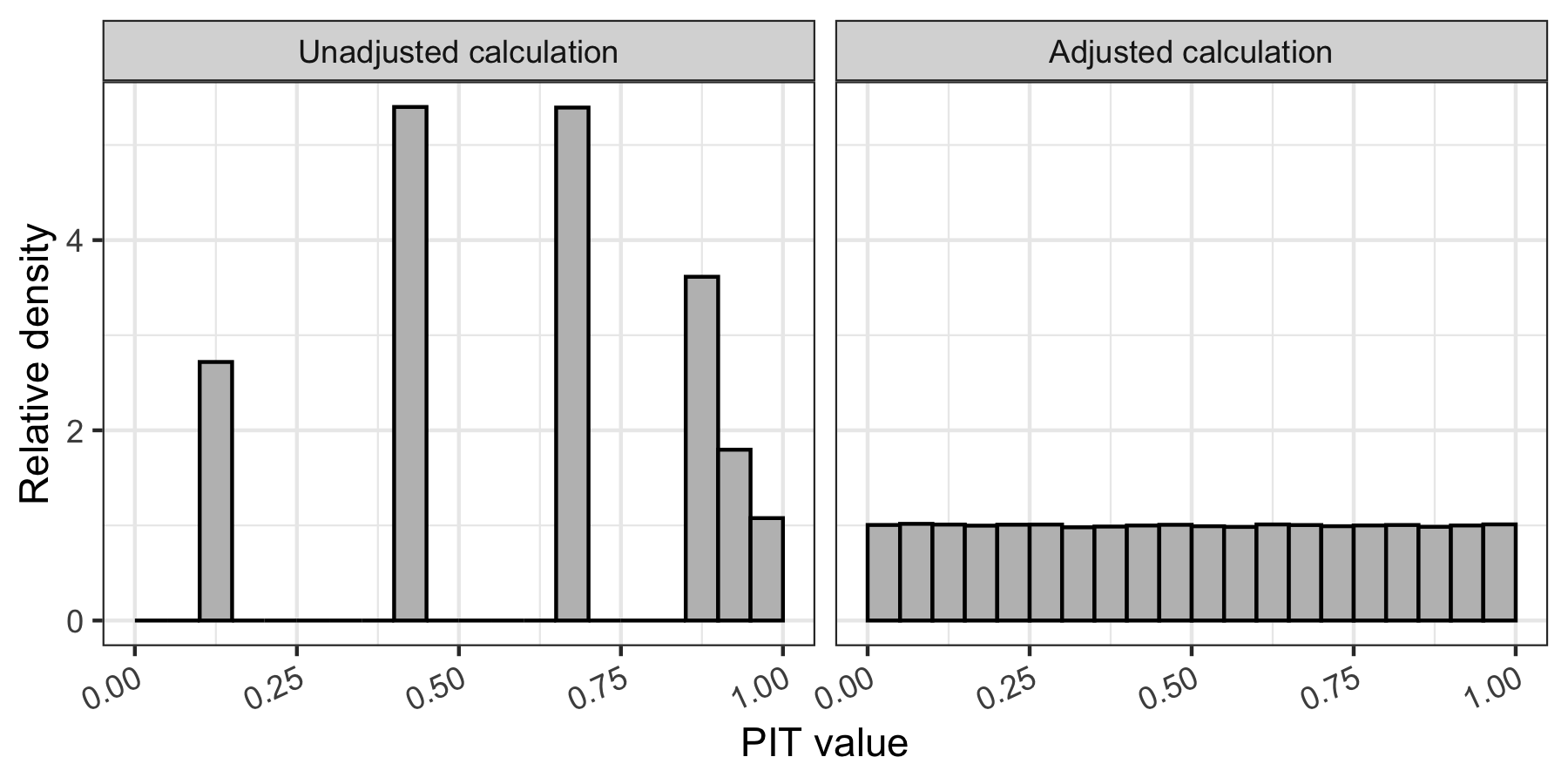}
	\caption{\small Unadjusted (left) and stochastically adjusted (right) PIT histograms of the Poisson distribution with mean 2. In each case, $N=100,000$ samples from the PIT distribution were drawn.}\label{fig:pit_discrete}
\end{figure}

Suppose we tried to perform the analog to the PIT calculation process from above using draws from this discrete distribution, i.e., sample lots of times from a Poisson distribution and plug those samples into the CDF $F_Y$ of that distribution. 
The output will \text{not} be a set of uniform draws between 0 and 1, because $F_Y$ can only take a discrete set of possible values. 
Consider, e.g., the Poisson distribution with mean $\lambda = 2$. Then the probabilities that $Y$ takes on values 0 through 3 are 0.14, 0.27, 0.27, and 0.18, respectively, and $F_Y(0)=0.14, F_Y(1)=0.14+0.27=0.41, F_Y(2)=0.14+0.27+0.27=0.68, F_Y(3)=0.14+0.27+0.27+0.18=0.86,$ and so on. 
That is, it is impossible for $F_Y$ to return any value less than 0.14, or between 0.14 and 0.41, etc. The left side of Figure \ref{fig:pit_discrete} shows the histogram of $F_Y(y_1), \ldots, F_Y(y_N)$ for $N=100,000$ when the $y_i$ are samples from a Poisson distribution with mean 2.

The standard PIT calculation requires that the random variables for which the PIT is calculated are continuous. 
If the random variables are discrete, however, the values the PIT can take will \textit{also} be discrete and thus not uniformly distributed between 0 and 1.
We adjust the PIT calculation for discrete random variables by taking a uniform sample between $F_Y(y-1)$ and $F_Y(y)$ to be the PIT value, rather than simply using $F_Y(y)$ itself as the PIT value (note if $y=0$, we use 0 instead of $F_Y(y-1)$ for the lower bound of the uniform) \cite{dunn1996randomized}.
This sampling step enables \textit{any} value between 0 and 1 to result from the PIT calculation, and creates discrete-adjusted PIT values that are uniformly distributed between 0 and 1.
The right side of Figure \ref{fig:pit_discrete} shows the histogram of $F^*_Y(y_1), \ldots, F^*_Y(y_N)$ for $N=100,000$ when the $y_i$ are samples from a Poisson distribution with mean 2, and $F^*_Y(y)$ is a uniform sample between $F_Y(y-1)$ and $F_Y(y)$ if $y>0$ and a uniform sample between $0$ and $F_Y(y)$ if $y=0$.
As expected, the distribution of the $F^*$ values appears Uniform(0,1).

\begin{figure}[thpb]
	\centering
	\includegraphics[width=0.7\textwidth]{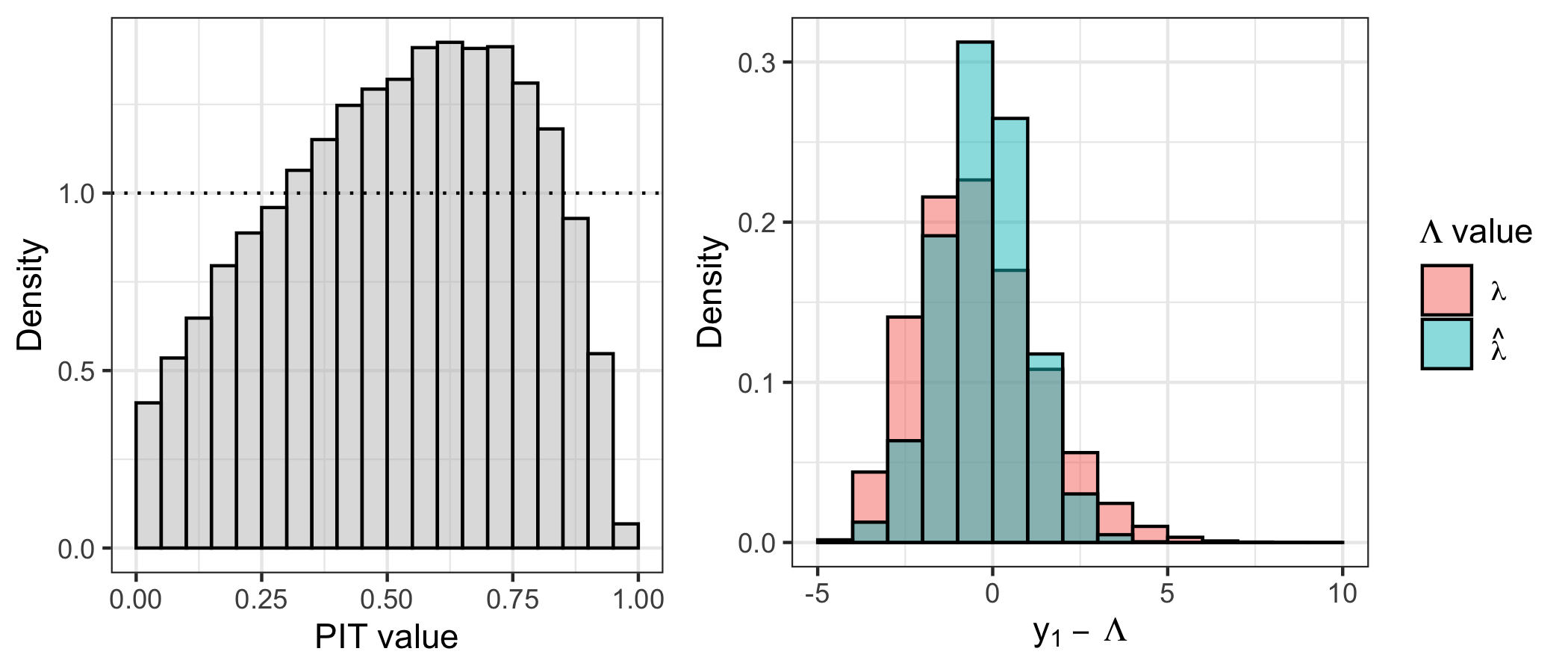}
	\caption{\small Left: The $N=100,000$ sampled stochastically adjusted PIT values for $y_1$ using $\hat{\lambda} = \frac{y_1+y_2}{2}$ in place of $\lambda$ in the calculation. The dashed horizontal line denotes the Uniform(0,1) density. Right: An overlay of the difference between the observed count and the true mean ($y_1 - \lambda$), and that between the observed count and the MLE ($y_1 - \hat{\lambda}$) for the $N$ replicates.}\label{fig:pit_mlePois}
\end{figure}

In the main text, we used the MLE of the shared signal rate from a pair of Poisson observations in the adjusted PIT calculations. As a simpler example illustrating the asymmetric mounding that occurs in this case, we do the following:
\begin{enumerate}
\item Sample $y_1, y_2 \sim \text{Poisson}(\lambda)$, where the true $\lambda=3.1$. 
\item Calculate $\hat{\lambda} = \frac{y_1+y_2}{2}$, the MLE of the mean based on the two samples.
\item Perform the discrete-adjusted PIT calculation for $y_1$ using $\hat{\lambda}$.
\item Repeat 1 through 3 above $N=100,000$ times.
\end{enumerate}
The $N$ sampled PIT values are shown on the left sub-plot of Figure \ref{fig:pit_mlePois}. We see both mounding and asymmetry relative to a Uniform(0,1) distribution. The right sub-plot of Figure \ref{fig:pit_mlePois} helps elucidate why this occurs. It shows the observed count minus the true mean ($y_1 - \lambda$) in pink, and the observed count minus the MLE ($y_1 - \hat{\lambda}$) in blue. The former is more skewed and higher variance relative to the latter. The reduced variance when using the MLE in place of the truth is what creates the PIT distribution mounding; you don't get as many observations $y_1$ ``far'' from $\hat{\lambda}$ as you do from $\lambda$, because the former uses $y_1$ in its calculation. The change in skewness relative to that when using the true mean is what creates the PIT distribution asymmetry. You can imagine additional area in the pink histogram bins below 0 in the right sub-plot of Figure \ref{fig:pit_mlePois} as pushing the PIT distribution to the right; that is, because you see \textit{fewer} $y_1$ values smaller than $\hat{\lambda}$ than are smaller than $\lambda$, you see fewer PIT values less than 0.5 and more above 0.5. Given all of the above, \textbf{we expect to see mounding and asymmetry in the PIT distributions in the main paper}, and we explore whether the shape and skew matches what we would expect under a true shared signal for the two specified Poisson distributed counts using simulation.

\clearpage

%%--------------------------------------------------------------------------------------------------

\section{Uncertainty about $b$ and $e$}\label{supp:uncertainty}

Note we drop the $i$-indexing in this section for notational convenience -- throughout assume each parameter is implicitly indexed by $i$. Although we treat both $b^{*}$ and $e^{bc}$ as known while performing the hypothesis testing calculations in our analysis, there is some amount of uncertainty about both terms. For the backgrounds, we are given the standard deviation $\sigma^{*}_{b}$ of each background term, with $\sigma^{*}_{b}$ relatively small compared to $b^{*}$. 

\begin{figure}[htpb]
\centering
\includegraphics[width=0.8\textwidth]{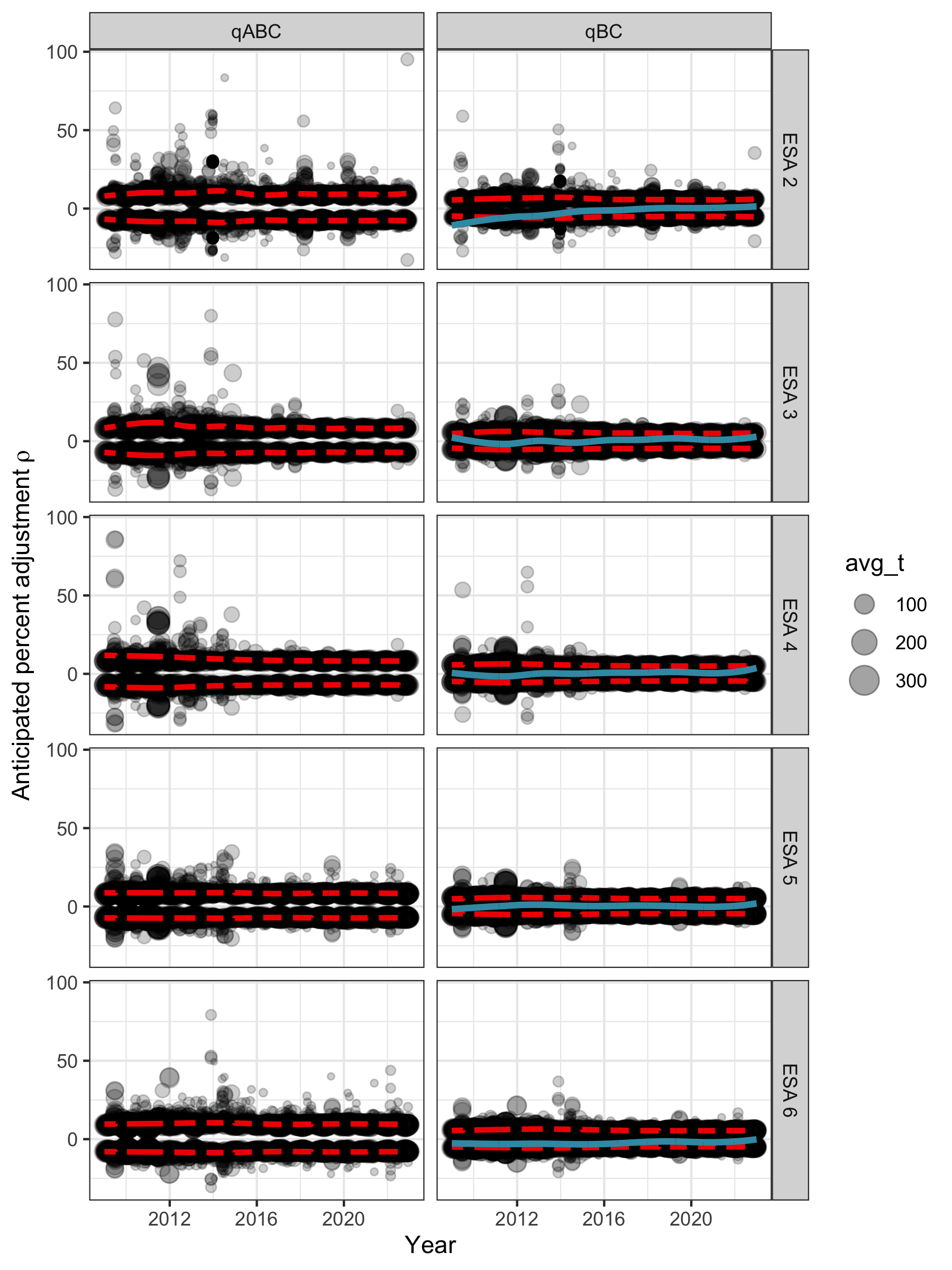}
\caption{\small Each positive and negative point corresponds to a single unique background from an ESA-orbit (most ESA-orbits have only one unique background, but some have multiple) and shows the percent error on said background suggested by plus or minus two times the background standard deviation $\sigma^{bc}_{b}$ provided by mission scientists. The points are sized according to average exposure time and mirrored to show both the positive and negative bounds. The temporally fitted bounds on this background adjustment are shown as two red lines, and the temporally fitted learned qBC background adjustment from the main text is shown as a blue line, as in Figure 10.}\label{fig:b2sd}
\end{figure}

\begin{figure}[htpb]
\centering
\includegraphics[width=0.98\textwidth]{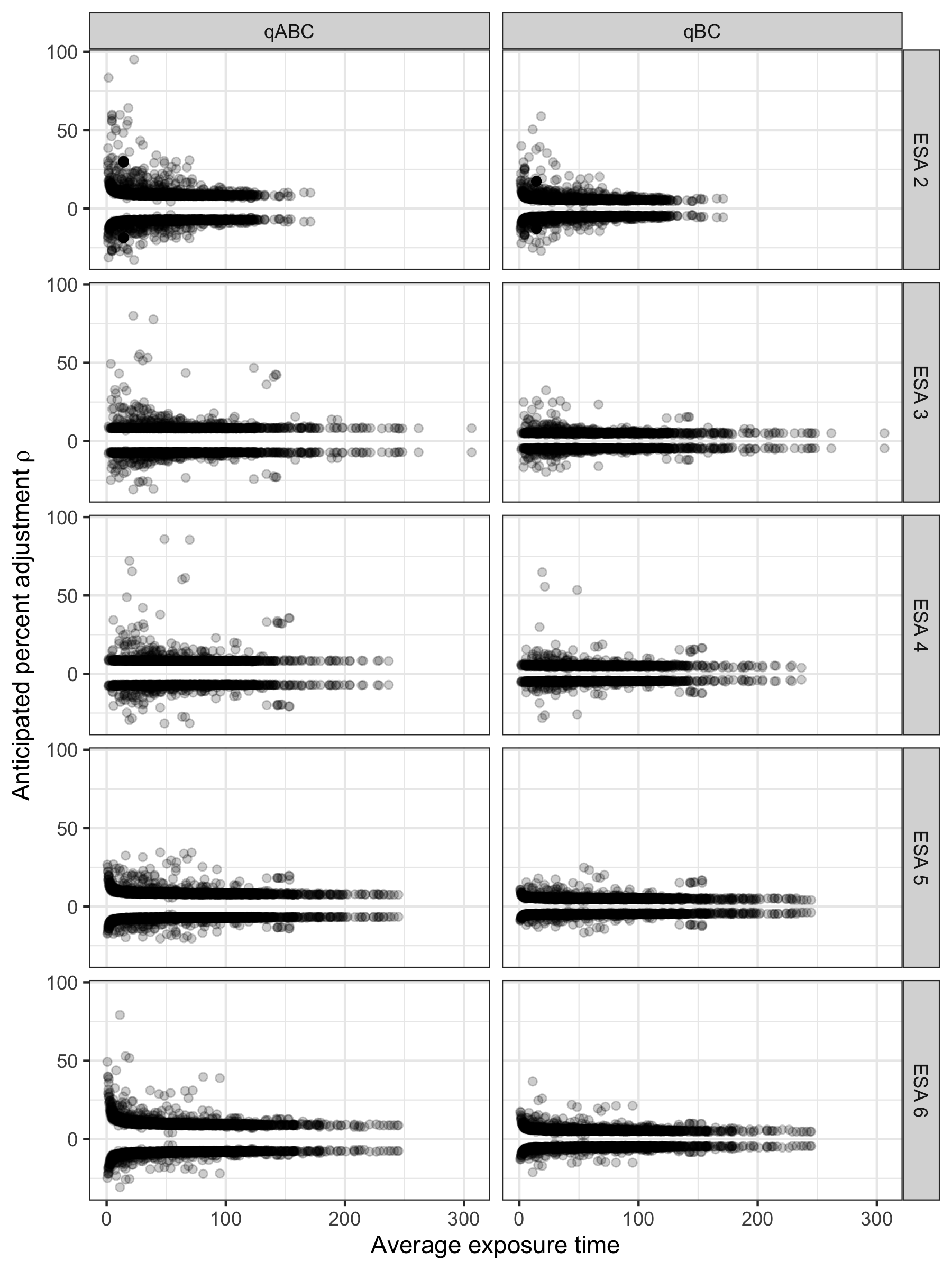}
\caption{\small Each positive and negative point corresponds to a single unique background from an ESA-orbit and shows the percent error on said background suggested by plus or minus two times the background standard deviation $\sigma^{bc}_{b}$ provided by mission scientists. The points are mirrored to show both the positive and negative bounds. These adjustments are plotted by average exposure time.}\label{fig:berr_time}
\end{figure}

Figure \ref{fig:b2sd} shows the unique values of $+/- 2 \times \sigma^{bc}_{b}$ by ESA-orbit, transformed to be in terms of implied percent adjustment of the qBC backgrounds. It also shows an exposure time weighted GAM fit to these percent adjustments, giving us an error bound within which any learned qBC background adjustment is within that expected by mission scientists. Figure \ref{fig:berr_time} shows the relationship between average exposure time and anticipated adjustment.

\begin{figure}[htpb]
\centering
\includegraphics[width=0.8\textwidth]{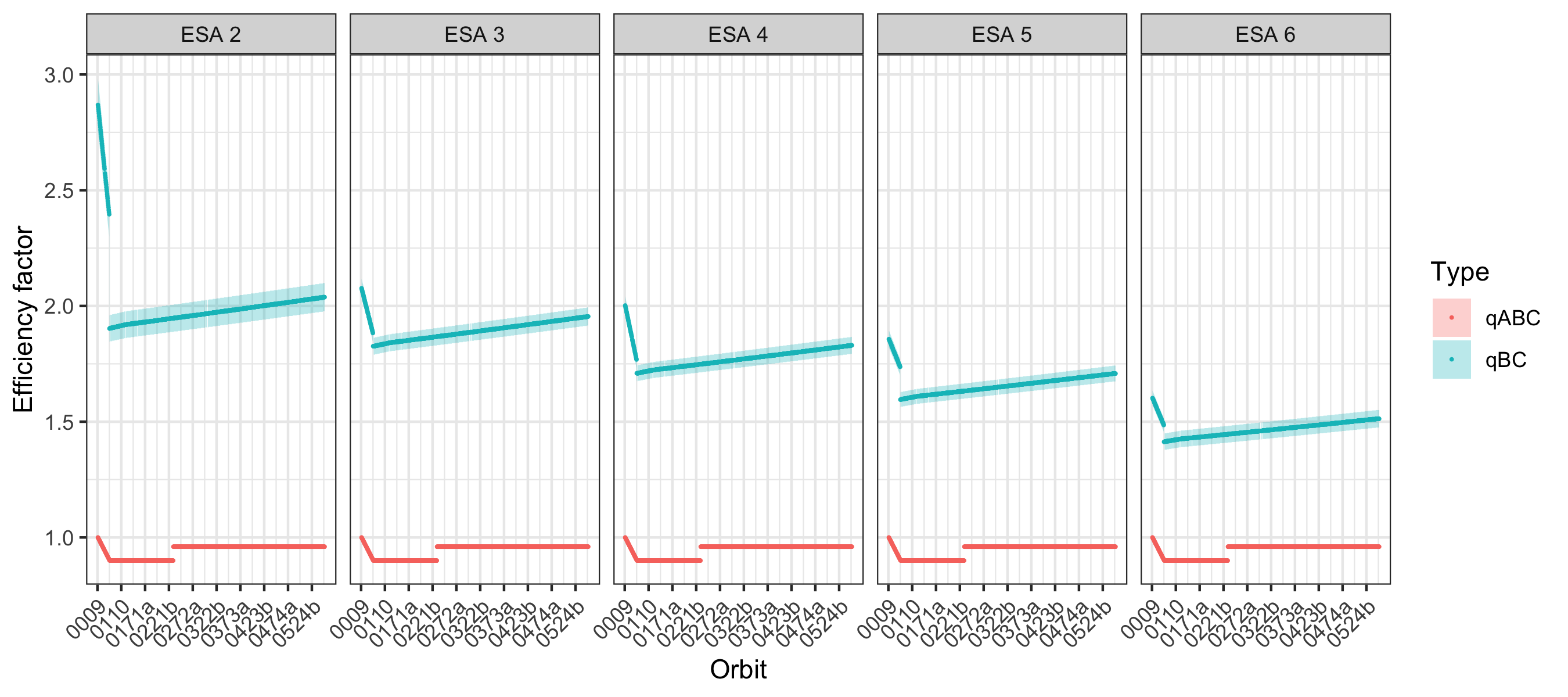}
\caption{\small Orbit- and ESA-specific values of the efficiency factor $e^{*}$ over time, with the center line showing the provided pointwise values. The bands show the uncertainty interval from $e_{ll}^{bc}$ and $e_{ul}^{bc}$ for each (orbit, ESA) pair; there is no uncertainty about $e^{abc}$ provided.} \label{fig:uncertainty_alpha}
\end{figure}

The $e^{bc}$ term, i.e., the multiplicative factor on the signal rate of the qBC counts, is comprised of known ESA-specific multiplier $G$ and ESA- and orbit-specific factors $c^{bc}$ (known within some interval, i.e., $c^{bc} \in [c_{ll}^{bc},c_{ul}^{bc}]$). Note $e^{abc}$ is known. 
That is, we have $e^{bc} \in [e_{ll}^{bc}, e_{ul}^{bc}]$ with $e_{ll}^{bc} = G c_{ll}^{bc}$ and $e_{ul}^{bc} = G c_{ul}^{bc}$. 
Figure \ref{fig:uncertainty_alpha} shows $e^{*}$ values over time (with uncertainty, for $e^{bc}$).

\begin{figure}[htpb]
\centering
\includegraphics[width=0.35\textwidth]{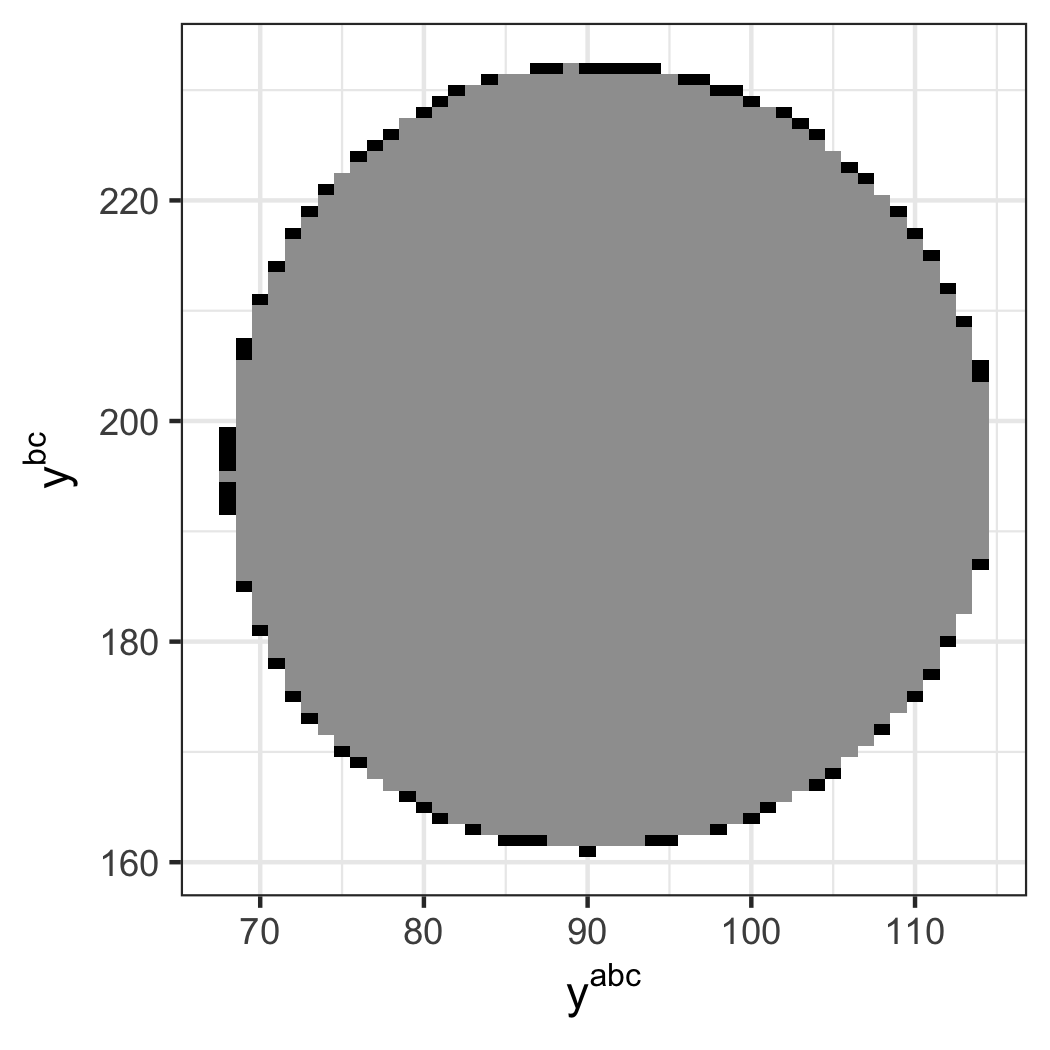}
\caption{\small The 95\% HDR formed using \textit{fixed} $b^{*}$ and $e^{bc}$ (grey) vs \textit{uncertain} $b^{*}$ and $e^{bc}$, where the backgrounds are sampled from a normal distribution having standard deviation $\sigma^{*}_{b}$ (black showing the additional count pairs included in the uncertain-based HDR relative to the fixed-based HDR). For the uncertain parameter HDR the black regions represent a total probability of only 0.00786; the black region is along the ``fringe'' of the HDR where the probabilities of a count pair taking those values are extremely low.} \label{fig:uncertainty_fixvsran}
\end{figure}

There are a number of benefits to treating $b^{*}$ and $e^{bc}$ as known through the analysis itself, namely:
\begin{itemize}
\item The hypothesis test can be performed analytically and computation is much faster than it would be under simulation-based hypothesis testing.
\item For a given orbit when $b^{*}$ and $\sigma^{*}_{b}$ are shared across observations, we expect the realized background to be common across those observations. The same is true for $e^{bc}$. Fixing these values automatically respects this constraint.
\item The uncertainty in $e^{bc}$ and $b$ leads to a negligible increase in the marginal uncertainty of the counts $y$ (i.e., the spread of $y$ is largely driven by inherent data noise and not uncertainty in $e^{bc}$ and $b$). See Figure \ref{fig:uncertainty_fixvsran}.
\end{itemize}
There is also very little risk in the choice to fix $b^{*}$ and $e^{bc}$. This choice leads to smaller uncertainty intervals for $(y^{abc}, y^{bc})$ than having uncertain $b^{*}$ and $e^{bc}$. Therefore, if we fail to reject $\text{H}_0$ with fixed parameters then we would also fail to reject $\text{H}_0$ in the uncertain-parameter case. That is, a known-parameter fail-to-reject automatically gets us an uncertain-parameter fail-to-reject too, and if we validate the qBC data with fixed parameters we would also validate it with uncertain parameters.
If we fail to validate the qBC data with fixed parameters (i.e., if we reject more than the proportion of count pairs rejected under simulation) we can go back and do a more complex analysis that incorporates the uncertainty
about $b^{*}$ and $e^{bc}$.

\clearpage

%%--------------------------------------------------------------------------------------------------
%%--------------------------------------------------------------------------------------------------

\section{Synthetic reference distribution creation}\label{supp:methods_simulation}

A given synthetic reference data set that assumes a shared signal is created by first using the real IBEX-Hi data to learn a reasonable estimate of a shared signal rate, and then resampling new qABC and qBC counts assuming this shared signal.
The code for producing a GAM fit and simulating data for a given orbit and ESA is shown below. Each GAM fit is done to a single orbit and ESA's worth of estimated signals, which are observed at 360 look directions at most. More precisely, the GAM inputs for a given ESA-orbit are the look direction indices, the outputs are $\tilde{s}_i$, and the weights are $w_i = \frac{t_i^{abc} + t_i^{bc}}{\sum_i( t_i^{abc} + t_i^{bc})}$ for $i=1,\ldots,360$ (although some look directions may be missing data). Note $\{t\}=\{t_{abc}\}=\{t_{bc}\}$ for our data set, so the weights simplify to $w_i = \frac{t_i}{\sum_i t_i}$.

\begin{code}
df = data.frame(y=s_neg, x=look_ind, w=t/sum(t))
gamfit = mgcv::gam(y~s(x,k=30), data=df, weights=w)
sdot_gam = mgcv::predict.gam(gamfit, newdata=data.frame(x=x))
sdot_smooth = sapply(sdot_gam, function(val) max(0,val))
lamdot_abc = t*(alpha_abc*sdot_smooth + b_abc)
lamdot_bc = t*(alpha_bc*sdot_smooth + b_bc)
ydot_abc = rpois(nrow(df), lamdot_abc)
ydot_bc = rpois(nrow(df), lamdot_bc)
\end{code}

%In simulations used for \hl{bootstrapping [more detail, and enumerate?]} a CvM statistic critical value, we add an extra simulation layer. 
%Specifically, we create a synthetic reference set according to the above, and we then treat this synthetic set as ``real'' data and simulate a new (doubly simulated) data set following the above process.
%The resulting pipeline allows us to see how much variability about the CvM statistics we could expect comparing ``real'' and ``synthetic'' data for which we know there was a true underlying shared signal in the mission data via this process.

For simulations used for assessing model component errors, we use the new ``corrected'' value of that model component when simulating data according to the above process. 
Doing so leads the synthetic data to in fact have that new model component as the true data generating process, e.g., for the qBC background error assessment we assume that the provided $b_i^{bc}$ actually has $\rho\%$ error, then simulate our reference data set using this new corrected qBC background level $\tilde{b}_i^{bc}$, where $\tilde{b}_i^{bc} = b_i^{bc} \cdot (1+\rho/100)$.

\begin{figure}[htpb]
	\centering
	\includegraphics[width=0.95\textwidth]{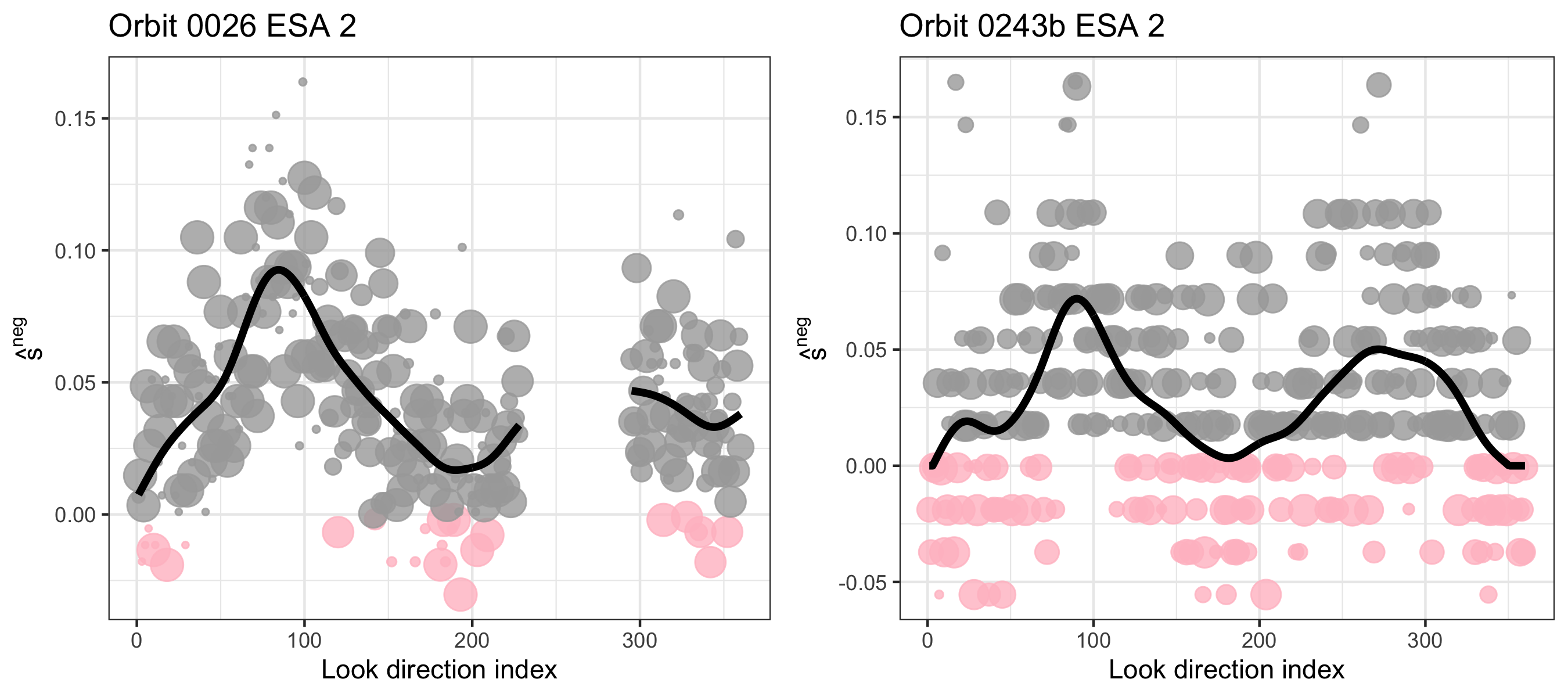}
	\caption{\small Weighted nonnegative GAM fit $\dot{s}_i^{smooth}$ (line) to the unconstrained MLE $\tilde{s}$ (points) for orbit 0026 (left) and orbit 0243b (right) at ESA 2. Negative $\tilde{s}$ values are shown in pink.} \label{fig:supp_simgam}
\end{figure}

Figure \ref{fig:supp_simgam} shows the smooth signal rate predictions, $\dot{s}_i^{smooth} = \max(0, \dot{s}_i^{GAM})$, for example orbits 0026 and 0243b at ESA 2. Note that the effect of the 0 truncation of the GAM fit (i.e., that $\dot{s}_i^{smooth} = \max(0, \dot{s}_i^{GAM})$) can be seen on the far right side, in the Orbit 0234b sub-plot.

\begin{figure}[htpb]
	\centering
	\includegraphics[width=0.95\textwidth]{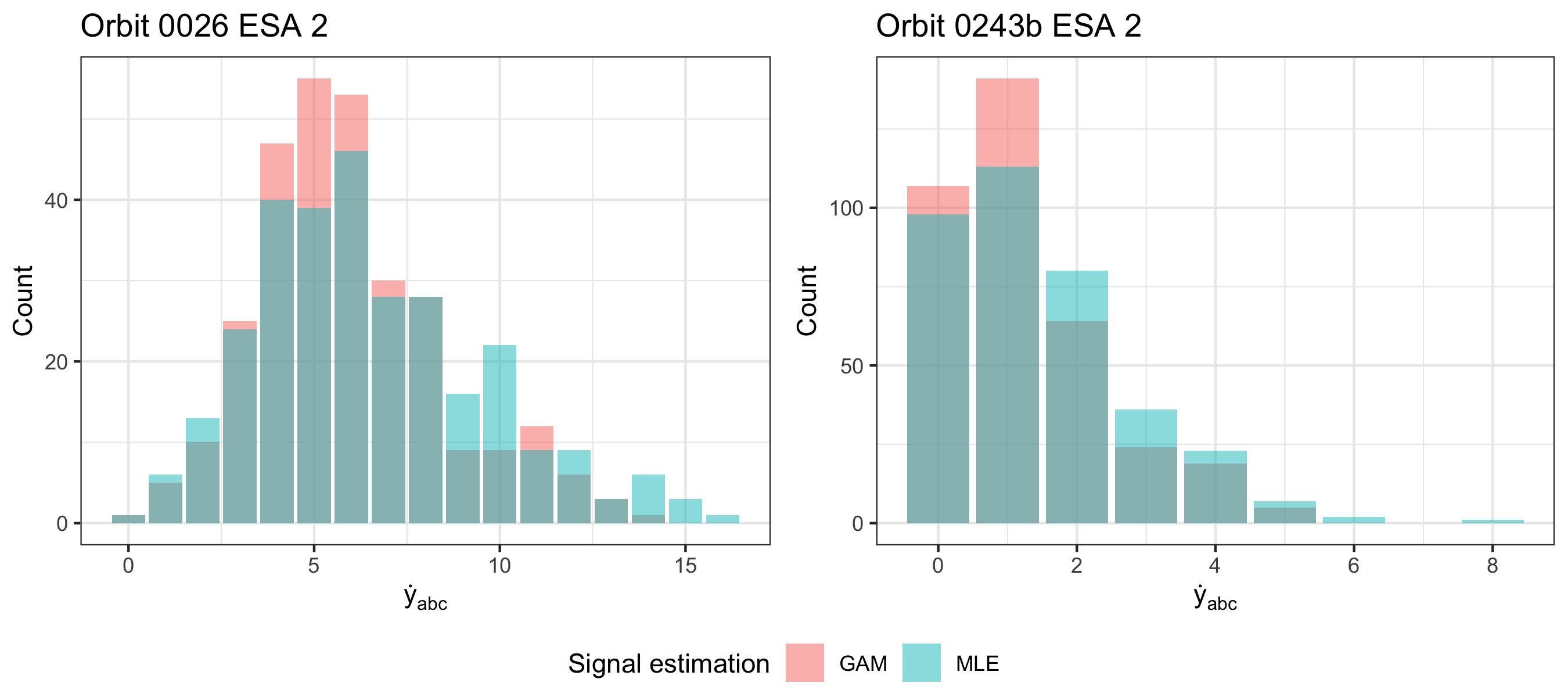}
	\caption{\small Realizations of simulated counts $\dot{y}_i^{abc}$ for example orbits 0026 and 0243b at ESA 2 using $\{\dot{\lambda}_i^{abc}\}$ as the mean (GAM estimation method) or $\{\hat{\lambda}_i^{abc}\}$ (MLE estimation method).} \label{fig:supp_simdata}
\end{figure}

Figure \ref{fig:supp_simdata} shows realizations of simulated counts $\dot{y}_i^{abc}$ for example orbits 0026 and 0243b at ESA 2 by using either the GAM-based method to calculate the expectation of the counts $\dot{\lambda}^{abc}$, or just using the MLEs from pairs of observations $\hat{\lambda}^{abc}$. The MLE-based method leads to more variability in the simulated counts -- note the wider spread relative to the GAM-based method. Because we know the MLE from a single count pair is an overfit estimate of the signal rate, this higher variability is pathologic. Using the GAM-based method gives synthetic data that are more reflective of the true data-generating process, which involves a signal rate that should be fairly smooth.

%\clearpage

%%--------------------------------------------------------------------------------------------------
%%--------------------------------------------------------------------------------------------------

\section{Additional validation results figures}\label{supp:validation_extra}

\begin{figure}[htpb]
	\centering
	\includegraphics[width=0.98\textwidth]{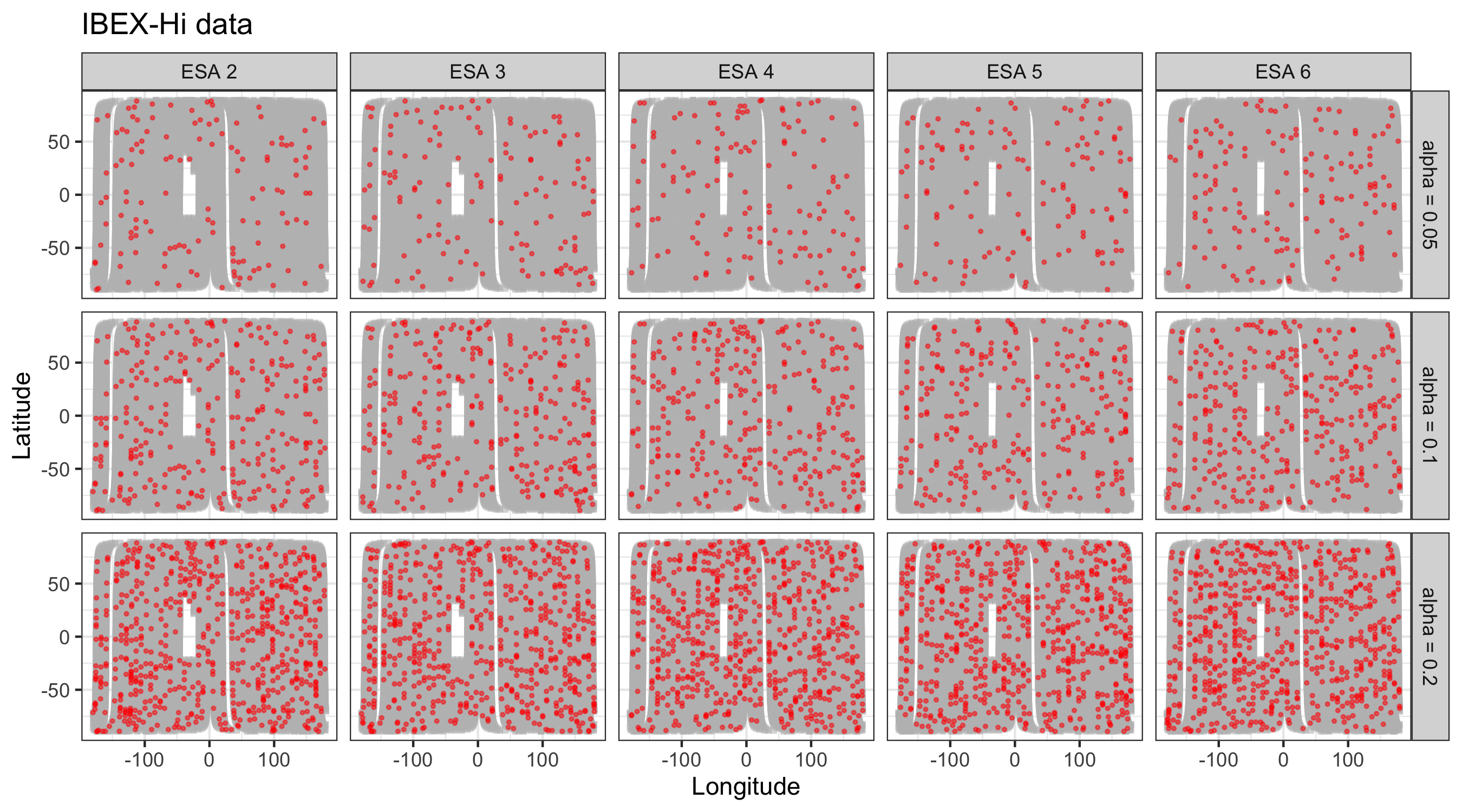}
	\includegraphics[width=0.98\textwidth]{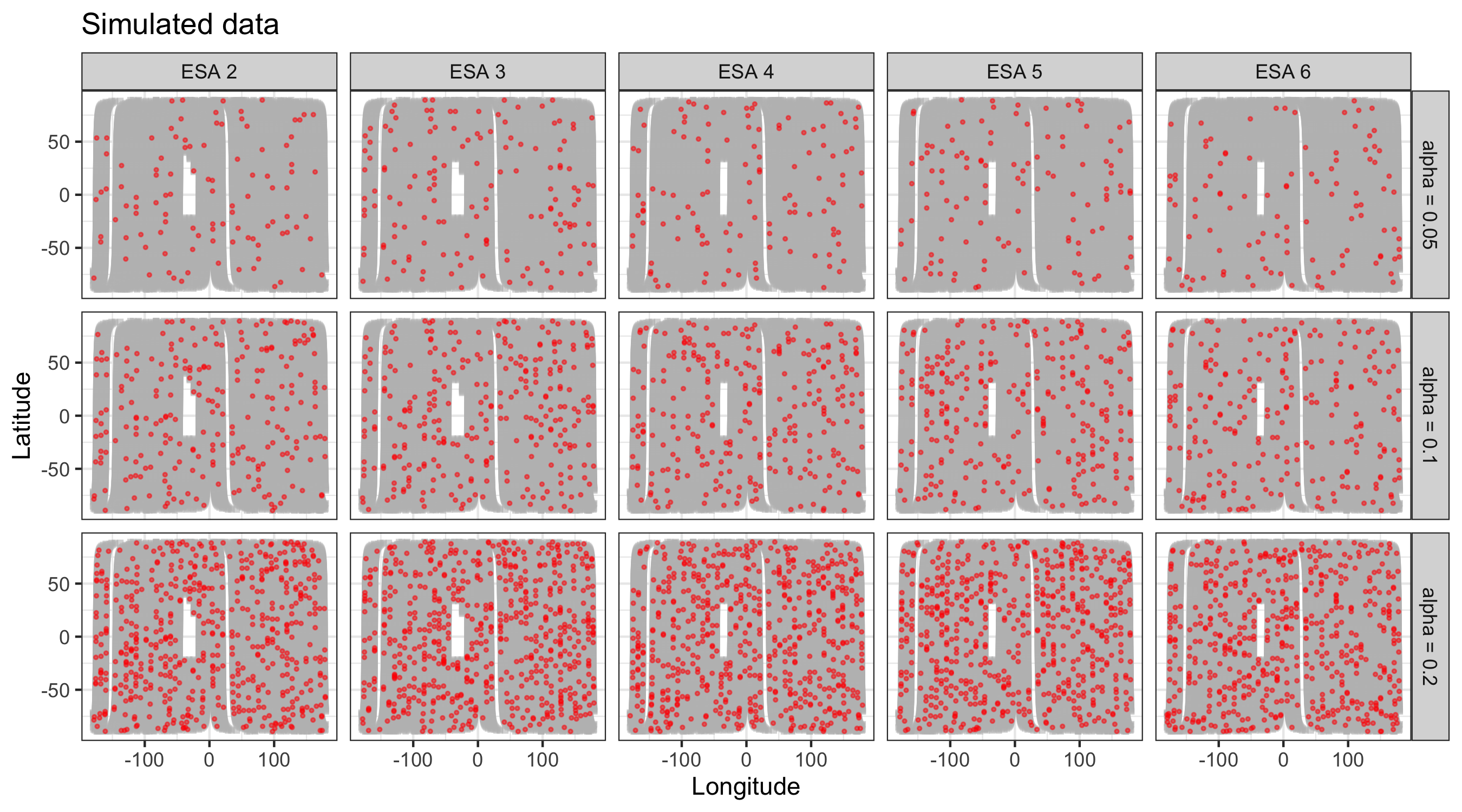}
	\caption{\small Look directions for the 2010A orbits for which $\text{H}_0$ was rejected under different significance levels alpha, plotted by latitude and longitude as red points against a background of grey non-rejected points. The top (bottom) sub-plots show the results for the IBEX-Hi data set (a synthetic data set).} \label{fig:supp_validation_mappoints}
\end{figure}

\begin{figure}[htpb]
	\centering
	\includegraphics[width=0.96\textwidth]{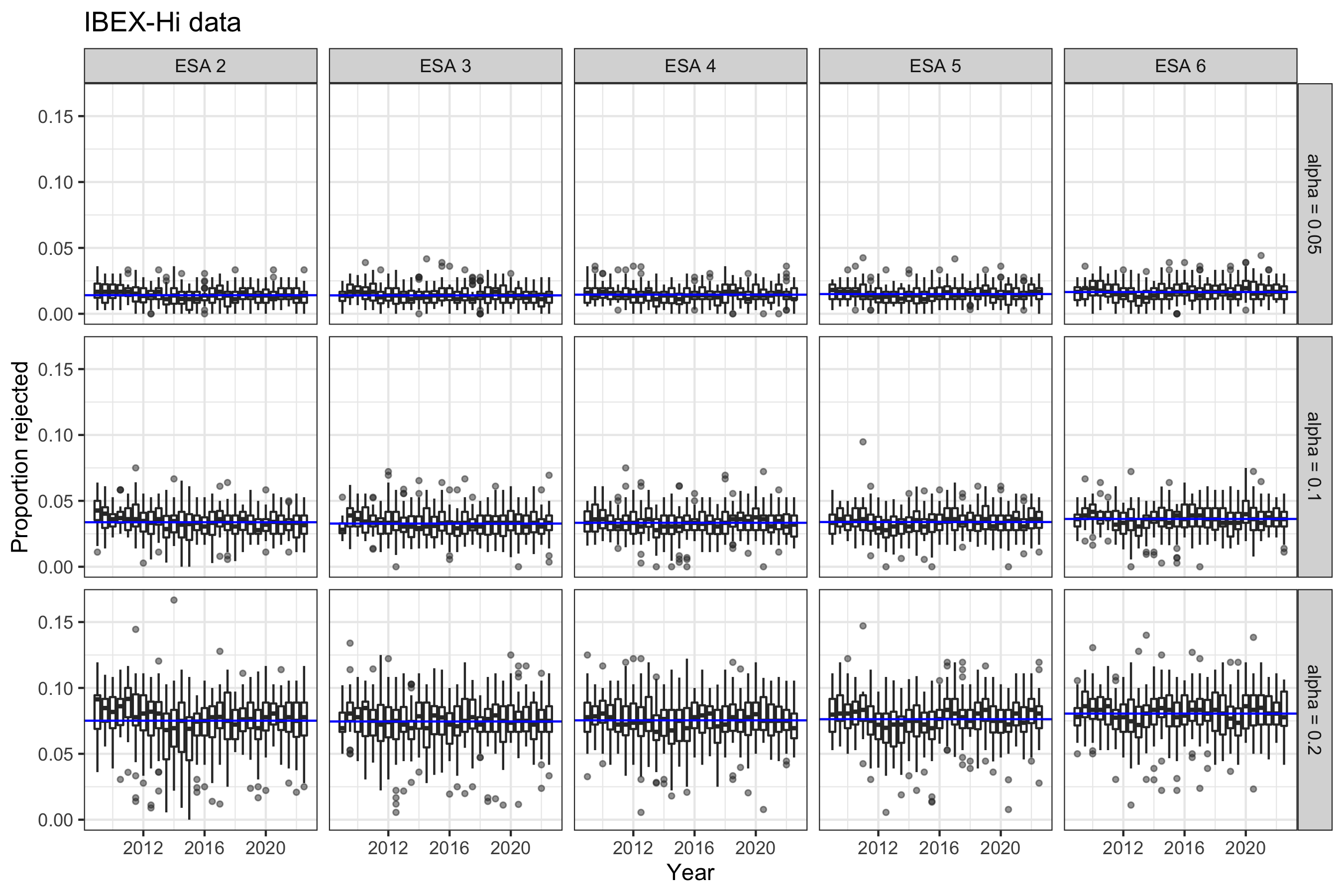}
	\includegraphics[width=0.96\textwidth]{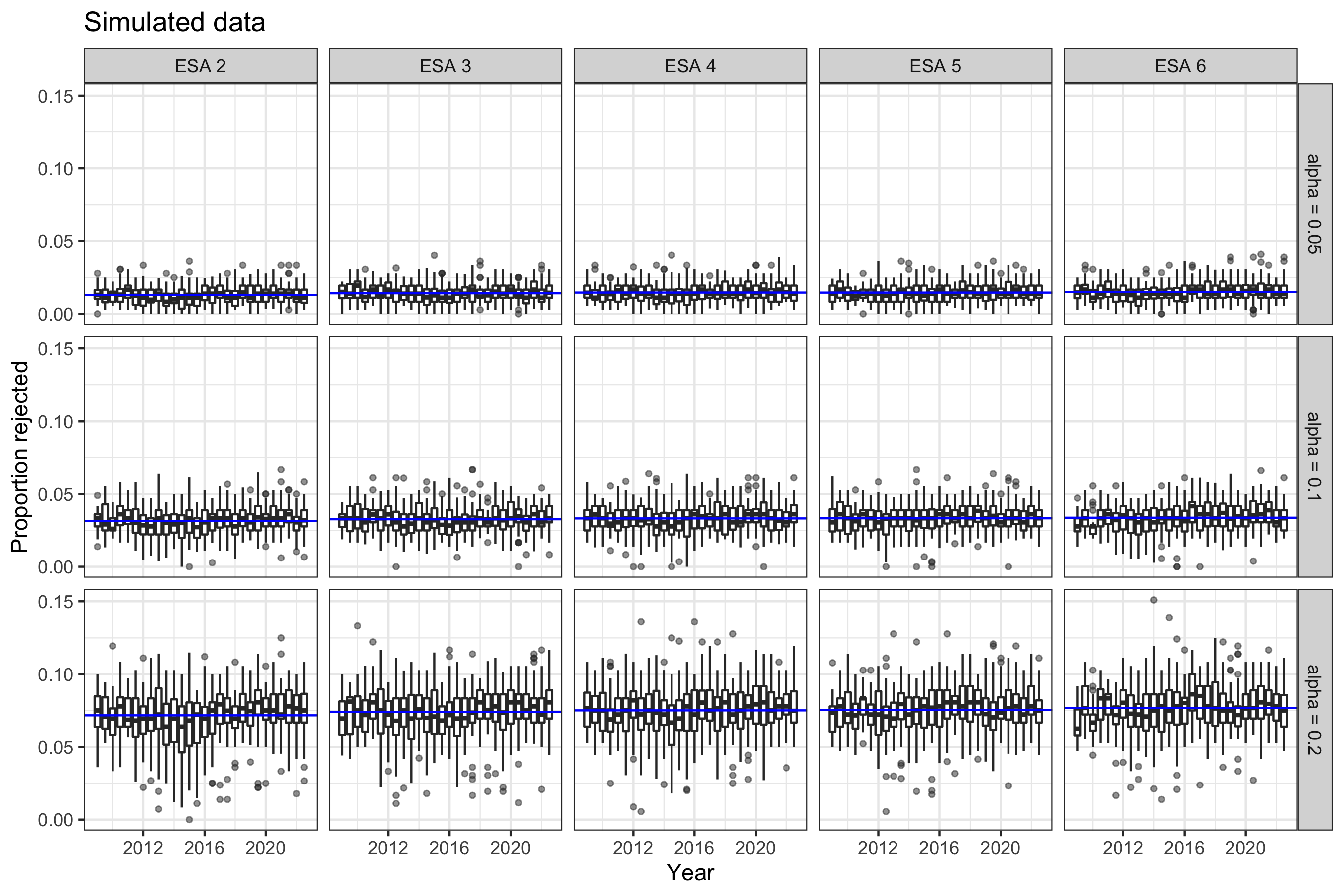}
	\caption{\small Proportion of points for a given orbit/ESA combination for which we fail to reject $\text{H}_0$ under different significance levels alpha. The top (bottom) sub-plots show the results for the IBEX-Hi data set (the synthetic data).} \label{fig:supp_validation_esa_yr}
\end{figure}

Figure \ref{fig:supp_validation_mappoints} shows the spatial rejection pattern by ESA for an example map, and Figure \ref{fig:supp_validation_esa_yr} shows the temporal- and ESA-level rejection patterns at a variety of significance levels alpha $\in \{0.05,0.1,0.2\}.$ 
There are no visible spatial, temporal, or ESA-level patterns in how many observations are rejected.

\begin{figure}[htpb]
	\centering
	\includegraphics[width=0.95\textwidth]{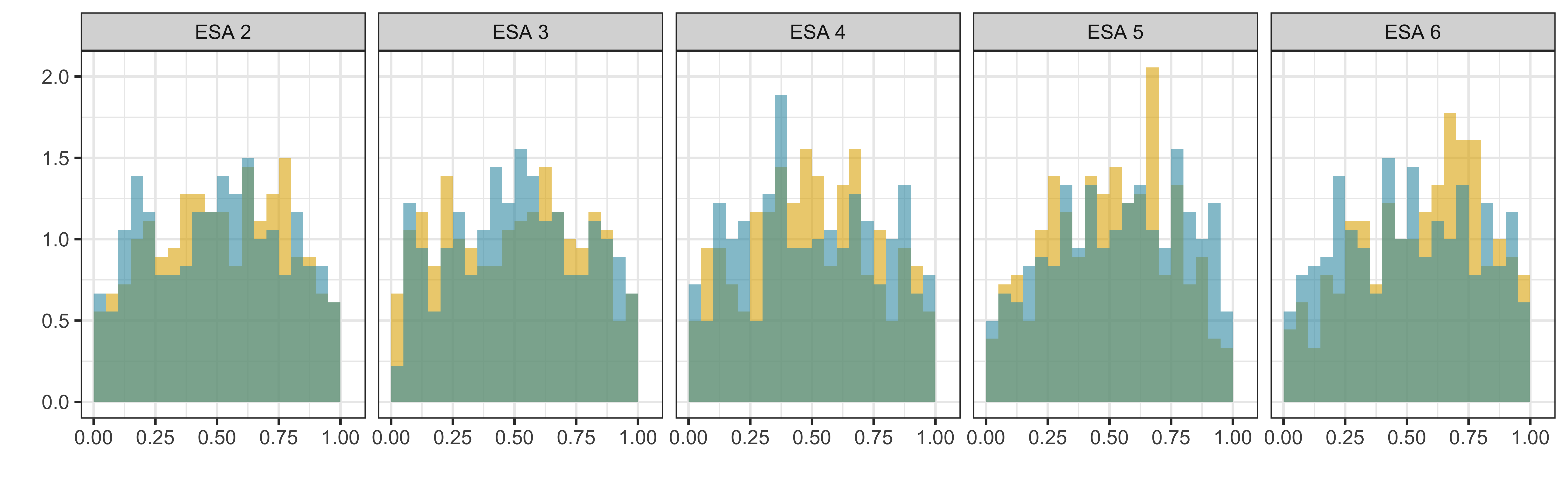}
	\includegraphics[width=0.95\textwidth]{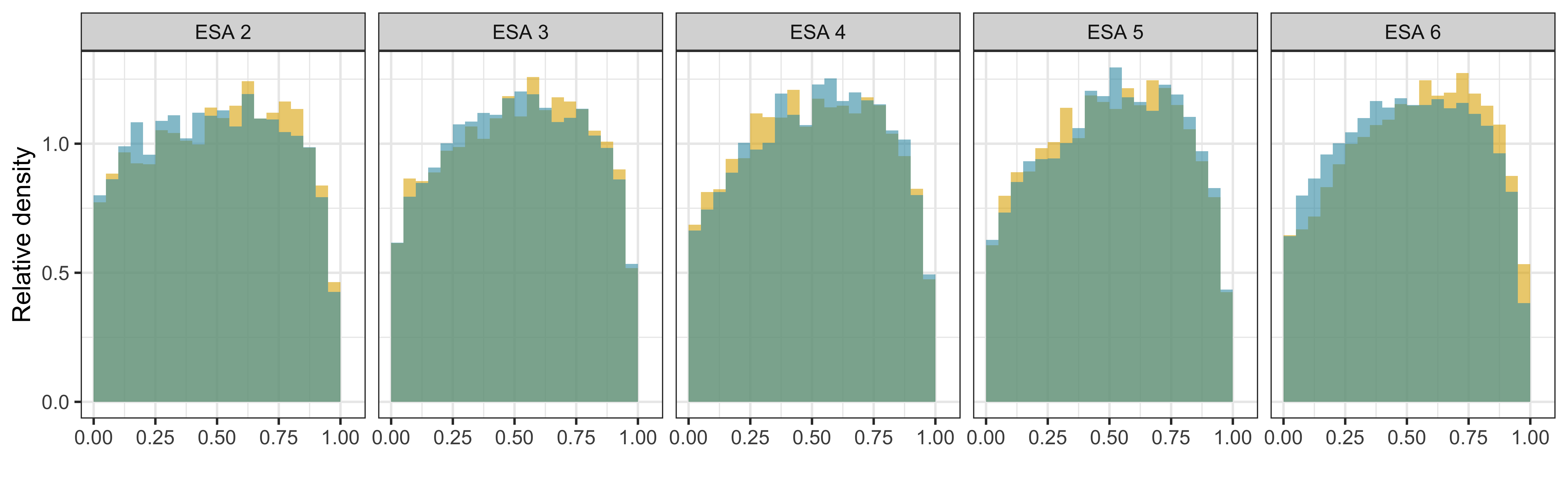}
	\includegraphics[width=0.95\textwidth]{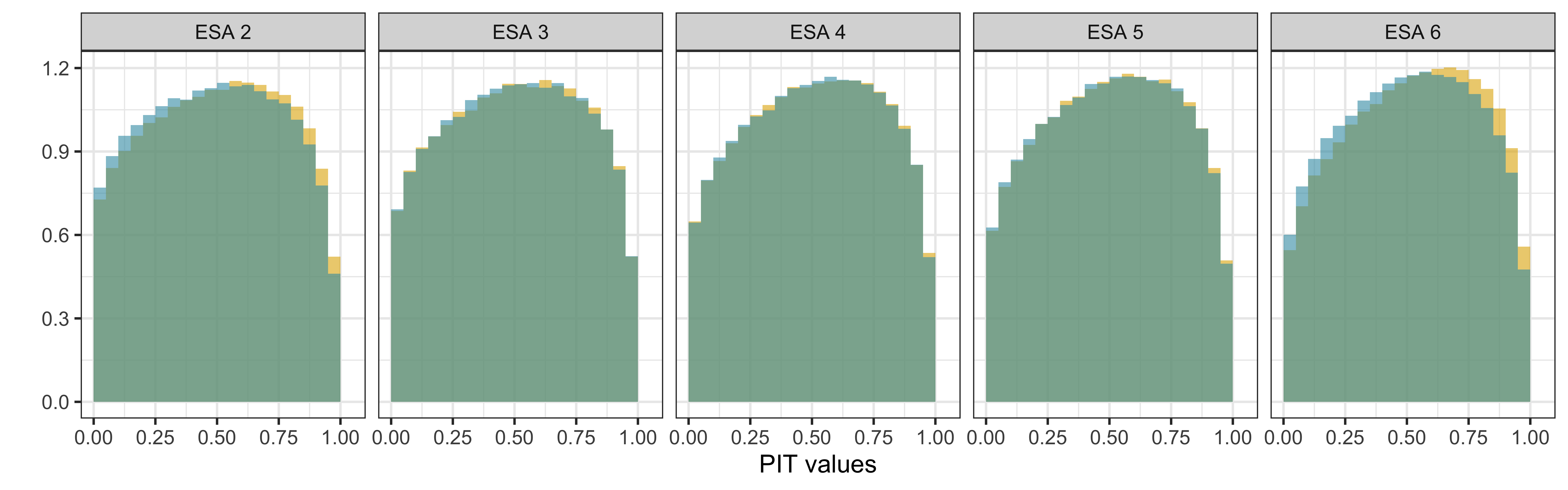}
	\includegraphics[width=0.6\textwidth]{figures/pvals/legend.pdf}
	\caption{\small PIT histograms for the qABCs from an example orbit (orbit 0276b, top), an example map (2010A, middle), and the entire mission (bottom). At the mission level, the KS statistic values for ESA 2 through 6 are 0.017, 0.004, 0.003, 0.003, and 0.025, respectively, with lower values indicating better agreement between observed and simulated data.} \label{fig:res_pvals_abc}
\end{figure}

Figure \ref{fig:res_pvals_abc} shows the observed and simulated qABC PIT histograms for each ESA from an example orbit, an example map, and the full mission. 
As the amount of observations increases, we see better resolved (i.e., less noisy) histograms. 
The largest deviations between observation and simulation occurs in ESA 6, as with the qBC PIT histograms, with ESA 2 also having relatively lower levels of agreement with simulation than the other ESAs.

Visualizations and numeric results for any given orbit, map, and/or ESA are available upon request.

\clearpage

%%--------------------------------------------------------------------------------------------------
%%--------------------------------------------------------------------------------------------------

\section{Additional background adjustment figures}\label{supp:bg_correct}

\begin{figure}[htpb]
	\centering
	\includegraphics[width=0.95\textwidth]{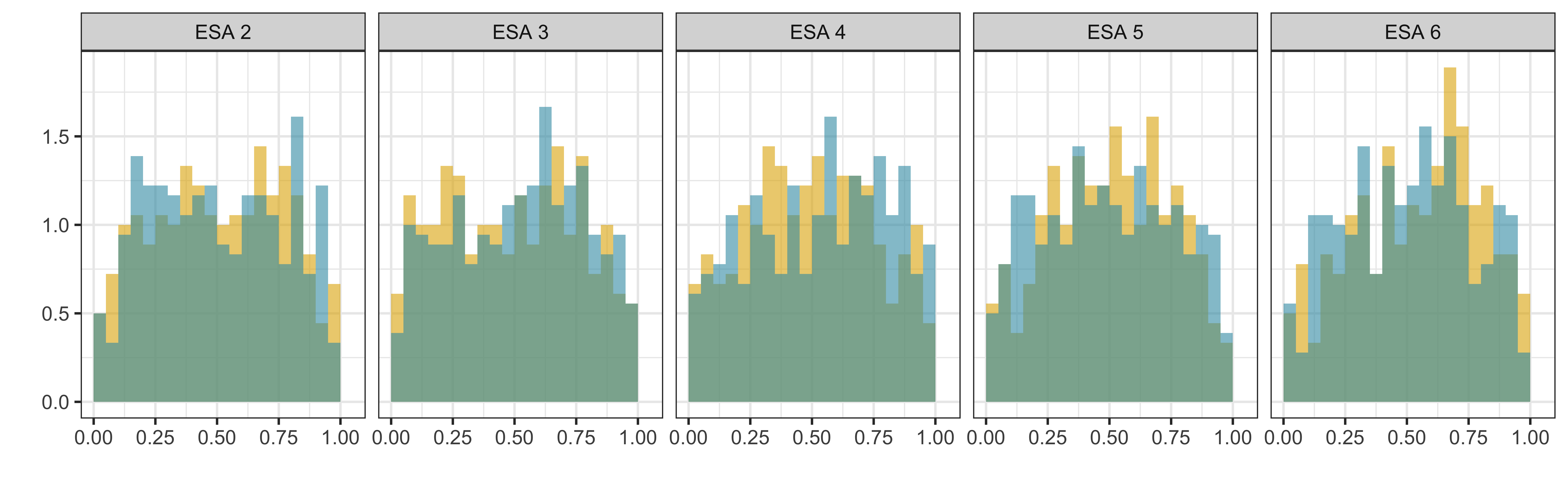}
	\includegraphics[width=0.95\textwidth]{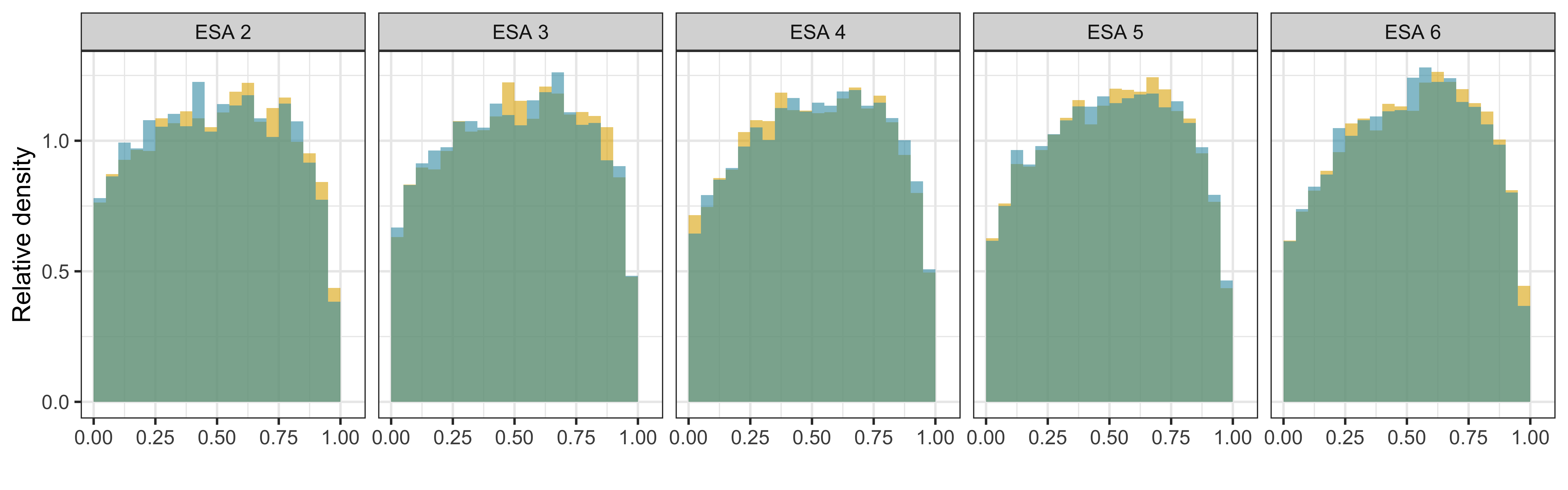}
	\includegraphics[width=0.95\textwidth]{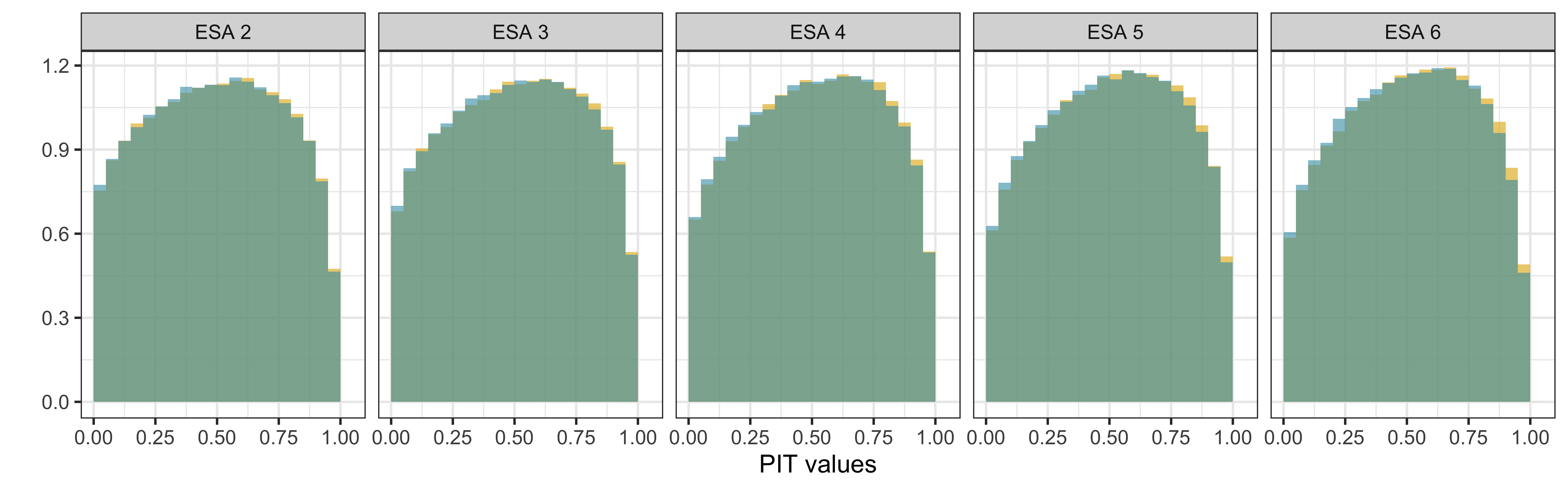}
	\includegraphics[width=0.6\textwidth]{figures/pvals/legend.pdf}
	\caption{\small PIT histograms for the qABCs from an example orbit (orbit 0276b, top), an example map (2010A, middle), and the entire data set (bottom) \textit{after} background adjustment and outlier removal. At the mission level, the KS statistic values for ESA 2 through 6 are 0.004, 0.005, 0.004, 0.006, 0.008, respectively, with ESAs 2 and 6 improving relative to the unadjusted data and the remaining ESAs having slightly higher values.} \label{fig:res_pvals_abc_fit}
\end{figure}

\begin{figure}[htpb]
	\centering
	\includegraphics[width=0.95\textwidth]{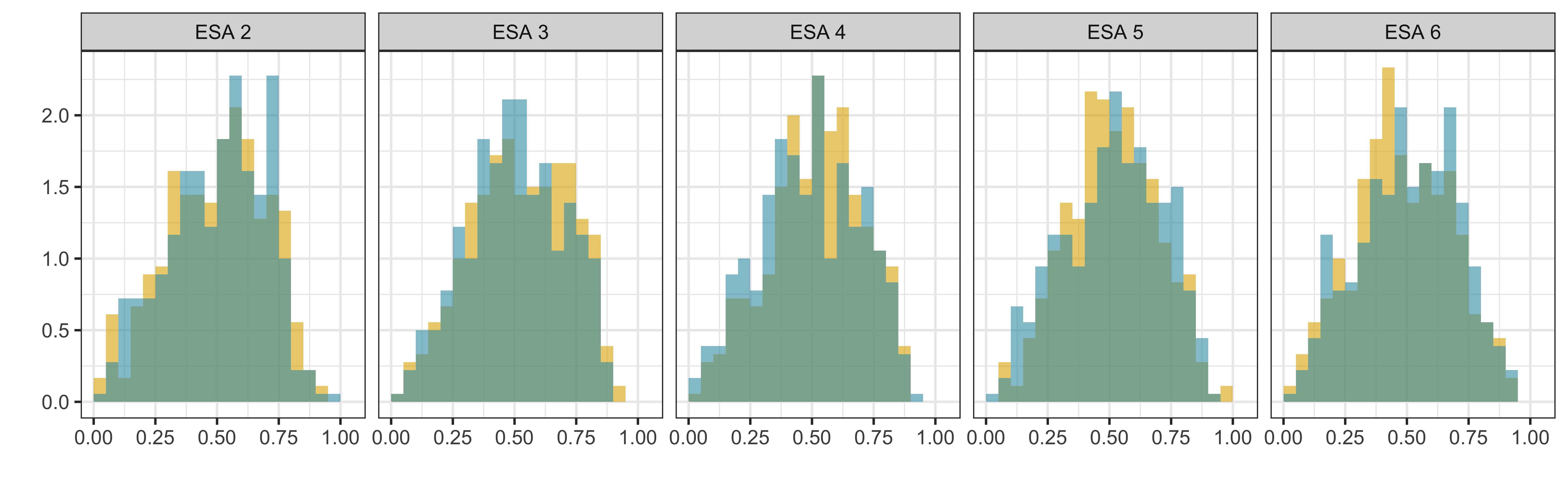}
	\includegraphics[width=0.95\textwidth]{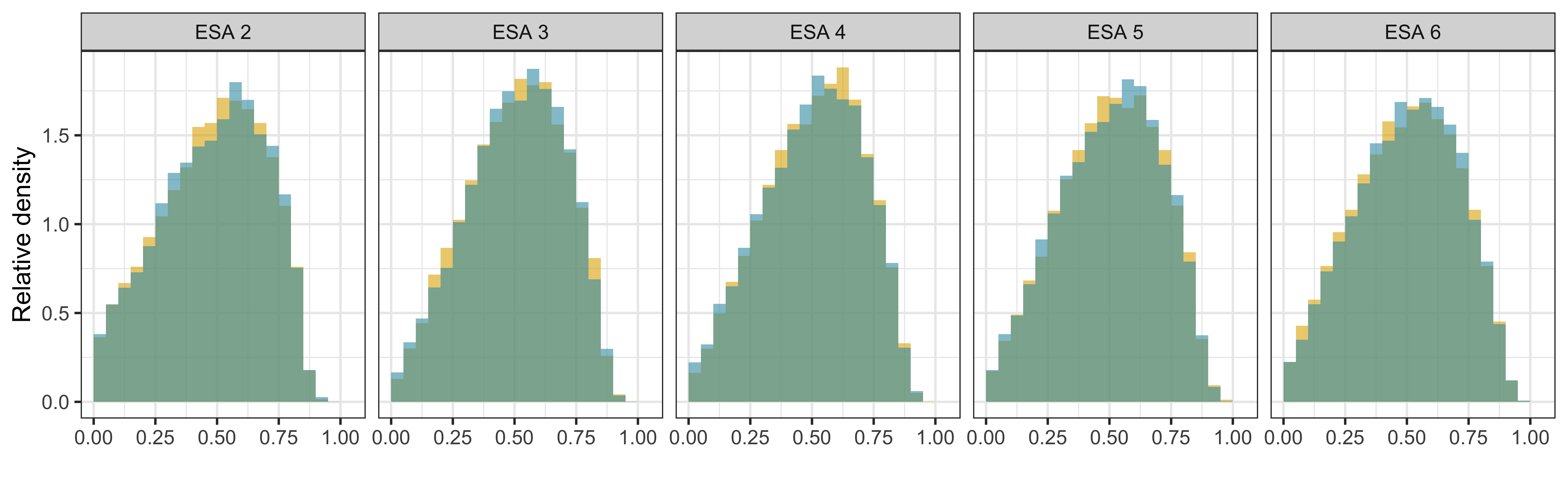}
	\includegraphics[width=0.95\textwidth]{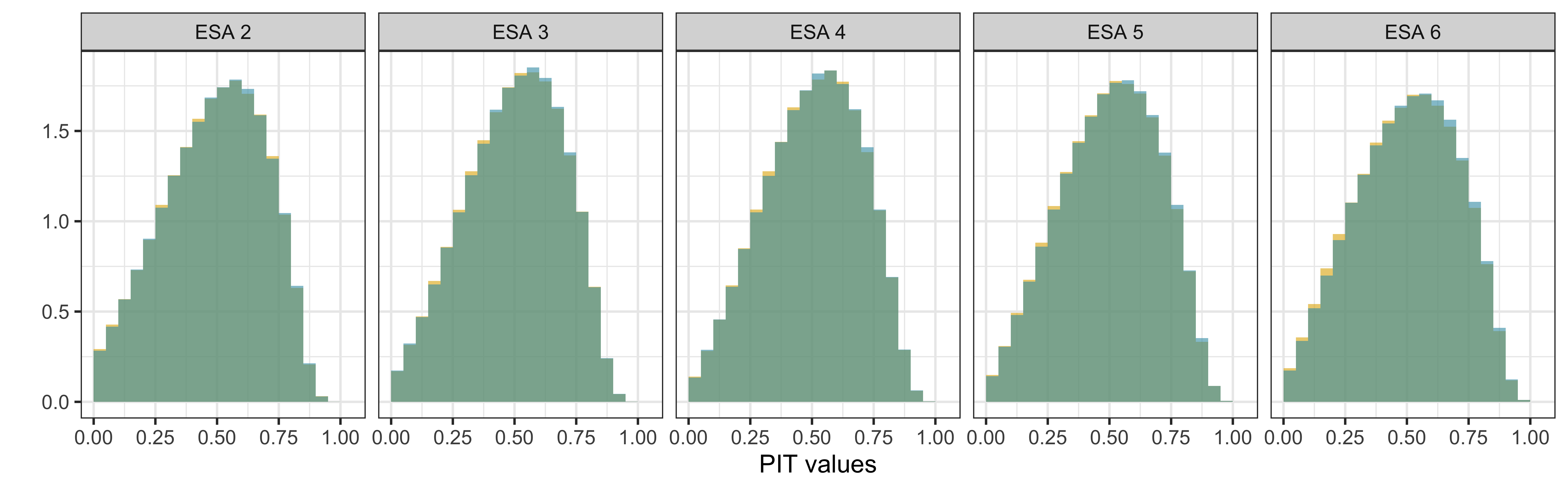}
	\includegraphics[width=0.6\textwidth]{figures/pvals/legend.pdf}
	\caption{\small PIT histograms for the qBCs from an example orbit (orbit 0276b, top), an example map (2010A, middle), and the entire data set (bottom) \textit{after} background adjustment and outlier removal. At the mission level, the KS statistic values for ESA 2 through 6 are 0.003, 0.004, 0.004, 0.006, 0.009, respectively, with ESAs 2 and 6 improving relative to the unadjusted data and the remaining ESAs having slightly higher values.} \label{fig:res_pvals_bc_fit}
\end{figure}

Figure \ref{fig:res_pvals_abc_fit} and \ref{fig:res_pvals_bc_fit} show the observed and simulated qABC and qBC PIT histograms, respectively, for each ESA from an example orbit, an example map, and the full mission after adjusting the background and removing ``outlier'' orbits per the procedure discussion in Section 4.2 of the main text. 
As the amount of observations increases, we see better resolved (i.e., less noisy) histograms.

\begin{figure}[htpb]
	\centering
	\includegraphics[width=0.98\textwidth]{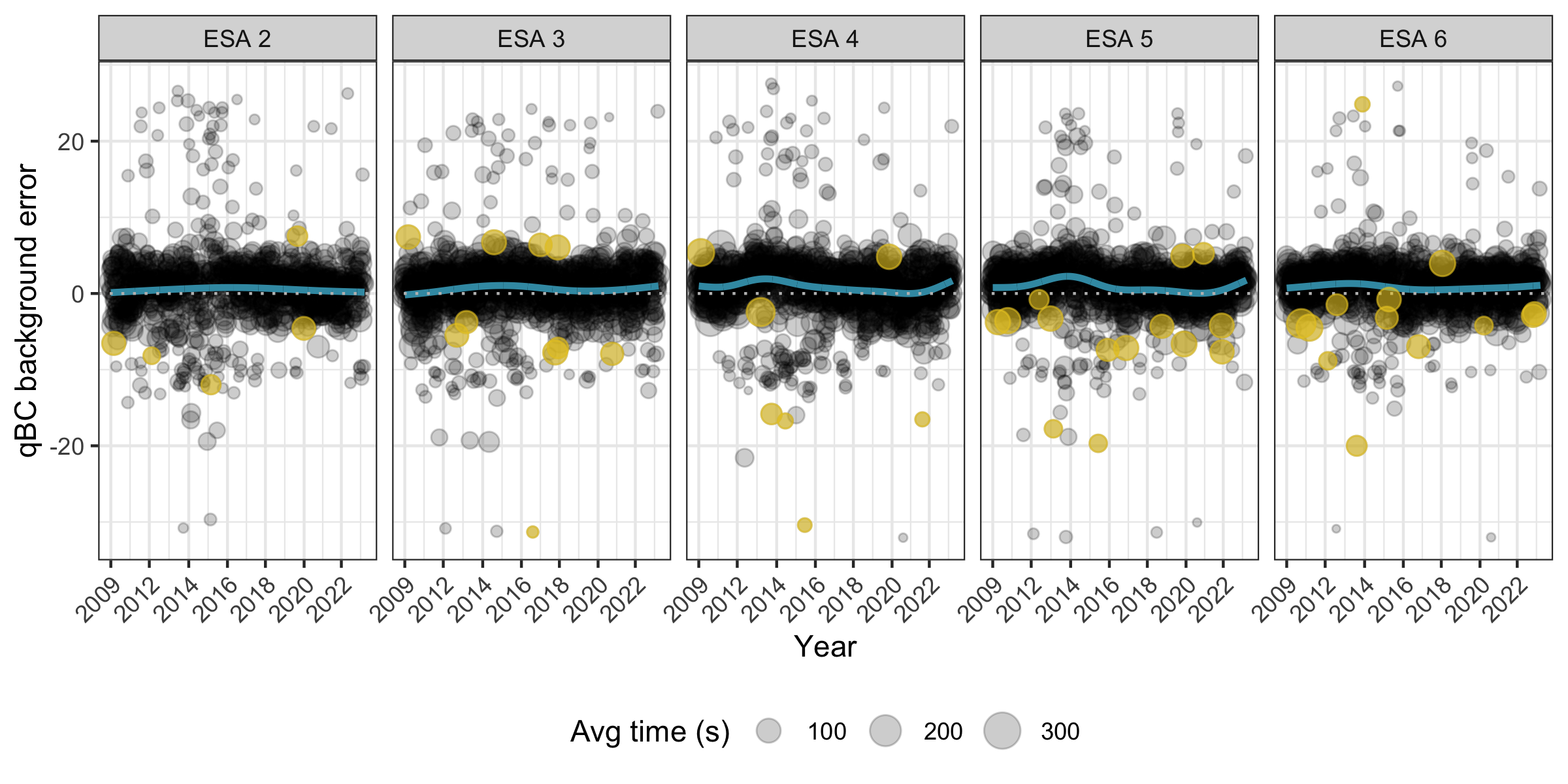}
	\caption{\small Each point corresponds to a single \textit{simulated} ESA-orbit and shows the identified ``best'' background adjustment, i.e., that minimizing a GAM fit to the CvM statistic across adjustments. The temporally fitted qBC background adjustment is shown as a blue line, and the flagged ``ESA-orbits of concern'' are shown as colored yellow points. The size of each point corresponds to the average exposure time in that ESA-orbit.}\label{fig:berr_fit_sim}
\end{figure}

Figure \ref{fig:berr_fit_sim} shows the ESA-specific temporal qBC background adjustment learned via a GAM fit  to the CvM statistic across adjustments $\rho$ as well as the flagged ``ESA-orbits of concern'' for \textit{synthetic data}; at worst, the fitted adjustment reaches 0.8, 1.1, 1.9, 2.3, and 1.3\% for ESAs 2 through 6, respectively and the absolute adjustment is on average 0.7\% across all ESAs.
This synthetic run is designed to give us a sense what magnitude of adjustment may be found via just because of fitting process itself; we have learned that adjustments within a percent or two that are not persistent across a long time scale may just be found due to chance.

\begin{figure}[htpb]
	\centering
	\includegraphics[width=0.98\textwidth]{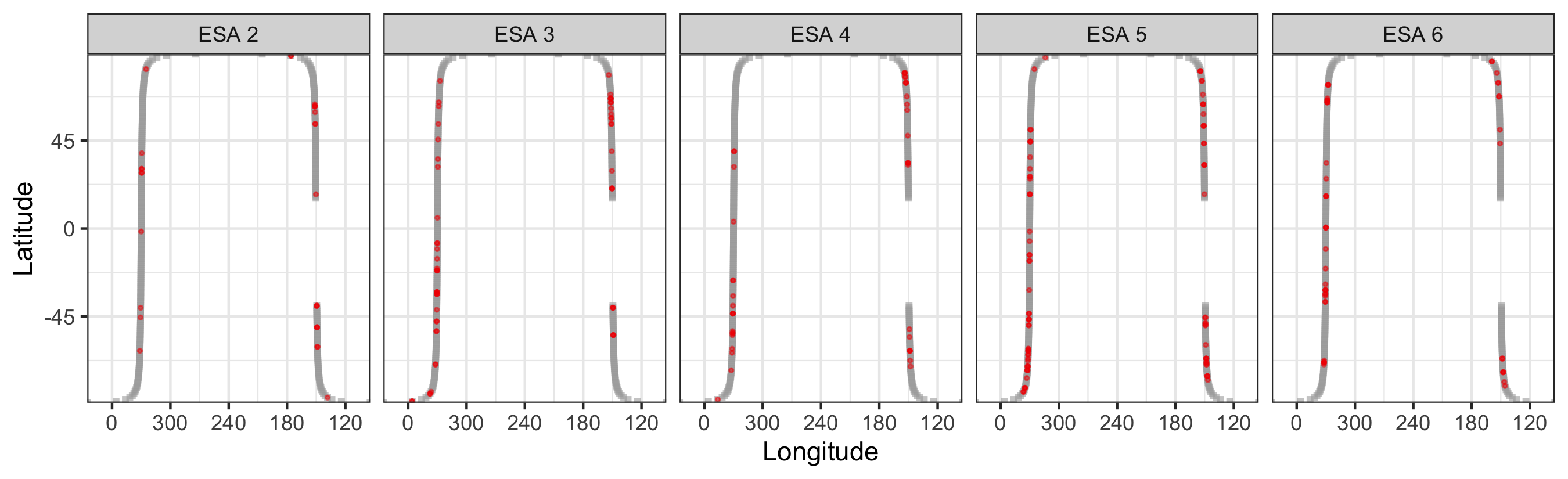}
	\caption{\small The hypothesis test results by ESA and look direction for orbit 0126.}\label{fig:spat_oneOrbit}
\end{figure}

\begin{figure}[htpb]
	\centering
	\includegraphics[width=0.6\textwidth]{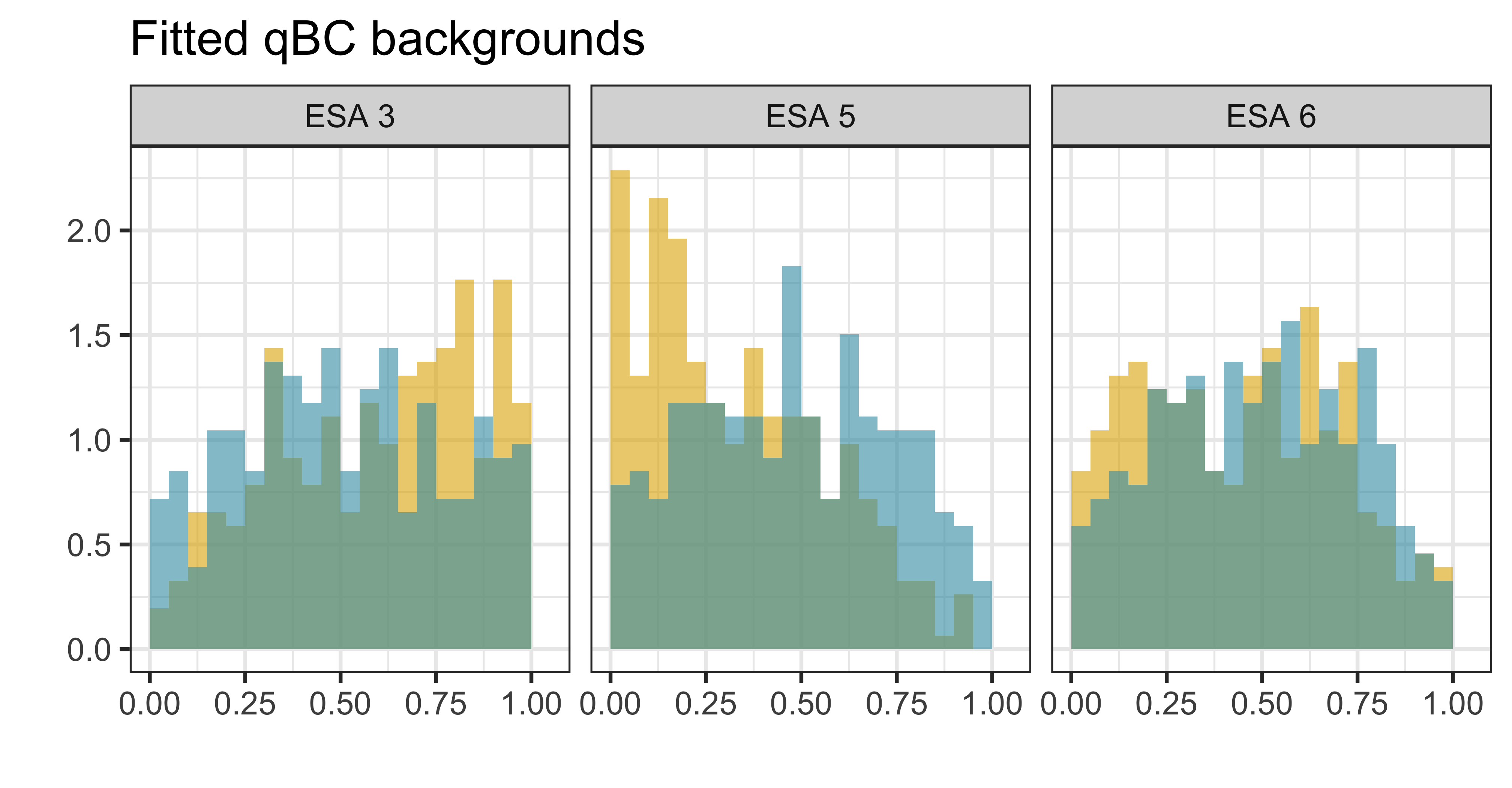}
	\includegraphics[width=0.6\textwidth]{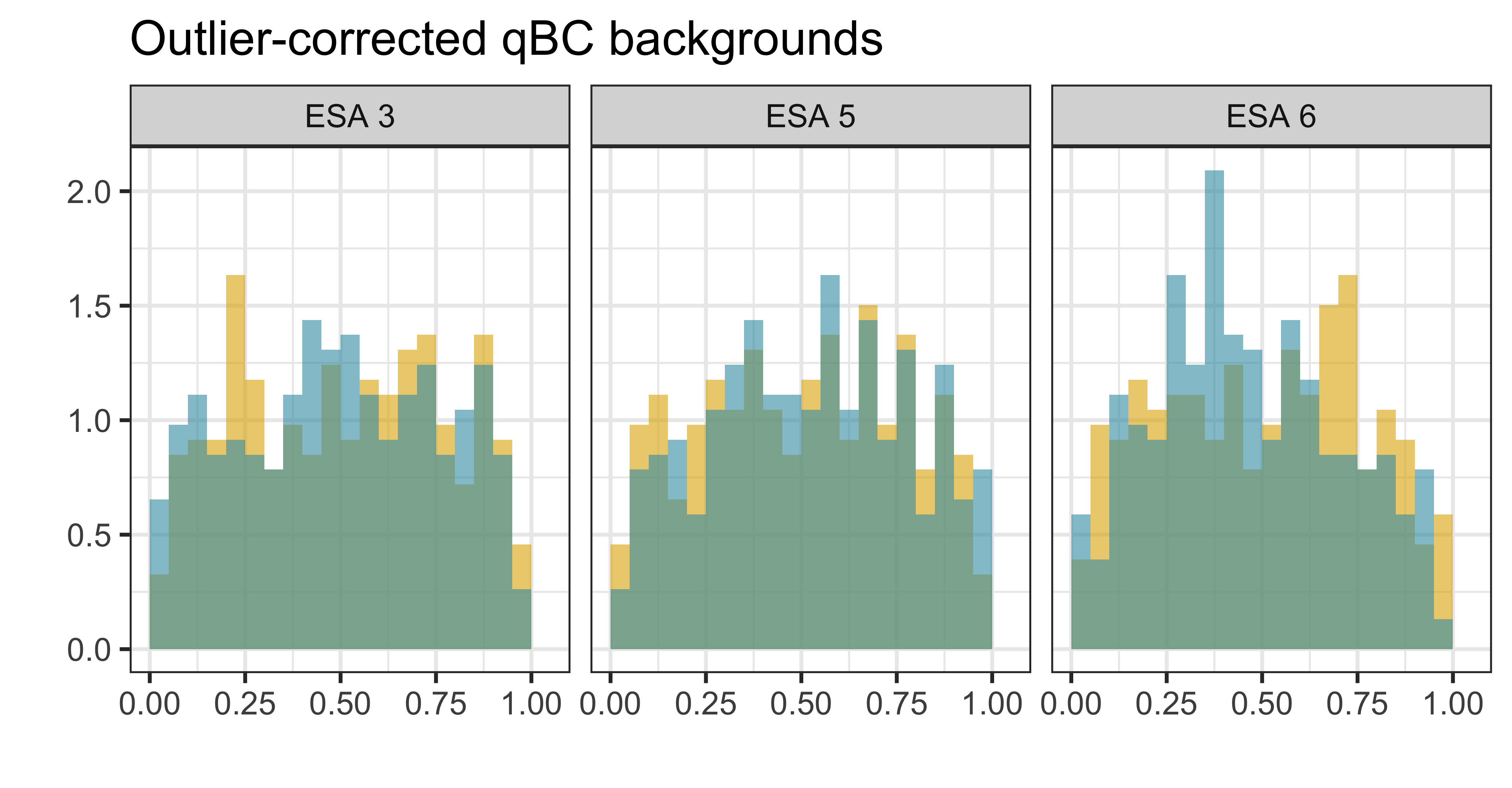}
	\includegraphics[width=0.6\textwidth]{figures/pvals/legend.pdf}
	\caption{\small PIT histograms for the qABCs from orbit 0126 with the fitted (top) and outlier-corrected (bottom) qBC background values.} \label{fig:pitABC_beforeafter}
\end{figure}

\begin{figure}[htpb]
	\centering
	\includegraphics[width=0.6\textwidth]{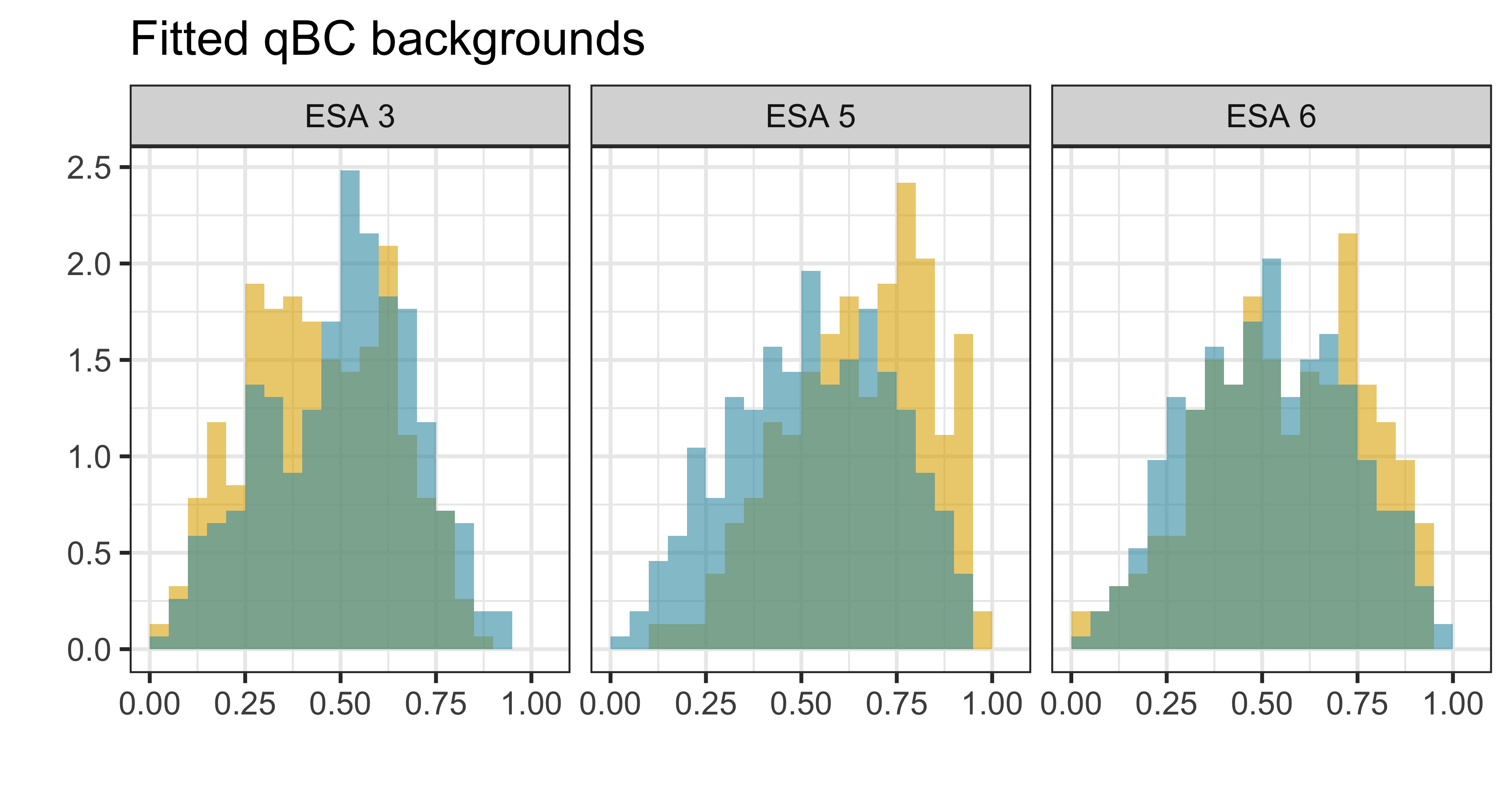}
	\includegraphics[width=0.6\textwidth]{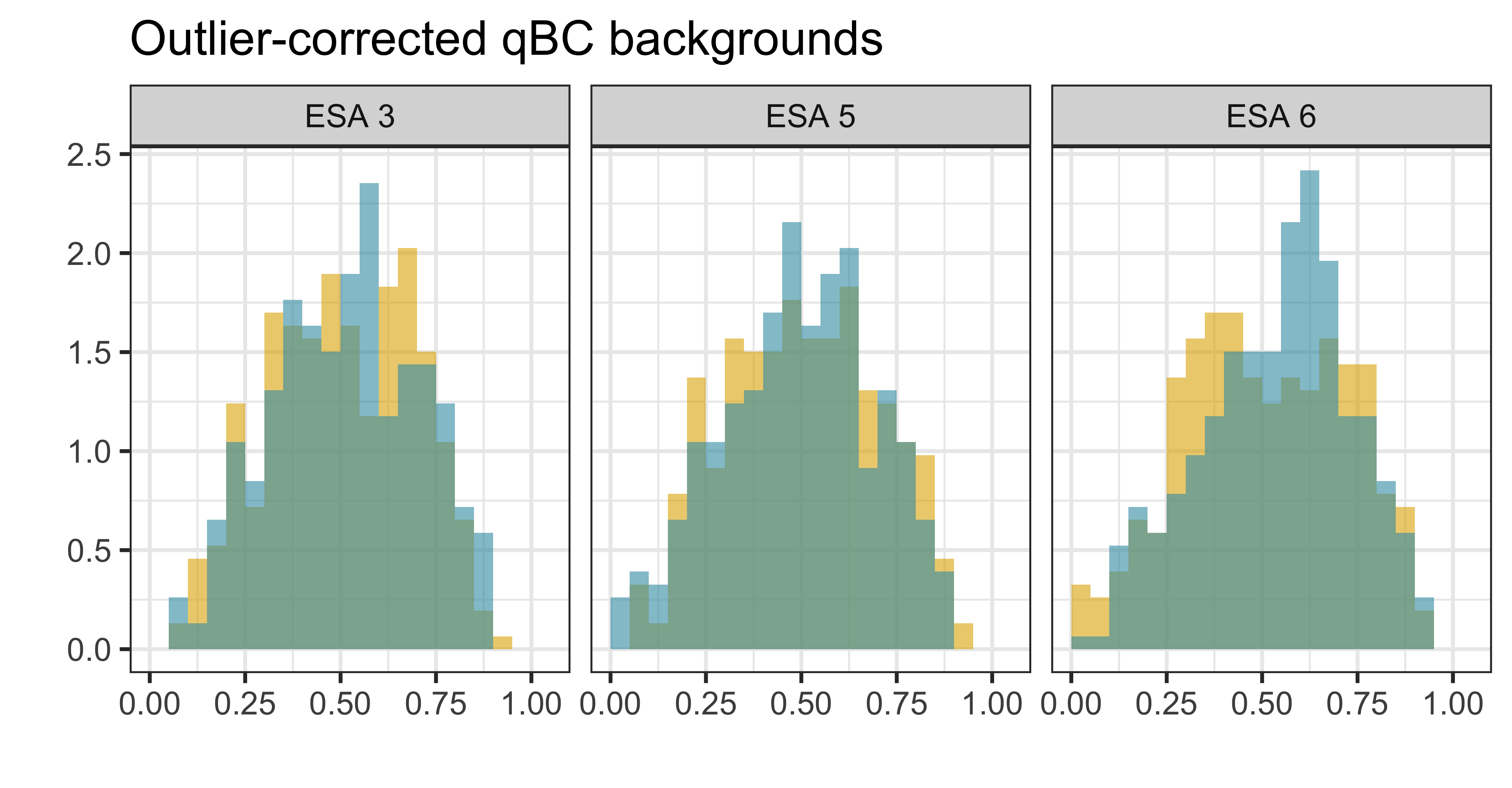}
	\includegraphics[width=0.6\textwidth]{figures/pvals/legend.pdf}
	\caption{\small PIT histograms for the qBCs from orbit 0126 with the fitted (top) and outlier-corrected (bottom) qBC background values.} \label{fig:pitBC_beforeafter}
\end{figure}

Figure \ref{fig:spat_oneOrbit} shows the hypothesis test results by look direction for all points in an example orbit, orbit 0126, for which ESAs 3, 5, and 6 were flagged as ESA-orbit of concern. 
Figures \ref{fig:pitABC_beforeafter} and \ref{fig:pitBC_beforeafter} show the observed and simulated qABC and qBC PIT histograms, respectively, for the outlier ESAs from this orbit at the ``fitted'' qBC background error (treated here as the baseline) and the ``outlier-corrected'' qBC background error.
More specifically, this fitted qBC background error value is the y-value at the fitted blue line from Figure 10, while the outlier-corrected qBC background error value is the y-value at the associated yellow point itself.
There is much better agreement between simulated and observed PIT histograms after outlier correction.

\clearpage

%%--------------------------------------------------------------------------------------------------
%%--------------------------------------------------------------------------------------------------

\section{Identifiability simulation}\label{supp:identifiability}

\begin{figure}[htpb]
\centering
\includegraphics[width=0.6\textwidth]{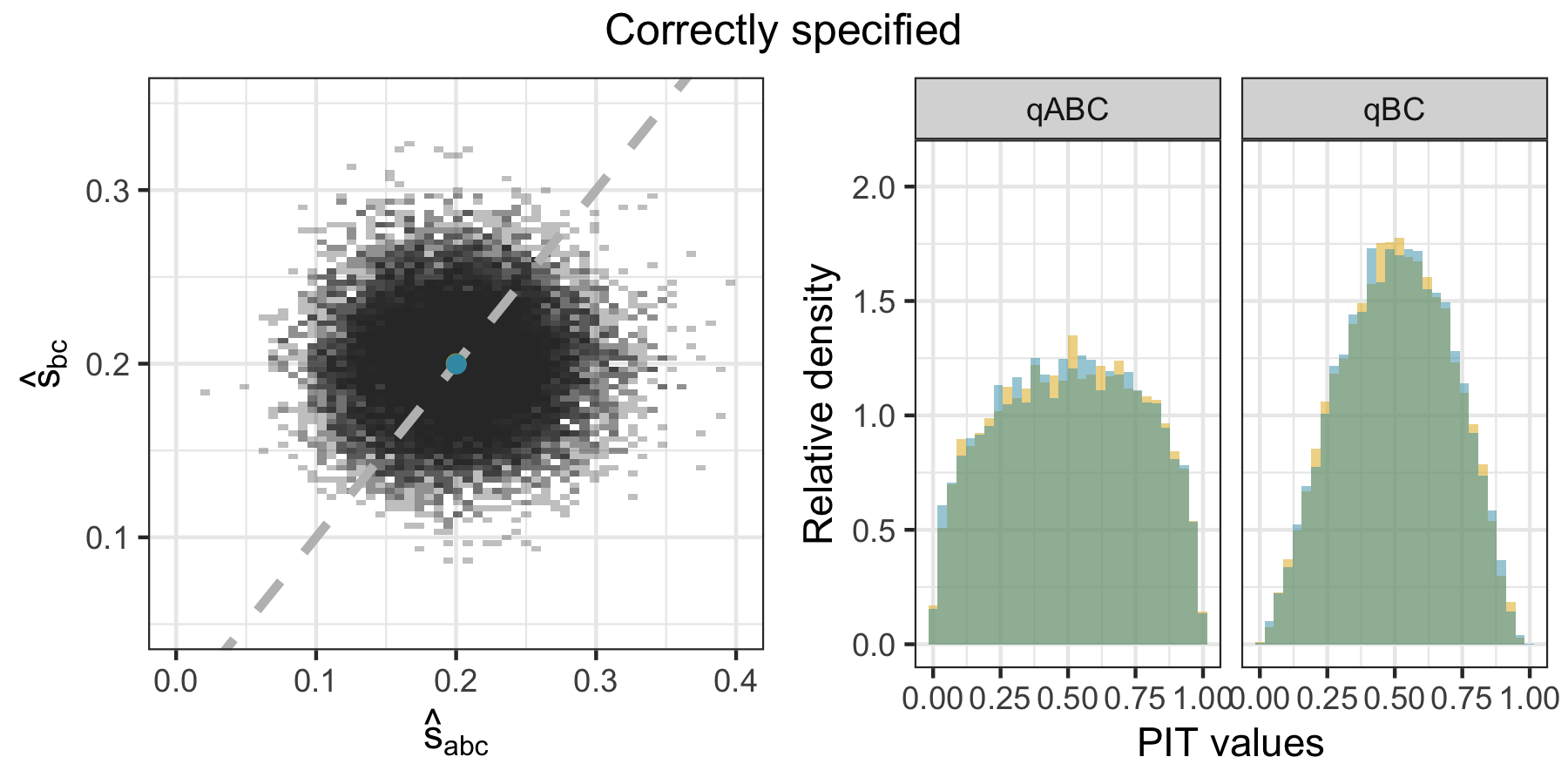}
\caption{\small Simulation under a true shared signal (i.e., $s_{abc} = s_{bc}$) where we specify all model components correctly. The dashed line shows equality, $\hat{s}_{abc} = \hat{s}_{bc}$, for reference.} \label{fig:supp_ident0}
\end{figure}

\begin{figure}[htpb]
\centering
\includegraphics[width=0.75\textwidth]{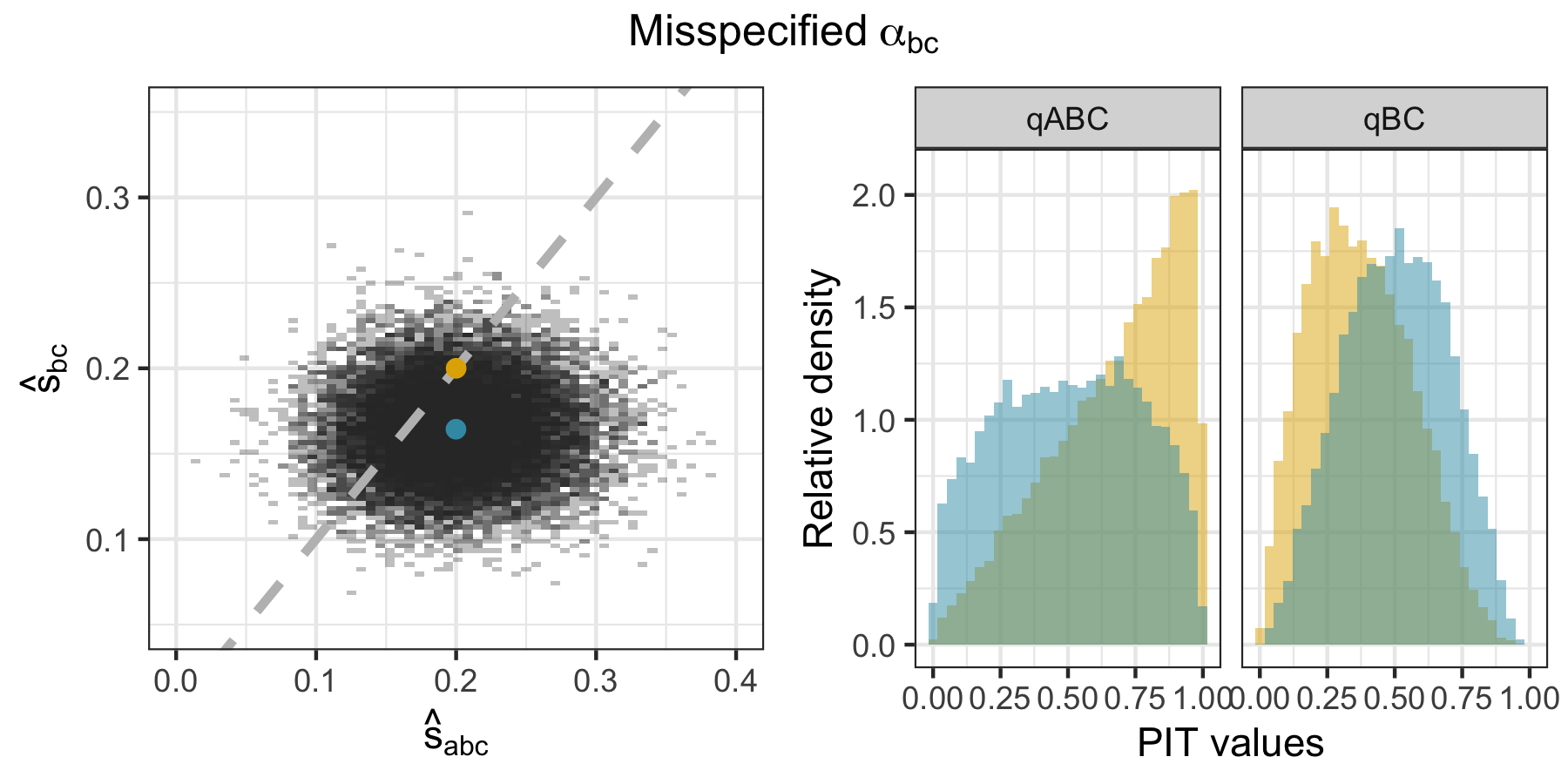}
\caption{\small Simulation under a true shared signal (i.e., $s_{abc} = s_{bc}$) where we misspecify the qBC efficiency by adjusting the assumed efficiency up by 21.4\%. The dashed line shows equality, $\hat{s}_{abc} = \hat{s}_{bc}$, for reference.} \label{fig:supp_ident1}
\end{figure}

\begin{figure}[htpb]
\centering
\includegraphics[width=0.75\textwidth]{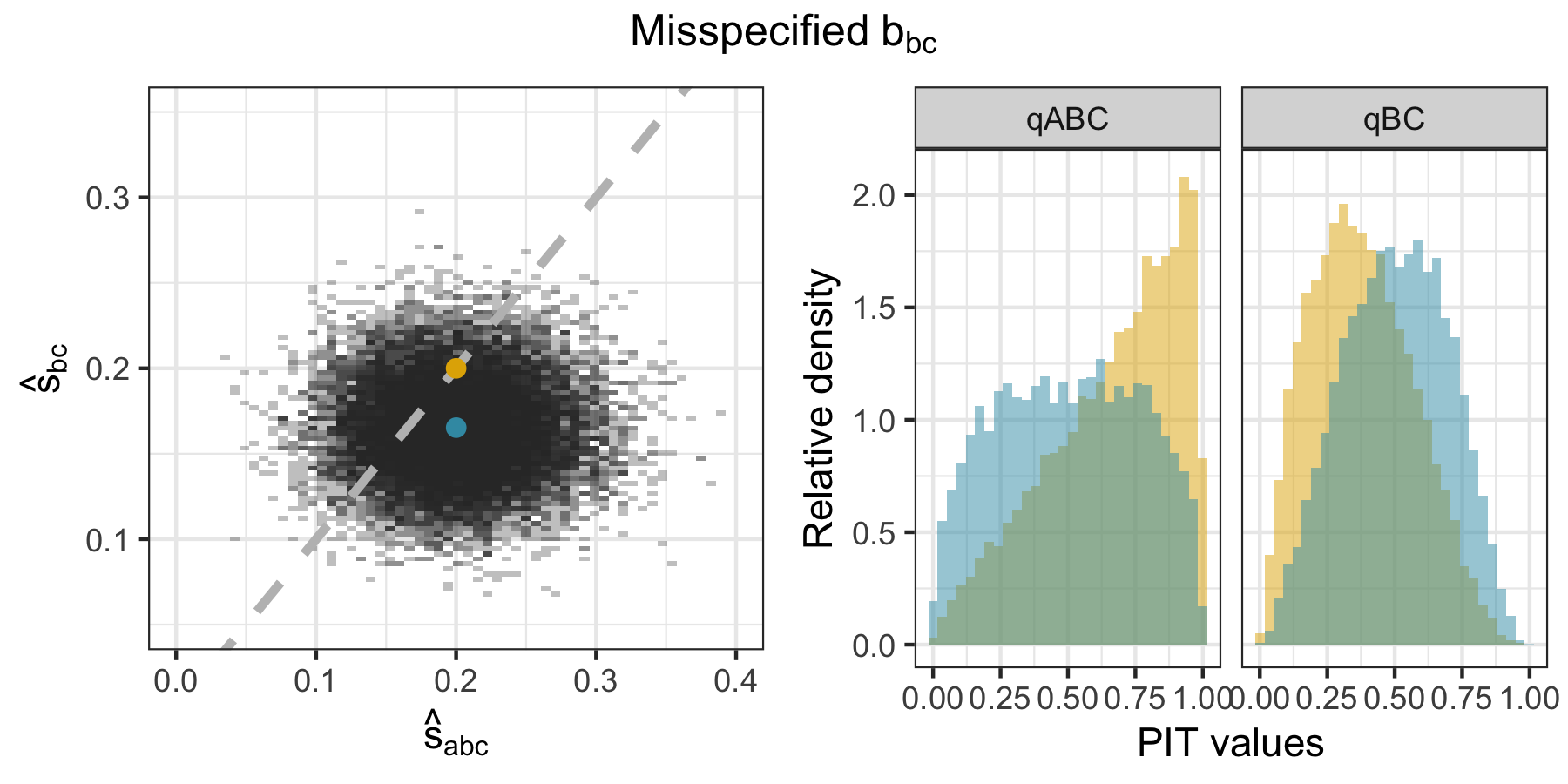}
\caption{\small Simulation under a true shared signal (i.e., $s_{abc} = s_{bc}$) where we misspecify the qBC background rate by adjusting the assumed background up by 13.2\%. The dashed line shows equality, $\hat{s}_{abc} = \hat{s}_{bc}$, for reference.} \label{fig:supp_ident2}
\end{figure}

\begin{figure}[htpb]
\centering
\includegraphics[width=0.75\textwidth]{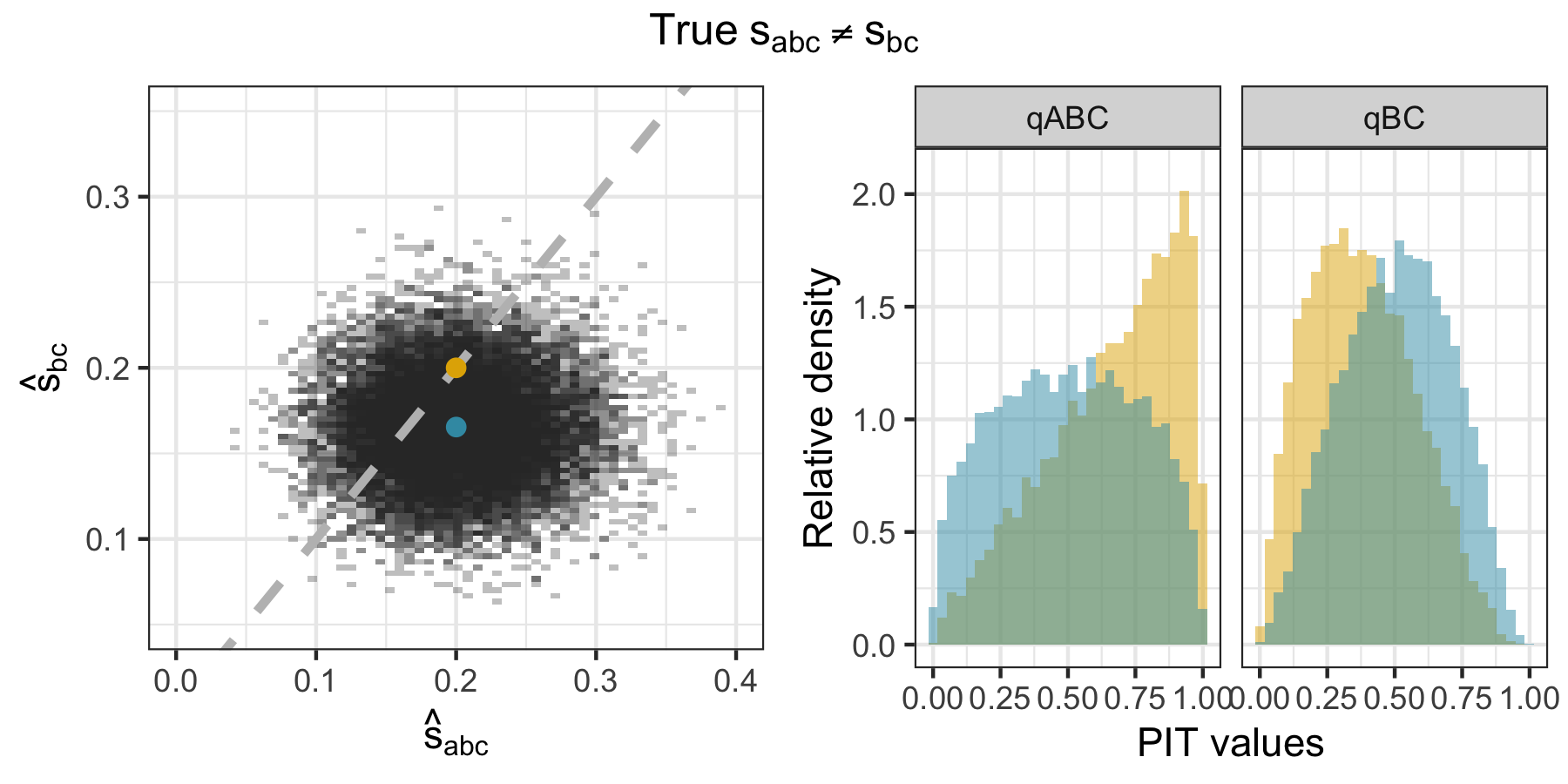}
\caption{\small Simulation where there is not a true shared signal (i.e., $s_{abc} \neq s_{bc}$) but we specify all other model components correctly. Here we have $s^{bc} = 0.165$, while $s^{abc} = 0.2$. The dashed line shows equality, $\hat{s}_{abc} = \hat{s}_{bc}$, for reference.} \label{fig:supp_ident3}
\end{figure}

The following simulations are designed to show the identifiability issue in diagnosing the source of model misspecification.
To begin, we chose values for the true signal rate ($s=0.2$), the true exposure time ($t=150$ seconds), and the qABC and qBC background and efficiency ($b^{abc}=0.1, b^{bc}=0.2, e^{abc}=0.96,$ and $e^{abc}=2.0$), and then simulate $14,400$ Poisson counts (analogous to 40 orbits' worth of data from a single ESA) using these values. 
We calculate $\hat{s}_{abc}, \hat{s}_{bc}$, and the PIT values for the qABC and qBC counts assuming there \textit{is} a shared signal. 
We consider four possibilities:
\begin{enumerate}
    \item The model is specified correctly (Figure \ref{fig:supp_ident0})
    \item There is a true shared signal, but we misspecify the qBC efficiency (Figure \ref{fig:supp_ident1})
    \item There is a true shared signal, but we misspecify the qBC background rate (Figure \ref{fig:supp_ident2})
    \item There is not a true shared signal (i.e., $s_{abc} \neq s_{bc}$), and we specify all other model components else correctly (Figure \ref{fig:supp_ident3})
\end{enumerate}

Misspecification in the qBC efficiency $e^{bc}$, in the qBC background $b^{bc}$, and a multiplicative bias in the qBC signal relative to the qABC signal (details of misspecification and visuals shown in Figures \ref{fig:supp_ident1}, \ref{fig:supp_ident2}, and \ref{fig:supp_ident3}, respectively) all lead to similar distributions of $\hat{s}^*$ and the PIT histograms. That is, you can tell that simulation and reality are not aligned, but it is hard to disentangle \textit{how} they are misaligned. Not shown are analogous diagnostics for misspecification of the qABC model components -- if one assumed that the qABCs may also be a cause of lack of model fit, the identifiability issue would be heightened further.

\begin{figure}[thpb]
\centering
\includegraphics[width=0.75\textwidth]{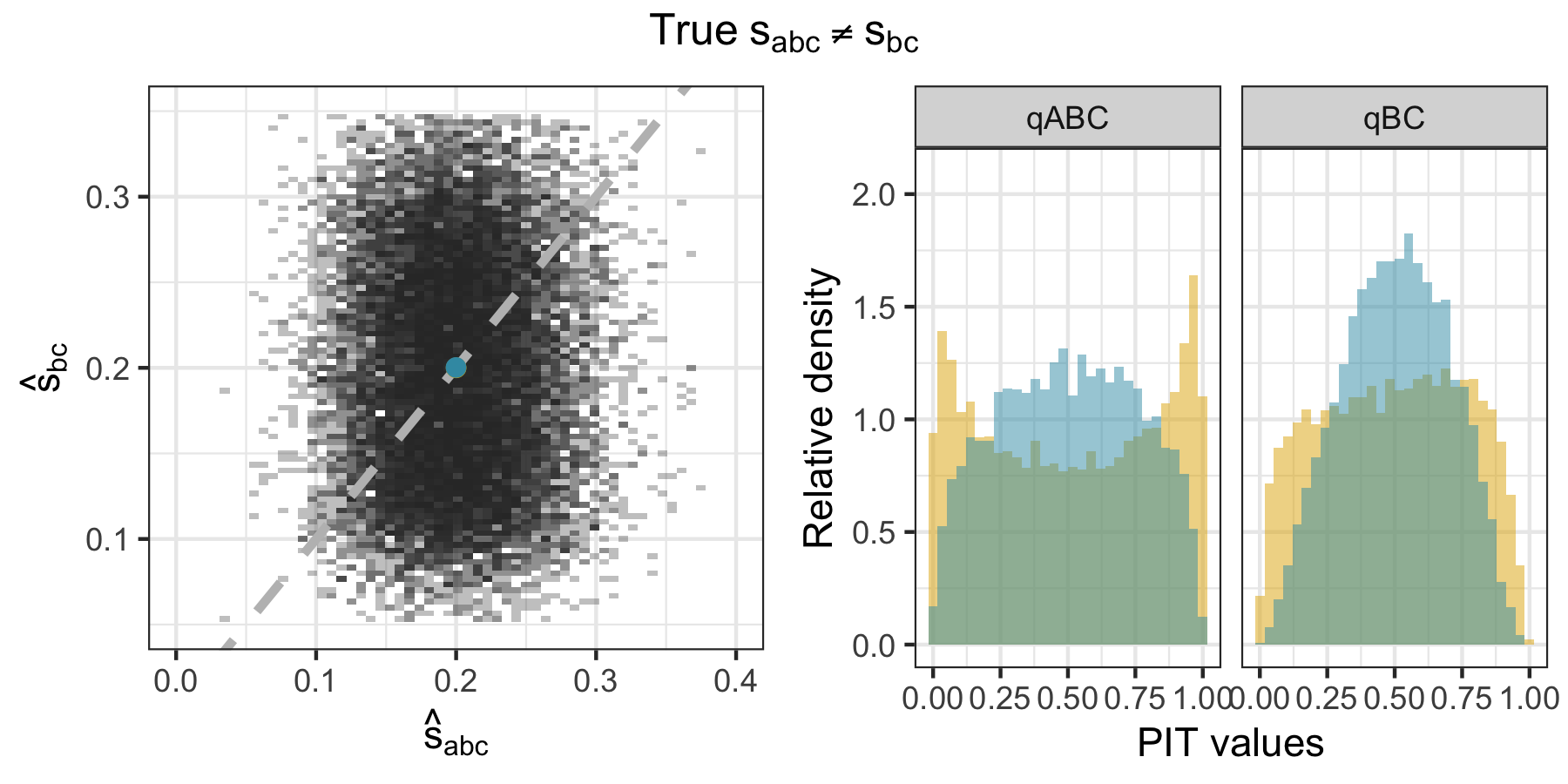}
\caption{\small Simulation where there is not a true shared signal (i.e., $s_{abc} \neq s_{bc}$) but we specify all other model components correctly. Here we have $s^{bc}$ sampled from a Uniform(0.1,0.3) distribution, while $s^{abc} = 0.2$. The dashed line shows equality, $\hat{s}_{abc} = \hat{s}_{bc}$, for reference.} \label{fig:supp_ident4}
\end{figure}

The previous examples were all for the case of a multiplicative bias in the misspecified parameter. 
For the efficiency factor and the background, this structure makes sense because these values are typically common across look directions for a given ESA/orbit combination. 
This structure makes less sense for the signal, which differs by look direction. If instead there was a non-constant differences between the qABC and qBC signal, this would look different than the constant biases in the efficiency factor and backgrounds did -- see Figure \ref{fig:supp_ident4} for an example of this case.

\clearpage

%%--------------------------------------------------------------------------------------------------
%%--------------------------------------------------------------------------------------------------

\section{Results assuming error is in shared signal}\label{supp:s_err}

\begin{figure}[htpb]
\centering
\includegraphics[width=0.3\textwidth]{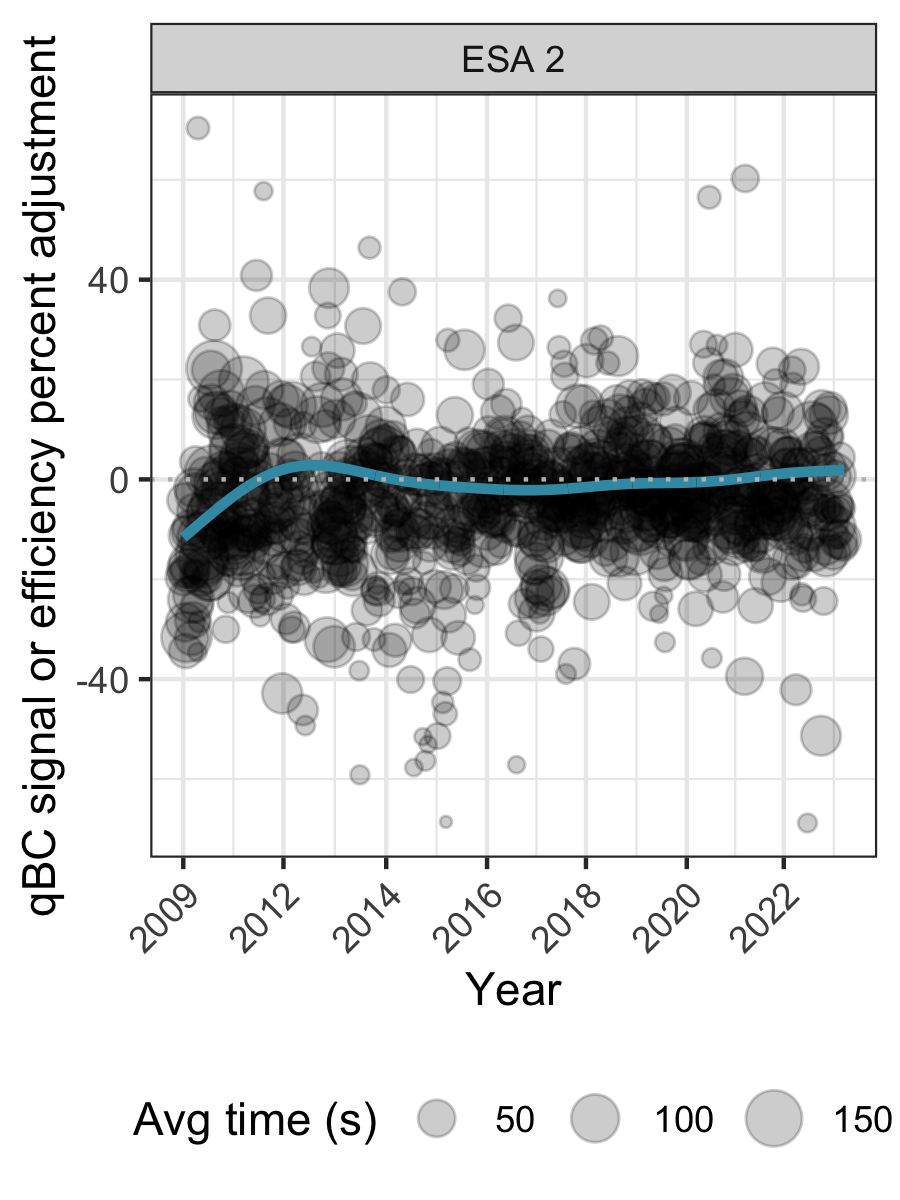}
\caption{\small Each point corresponds to a single ESA-orbit and shows the identified ``best'' qBC signal rate (analogously, efficiency factor)  adjustment, i.e., that minimizing a GAM fit to the CvM statistic across adjustments. The temporally fitted best adjustment is shown as a blue line. The size of each point corresponds to the average exposure time in that ESA-orbit.}\label{fig:combo_fit}
\end{figure}

Figure \ref{fig:combo_fit} shows the temporal qBC signal rate (analogously, efficiency factor) error learned via a GAM fit to the CvM statistic across adjustments $\rho$, allowing the qBC background to account for the data product disagreement up to the error bounds anticipated by mission scientists.
Only ESA 2 is shown, because no further adjustments beyond the background rate adjustment are needed for ESAs 3 through 6
The worst-case learned qBC efficiency factor adjustment value for ESA 2 is -11.6\% at the start of 2009, dropping within the anticipated qBC efficiency factor bounds of $+/-$3\% by mid-2010.

\begin{figure}[htpb]
	\centering
	\includegraphics[width=0.98\textwidth]{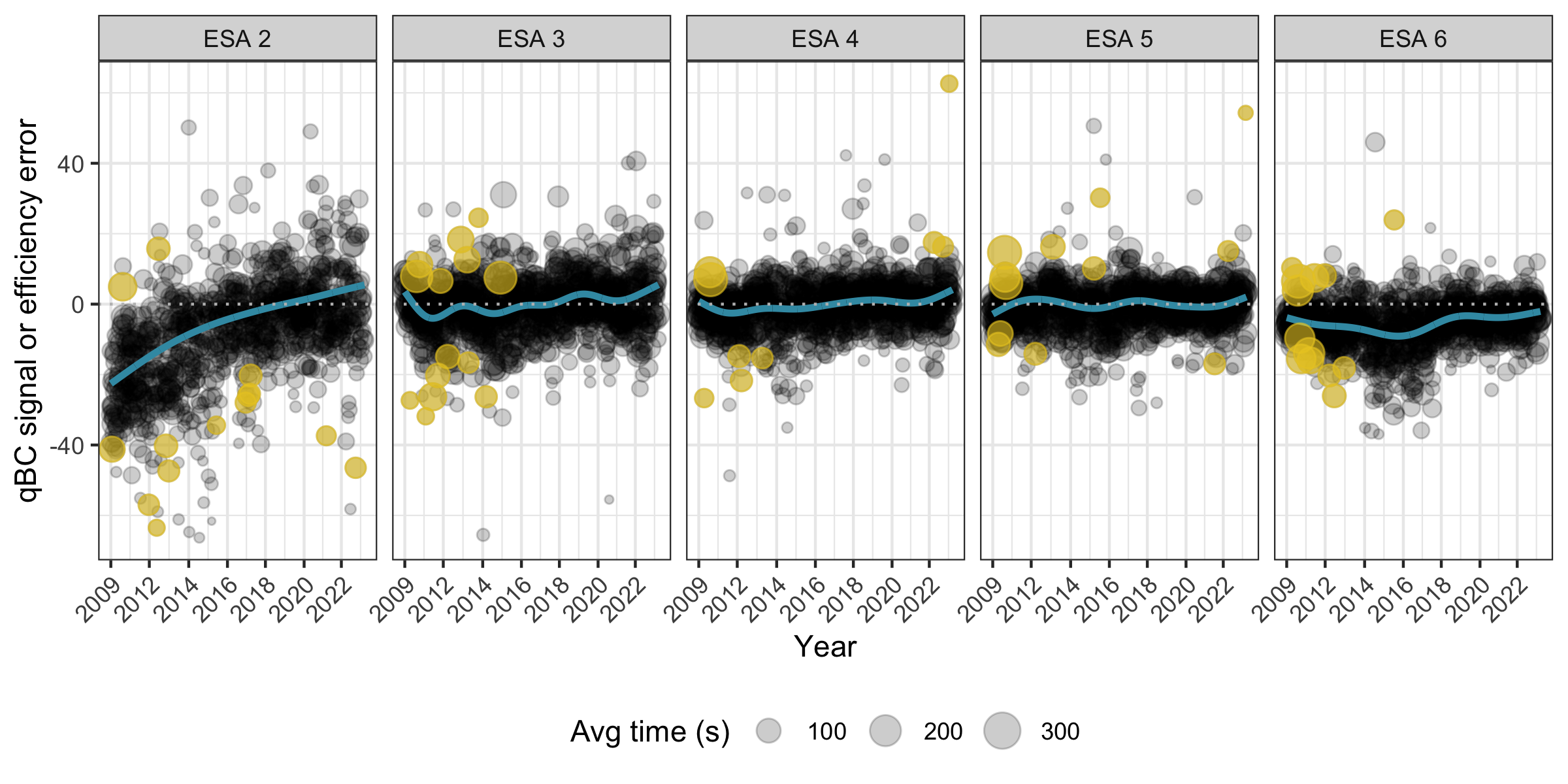}
	\includegraphics[width=0.98\textwidth]{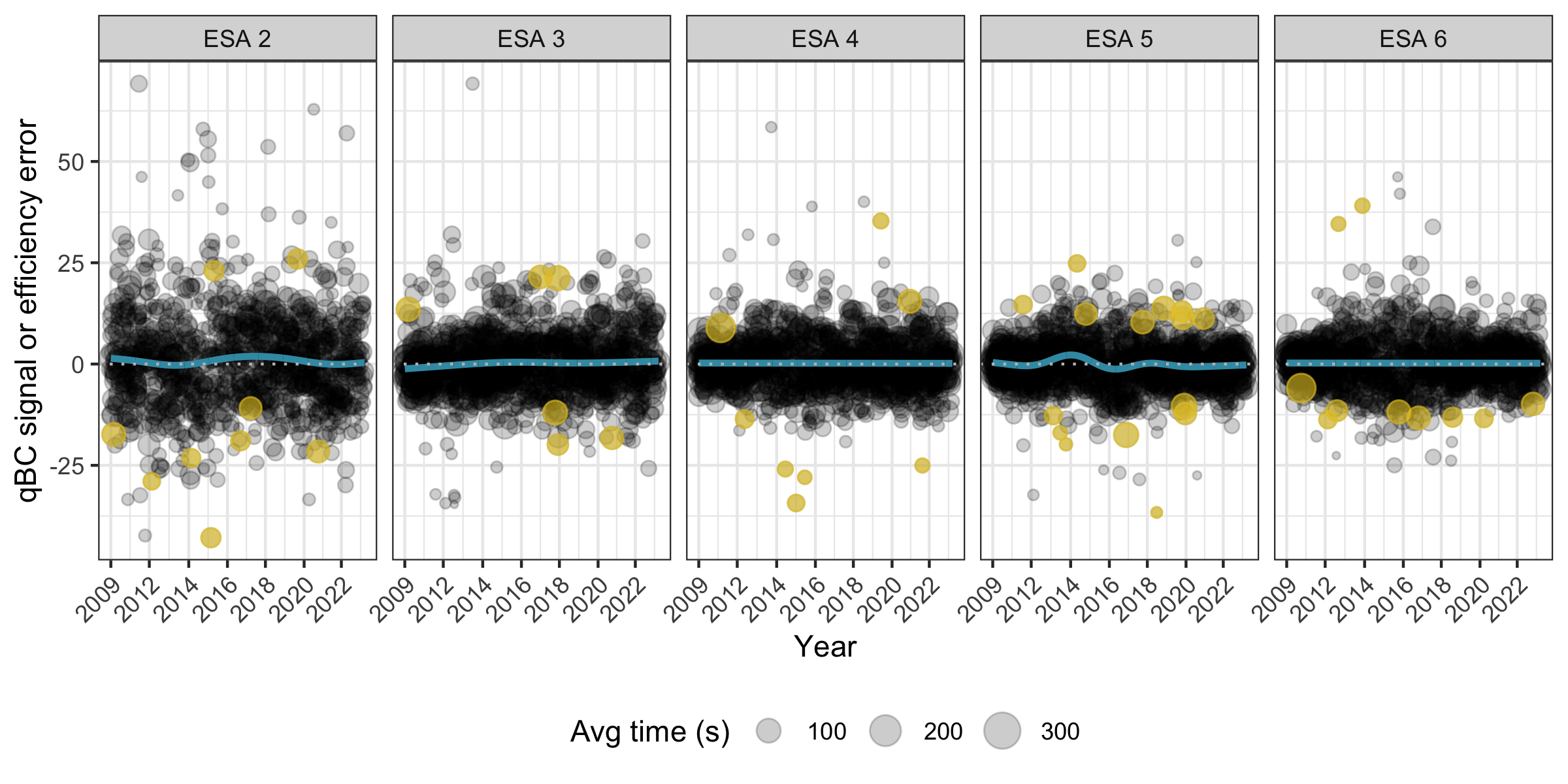}
	\caption{\small Each point corresponds to a single \textit{observed} (top) or \textit{simulated} (bottom) ESA-orbit and shows the identified ``best'' qBC signal rate (analogously, efficiency factor)  adjustment, i.e., that minimizing a GAM fit to the CvM statistic across adjustments. The temporally fitted best adjustment is shown as a blue line, and the flagged ``ESA-orbits of concern'' are shown as colored yellow points. The size of each point corresponds to the average exposure time in that ESA-orbit.}\label{fig:serr_fit}
\end{figure}

Figure \ref{fig:serr_fit} shows the ESA-specific temporal qBC signal rate (analogously, efficiency factor) error learned via a GAM fit to the CvM statistic across adjustments $\rho$ as well as the flagged ``ESA-orbits of concern'' for both the synthetic and mission data under the assumption that \textit{all} error lies in the the qBC efficiency factor (analogously, in the shared signal assumption). At worst, the learned error for the IBEX-Hi data reaches -22.6, 4.8, 4.8, -3.4, and -9.1\% for ESAs 2 through 6, respectively. 
The fit is consistent to a under half a percent across choice of model for all but ESA 2, with a LOESS fit yielding worst fitted errors of -25.8, 5.1, 5.1, -2.9, and -9.6\% for ESAs 2 through 6, respectively.
In the simulation-based run of this assessment, meant to diagnose how much error one may introduce via the fitting process itself, the learned absolute error stays under 2\% across all ESAs and is on average 0.5\%.
We find that 1.2\% of observed ESA-orbits should be flagged as potential ``ESA-orbits of concern'' for the science team to investigate under the signal rate deviation assumption, based on a 99\% quantile threshold learned via simulation.

\clearpage

%%--------------------------------------------------------------------------------------------------
%%--------------------------------------------------------------------------------------------------

\section{Improvement to map estimation}\label{supp:res_improve}

\begin{figure}[htb!]
\centering
\includegraphics[width=0.82\textwidth]{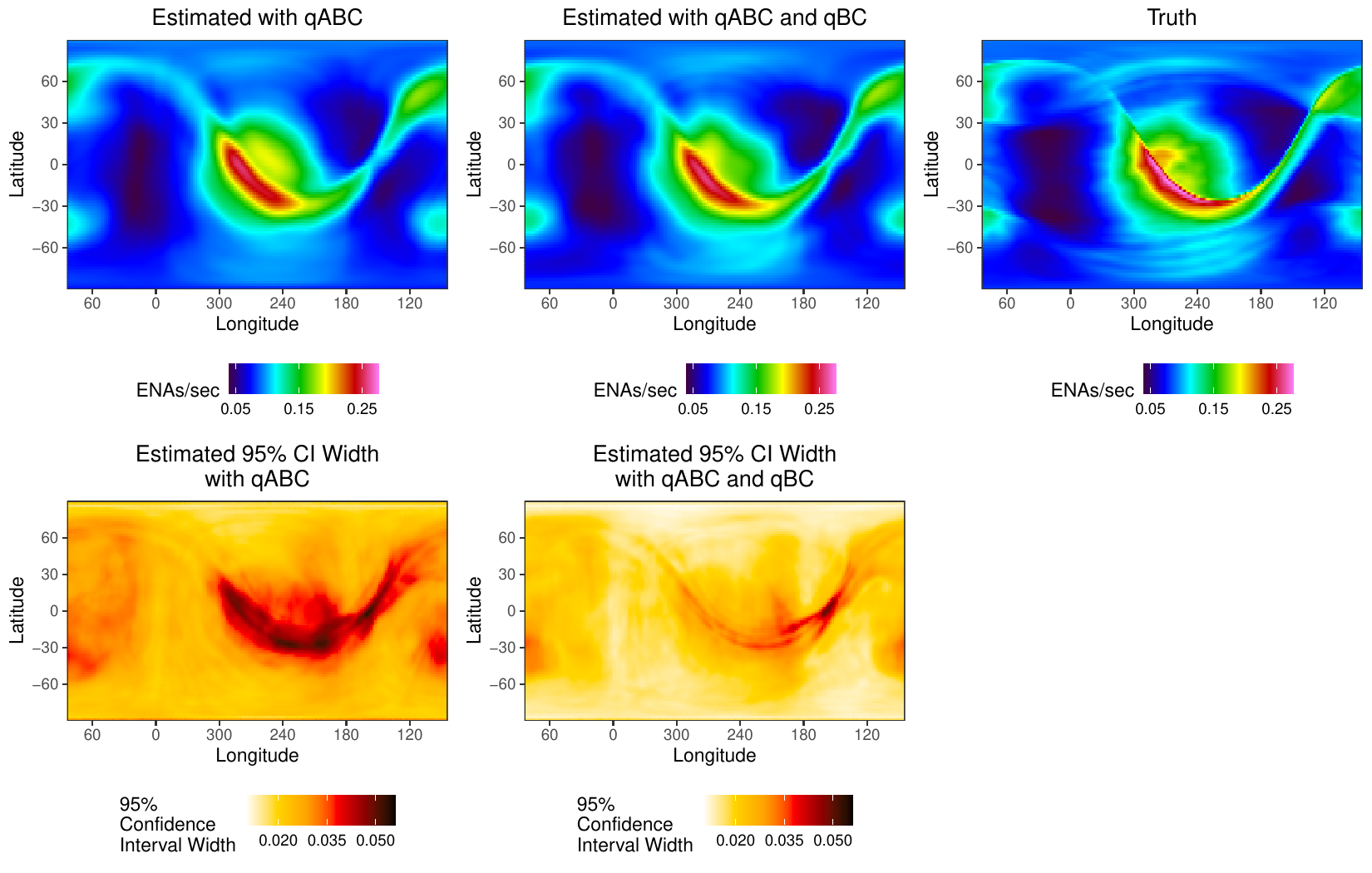}
\caption{\small Sky map estimates (top row) and 95\% confidence interval (CI) widths (bottom row) based on qABC data only (left column) and qABC and qBC data (middle column) simulated from a spatial retention model (right column). The estimated map with qABC and qBC data has better agreement with the simulated map and less uncertainty.}\label{fig:spatialretention}
\end{figure}

We consider simulated binned qABC and qBC direct event data (generated based on an ENA map derived from theoretical heliosphere and ribbon models \cite{zirnstein2018weak, zirnstein2019strong}). For each binned data set, we generate an estimated sky map using the Theseus procedure described in \cite{osthus2022towards}. Figure \ref{fig:spatialretention} plots the estimated sky map using qABC data only, the estimated map using both qABC and qBC data, and the simulated true spatial retention model sky map. Uncertainty maps are also shown.

The mean absolute percent error (MAPE) can be used to measure how well an estimated sky map agrees with a simulated true sky map. 
The MAPE for the estimated sky map using only qABC data is 6.7\% (meaning, on average, an estimated pixel's ENA rate deviates from the corresponding simulated pixel's ENA rate by 6.7\% of the simulated pixel), while the MAPE for the estimated sky map using both qABC and qBC data is 4.8\%. 
Said another way, the sky map estimated using qABC and qBC data is \emph{40\% more accurate} than using qABC data alone.
Furthermore, the 95\% confidence interval (CI) widths for the sky map estimated with qABC and qBC data are on average 40\% narrower compared to the estimated CI widths using qABC data only.
Thus, estimating sky maps with qABC \emph{and} qBC data result in better sky map estimates: they are more accurate and have smaller uncertainties.

\begin{figure}[htb!]
\centering
\includegraphics[width=0.82\textwidth]{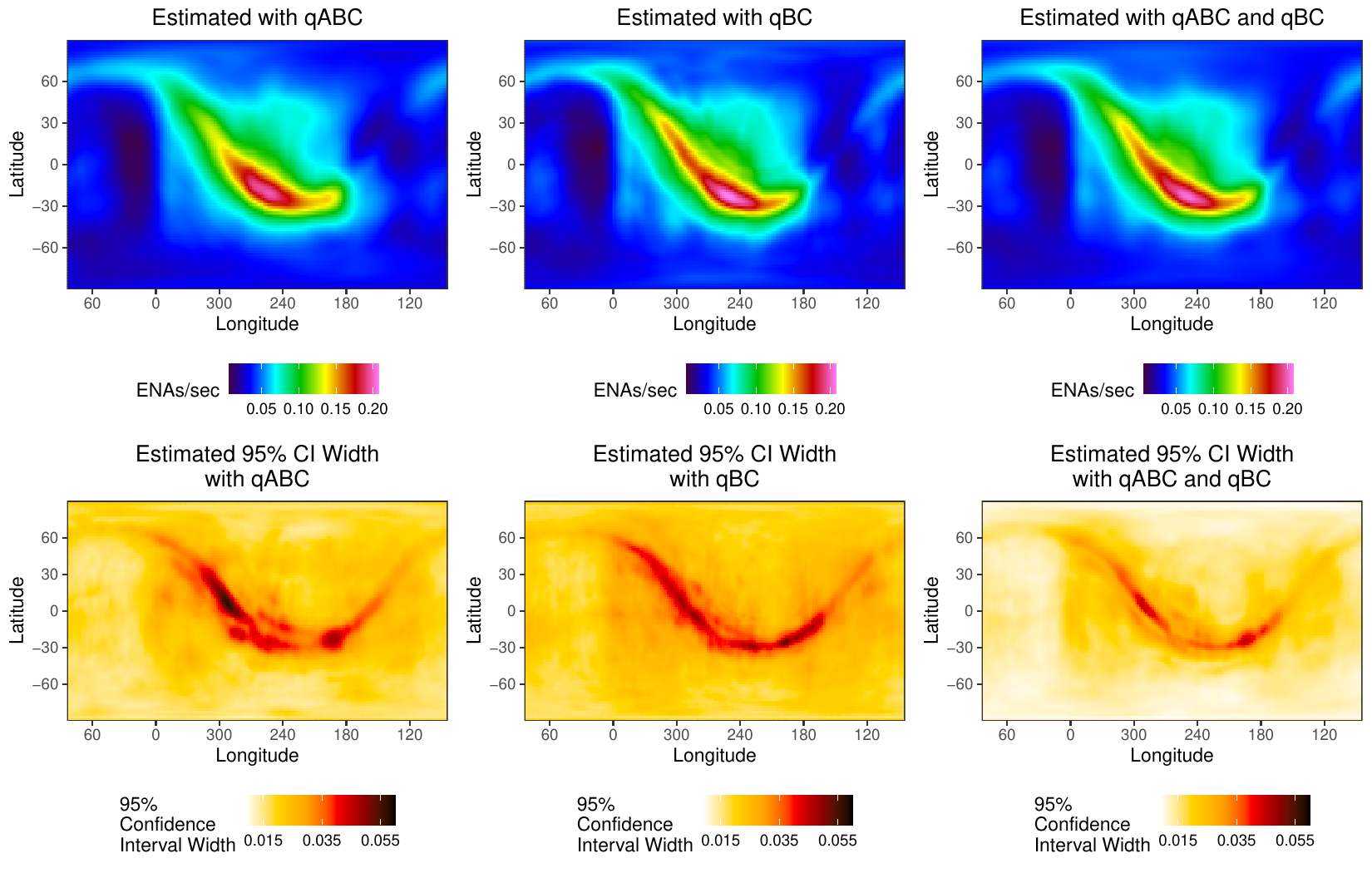}
\caption{\small 2019A, ESA 3 sky map estimates (top row) and 95\% confidence interval (CI) widths (bottom row) based on qABC data only (left column), qBC data only (middle column), and qABC and qBC data (right column). The estimated sky map with qABC and qBC data has fewer spurious visual features and has narrower 95\% CI widths.}\label{fig:realskymap}
\end{figure}

This improvement doesn't just hold for simulated data.
Figure \ref{fig:realskymap} shows the Theseus-estimated 2019A, ESA 3 sky maps based on qABCs only, qBCs only, and qABCs and qBCs. 
First, we notice that the estimated sky maps with qABCs only and qBCs only are quite similar to one another, indicating that the qABC and qBC data sets contain similar information (if they didn't contain similar information, we would expect the sky map estimates to look systematically different from each other). 
Second, while all three sky maps look broadly similar, we do see the sky map estimated with qABCs and qBCs removes some of the spurious artifacts (e.g., banding) that can be seen in the other two.
This is a because including more data in the map-rendering process helps the estimation procedure better discern between signal and noise.
Finally, the qABC \textit{and} qBC estimated sky map has CI widths that are, on average, 39\% narrower than their qABC \textit{or} qBC estimated counterparts. 
This 39\% reduction in uncertainty is almost the same as we observed with the simulated spatial retention model, bolstering confidence that the estimated sky map with qABCs and qBCs is not just more certain, but also better than the estimated sky maps made with only qABC data or qBC data.

%%--------------------------------------------------------------------------------------------------
%%--------------------------------------------------------------------------------------------------

\clearpage

%\newpage
%\bibliographystyle{plainnat}
%\bibliography{references.bib}

\end{document}